\documentclass[review,3p]{elsarticle}

\journal{Information Sciences}

\usepackage{amssymb}
\usepackage{amsmath}
\usepackage{algorithm}
\usepackage{algorithmic}
\usepackage[utf8x]{inputenc}
\usepackage{graphicx}
\usepackage{multirow}
\usepackage{url}
\usepackage{color}
\usepackage{subfigure}
\usepackage{epstopdf}

\newtheorem{example}{Example}

\usepackage{xspace} 

\newcommand{\csa}{$\textrm{CSA}$\xspace}
\newcommand{\icsa}{$\textrm{iCSA}$\xspace}
\newcommand{\tgcsa}{$\textrm{TGCSA}$\xspace}
\newcommand{\tgcsavvt}{$\textrm{TGCSA-3R}$\xspace}
\newcommand{\tgcsavb}{$\textrm{TGCSA-VB}$\xspace}

\newcommand{\cet}{$\textrm{CET}$\xspace}
\newcommand{\edgelog}{$\textrm{EdgeLog}$\xspace}
\newcommand{\iwt}{$\textrm{IWT}$\xspace}

\newcommand{\rank}{\mathsf{rank}}

\newcommand{\offprime}{\mathsf{off'}}
\newcommand{\offzero}{\mathsf{off_0}}
\newcommand{\offone}{\mathsf{off_1}}

\newcommand{\huffrle}{\mathsf{huff}\textrm{-}\mathsf{rle}\textrm{-}\mathsf{opt}}
\newcommand{\vbyte}{\mathsf{vbyte}\textrm{-}\mathsf{rle}}
\newcommand{\vbyteselect}{\mathsf{vbyte}\textrm{-}\mathsf{rle}\textrm{-}\mathsf{select}}

\newcommand{\ptrprime}{\mathsf{ptr'}}
\newcommand{\ptrzero}{\mathsf{ptr_0}}
\newcommand{\ptrone}{\mathsf{ptr_1}}

\newcommand{\vbytegapsm}{$\mathsf{vbytegaps}$}

\newcommand{\pfordelta}{\mathit{PForDelta}}
\newcommand{\rice}{\mathit{Rice~codes}}
\newcommand{\simple}{\mathit{Simple16}}

\newcommand{\no}[1]{}

\usepackage{todonotes}

\newcommand{\activeEdge}{\mathsf{activeEdge}}
\newcommand{\directNeighbor}{\mathsf{directNeighbor}}
\newcommand{\reverseNeighbor}{\mathsf{reverseNeighbor}}

\newcommand{\snapshot}{\mathsf{snapshot}}

\newcommand{\activedEdge}{\mathsf{activatedEdge}}
\newcommand{\deactivedEdge}{\mathsf{deactivatedEdge}}

\newcommand{\rangereport}{\mathsf{rangeReport}}

\newenvironment{code}
{\scriptsize \begin{tabbing}
\rule{\textwidth}{0.5mm} \\
xxxx\=xxxx\=xxxx\=xxxx\=xxxx\=xxxx\=xxxx\=xxxx \kill}
{\rule{\textwidth}{0.5mm}
\end{tabbing}}

\newenvironment{codebottom}
{ \scriptsize \begin{tabbing}
xxxx\=xxxx\=xxxx\=xxxx\=xxxx\=xxxx\=xxxx\=xxxx \kill}
{\rule{\textwidth}{0.5mm}
\end{tabbing}}

\newenvironment{codeLarge}
{\footnotesize \begin{tabbing}
xxxx\=xxxx\=xxxx\=xxxx\=xxxx\=xxxx\=xxxx\=xxxx \kill}
{ 
\end{tabbing}}

\begin{document}
\begin{frontmatter}	
\title{Using Compressed Suffix-Arrays for a Compact Representation of Temporal-Graphs\tnoteref{t1}}
\tnotetext[t1]{\footnotesize Funded in part by European Union's Horizon 2020 research and innovation programme
     under the Marie Sklodowska-Curie grant agreement No 690941 (project BIRDS).
  D. Caro is partially funded by the Chilean government initiative CORFO
13CEE2-21592 (2013-21592-1-INNOVA PRODUCCION).
  M. A. Rodr\'\i guez is partially funded by Fondecyt [1170497] and the Complex Engineering Systems Institute (CONICYT: FBO16).
  N. R. Brisaboa and A. Fari\~na were partially funded 
   by Xunta de Galicia/FEDER-UE [CSI: ED431G/01 and GRC: ED431C 2017/58]; 
   by MINECO-AEI/FEDER-UE [Datos 4.0: TIN2016-78011-C4-1-R and ETOME-RDFD3: TIN2015-69951-R]; and
   by MINECO-CDTI/FEDER-UE [CIEN: LPS-BIGGER IDI-20141259 and INNTERCONECTA:  uForest ITC-20161074].
   An early partial version of this article appeared in {\em Proc. SPIRE'14}  \cite{Brisaboa:spire2014}.}

	\author[udc]{Nieves R. Brisaboa}
	\ead{brisaboa@udc.es}

	\author[udp,tid]{Diego Caro}
	\ead{dcaro@udd.cl}

	\author[udc]{Antonio Fari\~na\corref{cor1}}
	\ead{fari@udc.es}

	\author[udec]{M. Andrea Rodriguez}
	\ead{andrea@udec.cl}

	\cortext[cor1]{Corresponding author}

	\address[udp]{Data Science Institute, Faculty of Engineering, Universidad del Desarrollo, Chile.\\}
	\address[tid]{Telefónica I+D Fellow, Chile\\}
	\address[udc]{Database Laboratory, University of A Coru\~na, Spain.\\}
	\address[udec]{Department of Computer Science, University of Concepci\'on, Chile.}		


\begin{abstract}

Temporal graphs represent  binary relationships that change along time. 
They can model the dynamism of, for example, social and communication networks. 
Temporal graphs are defined as  sets of contacts that are edges tagged with the 
temporal intervals when they are active. 
This work explores the use of the Compressed Suffix
Array (\csa),  a well-known compact and self-indexed
data structure in the area of text indexing, to represent large temporal graphs. The new structure,  called
Temporal Graph \csa (\tgcsa), is experimentally compared with the most
competitive compact data structures  in the state-of-the-art, namely,  \edgelog and \cet. 
The experimental results show that
\tgcsa obtains a good space-time trade-off. It uses a 
reasonable space and is efficient for solving 
complex temporal queries. Furthermore, \tgcsa has wider
expressive capabilities than \edgelog and \cet, because it is able  to
represent temporal graphs where contacts on an edge can temporally overlap. 

\end{abstract}


\end{frontmatter}

\section{Introduction}
\begin{figure}[htbp]
\begin{center}
\includegraphics[scale=0.7]{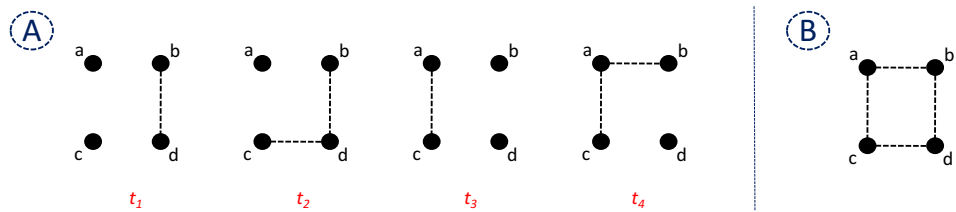}
\caption{A temporal graph composed of four vertexes with a lifespan of four time-instants: a) A snapshot-based representation
showing the available edges per time-instants, b) the time-aggregated graph of the temporal graph.}
\label{fig:timepath}
\end{center}
\end{figure}

\no{
Unlike static graphs, temporal graphs model real networks that exhibit a dynamic behavior,
that is, their edges can appear and disappear over time (see Figure \ref{fig:timepath}).
For example, consider the evolution of friendship relations when a user
adds or removes friends in online social networks,  or how
the network connectivity evolves when mobile devices
change their base station along time.
\no{ how the citation network grows when new scientific articles are published, 
or how links appear and disappear on the Web graph. }
\no{Even more, there is an increasing need in real applications to handle
large graphs  that exhibit changes along time and for which not only
the current state but also the past state is of interest. }
}

The main assumption of static graphs is that the relationship between two vertexes is always available.
However, this is not true in many real world situations. For example, consider how  friendship relations evolve in an online social network, or how
the connectivity in a communication network changes when users, with their mobile devices, move in a city. 
Temporal graphs deal with the time-dependence of relationships between vertexes by representing these relationships as a set of {\em contacts}~\cite{Nicosia:2013td}.
Each contact represents an edge (i.e., two vertexes) tagged with the 
time interval when the edge was active. For example, in a communication network, a contact
may represent a call between  users made from 4 pm to 4.05 pm.

%
The temporal dimension of edges adds a new constraint to the relationship between vertices not found in static graphs:
two vertexes can communicate only if there is a time-respecting path (also called journeys~\cite{Nicosia:2013td}) 
between them~\cite{Nicosia:2013td,Sizemore:2017ut,Wu:2014:PPT:2732939.2732945,Tang:2013gn,Holme:2012jo}.
For example, in Figure~\ref{fig:timepath}.b  (corresponding to the time aggregation of the edges in the temporal 
graph of Figure \ref{fig:timepath}.a), there are
two paths connecting the vertexes $a$ and $d$: one through the vertex $b$, and the other one through $c$. 
However, there is no such path when considering the temporal availability of the edges $(a,b)$ and $(a,c)$. 
Notice that the vertexes $b$ and $c$ are only reachable from the vertex $a$ because the edges reaching $d$ are not available.
Therefore, taking into account the temporal dynamism of  graphs
allows us to exploit information about temporal correlations and
causality, which would be unfeasible through a  classical analysis
of static graphs~\cite{Nicosia:2013td,Holme:2012jo,Michail:2016ed}.

\no{The alternative change-based approach \cite{Ferreira:2002vg} 
stores a log of time instants showing
when  and which relations between vertexes change from inactive to
active or vice versa. This approach has the advantages of storing
only what changes between time instants instead of storing the whole set of active edges
of the graph (snapshot) at each time instant. }

A direct approach to represent temporal graphs could be a
time-ordered sequence of snapshots (Figure \ref{fig:timepath}a), one for each time instant, showing
the state of the temporal graph at a time instant as a static graph.
Several centralized and distributed processing systems follow this approach 
(e.g. Pregel~\cite{Malewicz:2010:PSL:1807167.1807184}, 
Giraph\footnote{\url{http://giraph.apache.org/}}, 
Neo4J\footnote{\url{http://neo4j.org/}}, Trinity~\cite{Shao:2013:TDG:2463676.2467799}), but
without specific support for temporal extensions~\cite{Kosmatopoulos2016181}.

In temporal graphs where contacts are active during long time intervals (as in a social network), 
consecutive snapshots tend to become very similar. Thus, strategies based on a sequence 
of snapshots are space consuming because edges are duplicated in each snapshot.
An alternative change-based approach represents the temporal graph by the differences between snapshots; that is, 
by the set of edges that appear/disappear along time.
These differences can be calculated with respect to consecutive snapshots~\cite{Ferreira:2002vg}, or with 
respect to a derived graph that diminishes the number of stored edges~\cite{Ren:2011uk,Khurana:2012vn,Labouseur:2013vv,7498269}.

The change-based approach has also been used for pre-computing reachability queries~\cite{Semertzidis2016,7498269}, as
some paths may remain available for several time instants~\cite{Bannister:2013ul}.
Although these works improve the time performance of complex algorithms, they overlook the space cost, which becomes crucial for large temporal graphs. In this context, a compact representation can
keep larger sections or even the whole temporal graph in memory and,
in consequence, queries could become much more efficient by  avoiding disk transfers.

\no{There exist some proposals of data structures for temporal graphs
based on adjacency lists~\cite{Xuan:2002ts} and in distributed
environments~\cite{Khurana:2012vn,Labouseur:2013vv},  but they aimed at  improving 
the time performance for complex algorithms on
temporal graphs, overlooking their space cost. For large temporal graphs, however, 
space is crucial. In this context, a compact representation can
keep larger sections or even the whole temporal graph in memory and,
in consequence, queries could become much more efficient by  avoiding disk transfers.}

Recently, some compact approaches to represent  temporal graphs  have been proposed \cite{Caro:2015fa,DBLP:journals/is/CaroRB15}. 
%
The work in~\cite{Caro:2015fa} presents the $ck^d$-$tree$, a tree-shaped compact data 
structure based on the Quadtree~\cite{DBLP:journals/csur/Samet84}, which represents a temporal graph as a point in a four dimensional space. 
This data structure was designed to reduce space usage at the expense of time access in sparse temporal graphs.
\edgelog (Time Interval Log per Edge)~\cite{DBLP:journals/is/CaroRB15}
uses a compressed inverted index, which also provides fast answers to different  types of queries,
in particular,  when solving  adjacency queries involving the recovery of active neighbors of a vertex at a specific time instant.
\cet (Compact Events ordered by Time)~\cite{DBLP:journals/is/CaroRB15} uses a wavelet tree \cite{Navjda13,Grossi:2003vf} 
to represent temporal graphs and
is the best alternative in the state-of-the-art to answer queries
related to  time-instant events that change the state of an edge. 

\no{In \cite{DBLP:journals/is/CaroRB15}, \edgelog 
and \cet were presented and compared with other alternative structures, showing 
their good properties both in space usage and efficiency for a wide range of queries. }

Both \edgelog and \cet overcome the overload of storing a snapshot per each time instant 
by representing the temporal graph as a log of events. These events indicate when edges become active or inactive.
Then, the activation state of a given edge can be recovered by counting how many events occurred on that edge during a time interval.
If there is an even number of events, it means that the edge has been active and inactive several times.
Conversely, if the edge has an odd number of events, it means that the last state of the edge is active.
A detailed explanation of these data structures is available in Section~\ref{sec:preliminary}.

A main drawback of the log-based structures,  such as  \edgelog and \cet, is that they do not
allow the
representation of time-overlapping contacts of an edge. 
For example, if a contact represents the data communication between two machines $X$ and $Y$ during a 
time interval,  it is impossible to represent a second contact between $X$ and $Y$ during an 
overlapping time interval. This limitation arises because in these structures the event that represents the activation of the second
contact would be interpreted as the deactivation event of the first contact.

\no{of the graph, knowing as the snapshot or copy strategy~\cite{Ferreira:2002vg} and, 
therefore, they provide a fair comparison to evaluate a new structure for temporal graphs.}

\no{This definition implies that different contacts 
exist for a same edge that is active at different time intervals. In a general scenario, 
time intervals can overlap; however, a usual assumption is that an edge of the graph does not 
have contacts whose temporal intervals overlap. 
For example, if a contact represents a phone communication between $X$ and $Y$ during a 
time interval,  it is impossible to have another contact between $X$ and $Y$ during an 
overlapping time interval. 
However, in a general scenario where nodes are not individuals who communicate but nodes 
of a communication network, there may exist different calls from one to another node of the 
network at the same time.  A temporal graph is then a set of 4-tuples  of
the form $(u, v, t_s, t_e)$, indicating that an
edge from a vertex $u$ to a vertex $v$ is active during the time
interval $[t_s, t_e)$. As we will see later, for previous data structures, a basic assumption 
is that contacts for the same edge cannot temporally overlap.
}

The work in this paper presents and evaluates a data structure
named Temporal Graph \csa (\tgcsa). The \tgcsa is a compact and self-indexed
structure based on a modification of the well-known Compressed Suffix
Array (\csa)\cite{Sad03}, extensively used for text indexing. 
We focus on algorithms to process temporal-adjacency queries that recover
the set of active neighbors of a vertex at a given time instant. These queries  
are  basic blocks to solve  time-respecting paths~\cite{Michail:2016ed}, 
which
 can be useful in the context of moving-object 
data~\cite{DBLP:conf/kdd/MamoulisCKHTC04,DBLP:conf/gis/KroghPTT14}, 
and also when analyzing activity patterns as  temporally ordered sequences of 
actions  occurring at specific time instances or time 
intervals~\cite{DBLP:journals/corr/LiuNHZR16,DBLP:journals/ijon/LiuNLR16}. 

We also present algorithms for answering queries that recover the snapshot of the graph
at a time instant, as well as queries to recover the state of single edges. In addition, we include
a complete experimental evaluation with real
and synthetic data that compares \tgcsa with \edgelog and \cet  in terms of both  space and time usage. 
The results of this evaluation show that \tgcsa opens new opportunities for the
application of suffix arrays \cite{MM93,Sad03} in the
context of graphs in general, and of temporal graphs in particular.
\medskip

As discussed above, there are different fields where the application of our \tgcsa, or other compact existing alternatives from the 
state of the art such as \edgelog or \cet, can be of interest. Among others, we can mention~\cite{DBLP:conf/sbp/ShmueliAP14,doi:10.1093/bioinformatics/btv227}: (i) Social networks, where friendships establish
connections between nodes that can vary along time. (ii) Biological networks, where function brain connections are dynamic.
(iii) Communication networks, where nodes are connected while their exchange information.
This applies to person-to-person and machine-to-machine communication. (iv) Transportation networks, where the connectivity between nodes  can change due to scheduling and traffic conditions. In this context, one could also model movements on a network by considering that two nodes are connected  if there exists an object that moves from one to the other node during a time interval. 

The structure of this paper is as follows.
Section~\ref{sec:preliminary} presents preliminary concepts about
temporal graphs and  relevant queries on them. To make the paper self-contained, 
Sections~\ref{sec:edgelog} and \ref{sec:cet} provide a brief overview  of
both  \edgelog  and \cet. These are the state-of-the-art techniques we compare \tgcsa with.   
Section~\ref{sec:TGCSA} introduces  \tgcsa by showing how to modify a traditional \csa\ to
create  \tgcsa. It also describes how \tgcsa  solves  relevant queries for temporal
graphs and provides pseudocode for such operations. Finally, this section presents
a new representation of the $\Psi$ array from \csa\cite{GV00,FBNCPR12}, called in this work
$\vbyte$, which increases  the query performance of \tgcsa. 
Section~\ref{sec:experiments} provides the experimental evaluation that uses
real and synthetic data. Final conclusions and future research
directions are given in Section~\ref{sec:conclusions}.

\section{Preliminary concepts} \label{sec:preliminary}
In this section we introduce temporal graphs and  a classification
of the relevant basic queries that could be of interest for most
applications. We also revise previous compact
representations of temporal graphs.
%


\subsection{Temporal graph definition}
Formally, a temporal graph is a set $\mathcal{C}$ of contacts that connect
 pairs of vertexes in a set $V$ during a time interval defined over the set   $\mathcal{T}$ that represents the \textit{lifetime} of the graph. A
{\em contact} in  $\mathcal{C}$  of an edge $(u,v) \in E \subseteq V \times V$ is a 4-tuple
$c = (u,v,t_s,t_e)$, where $[t_s,t_e) \in \mathcal{T} \times
\mathcal{T}$ is the time interval when the edge $(u,v)$ is
active~\cite{Nicosia:2013td}. We say that an edge $(u,v)$ is
\textit{active} at time $t$ if there exists a contact $(u,v,t_s,t_e)
\in \mathcal{C}$ such that $t \in [t_s,t_e)$.
Note that this definition applies for directed graphs as we consider ordered pairs of vertexes.

\begin{figure}[htbp]
\centering

\subfigure[Set of contacts]{\includegraphics[width=0.35\textwidth]{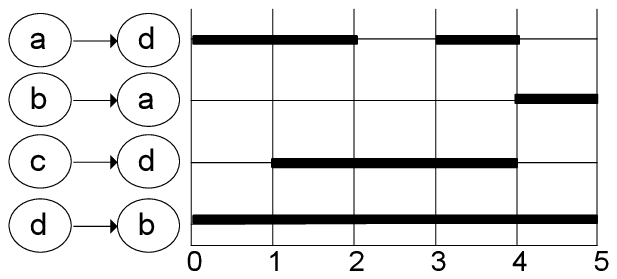}}
\hspace{1.5cm}
\subfigure[\edgelog\ representation]{\includegraphics[width=0.33\textwidth]{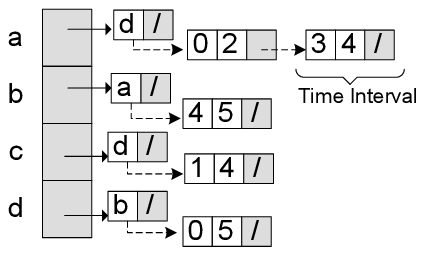}}

\caption{A temporal graph of $4$ vertexes and its \edgelog representation. The reverse aggregated graph is omitted in (b). }
\label{fig:tvgraph}
\vspace{-0.3cm}
\end{figure}

We classify operations on temporal graphs into two categories: queries for checking the connectivity between vertexes 
and queries for retrieving the changes on the connectivity occurred along time.
For the first category of queries, we define four operations: (1) $\activeEdge$ checks if an edge is active. 
(2) $\directNeighbor$ returns the active direct neighbors of a vertex. 
(3) $\reverseNeighbor$ gives the active reverse neighbors of a vertex. 
(4) $\snapshot$ returns all the active edges. 
For example, in the temporal graph of Figure~\ref{fig:tvgraph}.a, we know that at time instant $t=1$ the edge $(a,d)$ is active, 
the set of direct neighbors of $c$ is $\{d\}$ and the set of reverse neighbors of $d$ is $\{a,c\}$; whereas
the snapshot at  time $t=3$ corresponds to the edges $\{(a,d), (c,d), (d,b)\}$. 

For queries retrieving the changes on connectivity, we defined two operations:
(1) $\activedEdge$ returns the set of edges that were  activated. 
(2) $\deactivedEdge$ returns the set of edges that were deactivated.
For example, given Figure~\ref{fig:tvgraph}.a at time instant $t=4$, the edge $\{(b,a)\}$ was activated, and the
edges $\{(a,d),(c,d)\}$ were deactivated.

Note that all previous queries have a time-instant or a time-interval version. In what follows, we concentrate on time-instant queries, which can be 
easily extended to answer time-interval queries, and they also serve 
as the building blocks for more complex temporal measures that are based on recovering time-respecting paths~\cite{Michail:2016ed}.

\subsection{\edgelog: Baseline representation} \label{sec:edgelog}

A simple temporal graph representation~\cite{Xuan:2002ts} stores the
aggregated graph\footnote{The static graph including all the edges that
were active at any time during the lifetime of the temporal graph.}
as $|V|$ adjacency lists, one per each vertex, with
a sorted list of time intervals attached to each neighboring vertex indicating
when that edge is/was active. Figure~\ref{fig:tvgraph}.b shows a conceptual example.

\no{A simple mechanism for representing temporal graphs is to store the
aggregated graph as $|V|$ adjacency lists, one per vertex, with
a sorted list of time intervals attached to each neighbor indicating
when that edge is active~\cite{Xuan:2002ts}. An example of
this basic structure is conceptually represented in Figure~\ref{fig:tvgraph}.b.
}



    \no{To check if an edge $(u,v)$ is active at time $t$, the first step is
    to check if $v$ appears on the adjacency list of vertex $u$. If the
    edge is found, then we need to look at the time-point list of that edge to check if $t$ falls into one of the
    time intervals related to $(u,v)$. Note that the edge will be active at
    time $t$ if there exist two consecutive time points in the list,
    $t_i$ and $t_j$ such that $ t_i ~\leq~ t ~<~t_j$ and $t_i$ is at
    an odd position.
    }

To check if an edge $(u,v)$ was active at time $t$, we first check if $v$ appears 
within the adjacency list of vertex $u$. If $v$ is found, then we need to check if $t$ falls into one of the
time intervals related to $(u,v)$ that are represented in the
time-interval list of that edge. 
\no{Note that the edge is active at
time $t$ if there exist two consecutive time points in the list,
$t_i$ and $t_j$ such that $ t_i ~\leq~ t ~<~t_j$ and $t_i$ is at
an odd position.}
Direct neighbors of vertex $u$ at time $t$ are
recovered similarly. For each neighbor $v$ in the adjacency list of $u$, we check if $t$ is within the time intervals of the edge $(u,v)$.

\no{
  This simple representation has two main drawbacks: {(1)} it could use much space; 
  and {(2)} the reverse
  neighbors operations require traversing all adjacency lists. Both issues
  are overcome in what we call \edgelog. It compresses
  adjacent vertexes of the aggregated graph and time-interval lists
  using usual inverted list representations. Also, to avoid
  traversing all adjacency lists in reverse-neighbor queries, it
  stores the reverse aggregated graph containing  adjacency
  lists with the reverse neighbors of each vertex.
}

A simple representation of the aggregated graph and the temporal labels attached to vertices
 has two main drawbacks: {(1)} it uses much space; 
and {(2)} operation $\reverseNeighbor$ 
requires traversing all the adjacency lists. The data structure \edgelog~\cite{DBLP:journals/is/CaroRB15} 
addressed these  weaknesses. On the one hand, since both the adjacency list and 
the time-interval list are sorted (i.e., they are
of the form $\langle t_1,t_2, t_3, ..., t_l \rangle$, with $t_i <
t_{i+1}$), they can be represented as d-gaps  $\langle t_1,
t_2-t_1, t_3-t_2, ..., t_l-t_{l-1} \rangle$, and those differences can be compressed using a variable-length
encoding  (e.g.,
$\pfordelta$~\cite{DBLP:conf/icde/ZukowskiHNB06},
$\simple$~\cite{DBLP:conf/www/ZhangLS08}, $\rice$ \cite{WMB99}).
On the other hand, to avoid
traversing all the adjacency lists in $\reverseNeighbor$ queries, \edgelog\
stores a reverse aggregated graph containing an adjacency
list with all the reverse neighbors of each vertex. Therefore, to get the reverse
neighbors of vertex $v$ at time $t$,  we first use the reverse adjacency list
to obtain the candidate reverse neighbors of $v$. Then, for each candidate 
reverse neighbor $u$, we search for $v$ in its adjacency list and, finally, 
check if the edge $(u,v)$ is active at time $t$ (using the time-interval list of the edge).

\no{
As time on reverse neighbor operations are improved storing the
reverse aggregated graph, we just need to check the time-interval
lists of the edges that could be active at the query time, instead
of looking at all the adjacency lists. For example, to get the reverse
neighbors of vertex $v$ at time $t$,  we first need to retrieve the
possible reverse neighbors using the inverted adjacency list. Then,
we decompress the adjacency list of the possible neighbors and we
check if $t$ belongs to the time-interval list of the possible
reverse neighbors.

To obtain the state of and edge $(u,v)$ at time $t$, we need to
reconstruct the adjacency list of $u$ and the time-interval list of
$(u,v)$. Since the time-interval lists related to vertex $u$ are all
concatenated, we need to find the range of the list related to the
edge $(u,v)$. This can be done obtaining the $i$-th position  of
neighbor $v$ in the adjacency list. Then the range of time intervals
related to the edge $(u,v)$ corresponds to the positions $i-1$ and
$i$ in the offset list of vertex $u$. Finally, we need to check if
$t$ falls into one of the time intervals of edge $(u,v)$.
Direct-neighbors operations follow the same idea, but for all
neighbors instead of a single edge.
} 

\subsubsection{Strengths and weaknesses of   \edgelog}

Although \edgelog\ is a simple structure using well-known technology,
it is expected to be extremely space-efficient when the temporal graph has a low
number of edges per vertex and a large number of contacts per edge. 
In the opposite way, a low number of contacts per edge will have a
negative impact on the compression obtained by \edgelog\ (as d-gaps become large).
Note also that, even with the reverse aggregated graph to find reverse neighbors, the
performance is expected to be poor if the number of edges per vertex is high
 because all their adjacency lists  will have to be checked.

\edgelog\ was designed to be efficient for $\activeEdge$, $\directNeighbor$, 
and $\reverseNeighbor$ queries,
but it could not efficiently answer  queries such as: {\em ``Find all
the edges that have active contacts at time $t$''} or {\em ``Find all the
edges that have been  active only once''}.
This is because in such operations, all the adjacency lists must
be processed.
Also, the applicability of  
\edgelog\ is limited to temporal graphs whose contacts do not temporally overlap;  that is, it assumes that a
  contact of an edge ends before another contact of the same edge starts.

\no{
  Finally, it must be pointed out that the applicability of the 
  \edgelog\ is limited to temporal graphs where
  edges can not have overlapping contacts in time; that is, it assumes that a
  contact of an edge ends before another one starts. For example, if the
  temporal graph represents when a type of product is sent and  delivered to a given customer, 
  it can happen that a second shipment of the same product may be also sent to the same customer 
  before the previous one arrives. 
  In such cases, the list of time intervals of  contacts  could not be represented by differences and encoded as
  efficient as they currently are.
}

\subsection{\cet: Compact Events ordered by Time} \label{sec:cet}
\begin{figure}[]
\centering
\includegraphics[width=0.9\textwidth]{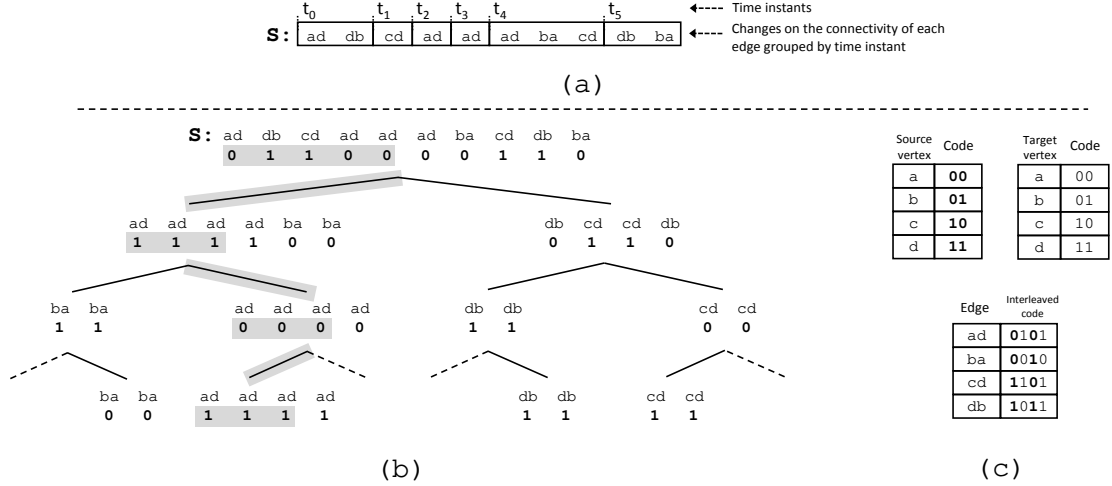}

\caption{The \cet data structure representing the temporal graph in Figure~\ref{fig:tvgraph}.a. The top part shows the sequence 
of changes $S$. The bottom-left part shows the Interleaved Wavelet Tree (\iwt) representation of $S$. The bottom-right part shows the interleaving bits used to represent pairs of symbols in the \iwt.  }
\label{fig:iwt}
\vspace{-0.3cm}
\end{figure}

In~\cet a temporal graph is a sequence of symbol pairs that represent the changes on the 
connectivity between vertexes. 
Each pair represents either the activation or deactivation of an edge along time. Note that a contact of the form $(u,v,t_s,t_e)$ generates
two changes: an activation of the edge $(u,v)$ at time $t_s$, and a deactivation at time instant $t_e$.
The sequence of pairs ($S$) is composed of the changes on the connectivity of edges (i.e., activations or deactivations produced by all the contacts
in the temporal graph)  grouped by time instant in increasing order. 
In Figure~\ref{fig:iwt}.a, we show how the sequence of changes of the temporal graph from Figure \ref{fig:tvgraph}.a is built. We can see that the first two entries of $S$ correspond to the edges $(a,d)$ and $(d,b)$ that are activated at time instant $t_0$. Next entry corresponds to the activation of the edge $(c,d)$ at time instant $t_1$. The fourth and fifth entries of $S$ are related to the edge $(a,d)$, which is deactivated at time instant $t_2$ and activated again at $t_3$, respectively. The next three entries reflect the changes produced at $t_4$ when the edges $(a,d)$ and $(c,d)$ are deactivated and $(b,a)$ is activated. Finally, the edges $(b,a)$ and $(d,b)$ are deactivated at time instant $t_5$.

The activation state of an edge at time instant $t$ is computed by {\em counting} how many times the pair encoding the edge appears
in the subsequence of changes within the time interval between $0$ and $t$ (in the closed time interval). 
As we assume that all edges are inactive at the beginning of the lifetime, the first occurrence of the pair means that the edge 
becomes active, the second occurrence means that the edge becomes inactive, and so on.
In consequence, if the pair appears an odd number of times, it means that the state  of the edge is active; otherwise, it is inactive.
For example, we can see in Figure~\ref{fig:iwt}.a that, because the pair $ad$ occurs three times within interval $[t_0,t_3)$, the edge $(a,d)$ is active at time instant $t_3$.
The direct neighbors of a vertex $u$ at time $t$ are also recovered using the counting strategy, but checking the frequency of 
the form $(u,*)$, i.e., the pairs whose first component is $u$. 
Similarly, the reverse neighbors of $v$ are obtained by counting the pairs that end with $v$.


The sequence of pairs that composes $S$ is represented in an Interleaved Wavelet Tree (\iwt)~\cite{DBLP:journals/is/CaroRB15}, a variant of the Wavelet 
Tree~\cite{Grossi:2003vf,Grossi:2011gy} capable of counting the number of occurrences of multidimensional 
symbols in logarithmic time, while keeping a reduced space.
The Wavelet Tree is a balanced binary tree, whose leaves are labeled with symbols in an alphabet 
$\Sigma$, and whose internal nodes handle a range of the alphabet. Each node of the Wavelet 
Tree represents the sequence as a bitmap with 0s and 1s, 
depending on the binary code
used to represent each symbol in the alphabet $\Sigma$. 
Figure~\ref{fig:iwt}.b shows the \iwt representation for the sequence of changes $S$ of the temporal graph in Figure \ref{fig:tvgraph}.a. 
(For more details on the Wavelet Tree and its applications, refer to \cite{Navjda13}). 
%
%


In the \iwt, the pairs of symbols in $S$ are represented by an {\em interleaved} code that is the result of interleaving the bits (Morton Code~\cite{Samet:2006va}) of the codes corresponding to the source and target vertexes of each pair. Figure~\ref{fig:iwt}.c shows the interleaved bits for the pairs (corresponding to the edges) of the temporal 
graph in Figure~\ref{fig:tvgraph}.a. Note that the symbols in pair \texttt{ad} are given the codes \texttt{\underline{00}} 
and \texttt{11} respectively. Therefore, the interleaved code for pair  \texttt{ad} is
\texttt{\underline{0}}\texttt{1}\texttt{\underline{0}}\texttt{1}, and those four bits are represented along the
wavelet tree by starting in the root node with the first \texttt{\underline{0}}. Because that bit is a zero, we move to the left child in the 
next level where we use the second bit of such code. This second bit is  \texttt{1} and appears at the first position in the bitmap.
Subsequently, we move to the right child in the next level, and use the third bit of the code, which is  the
\texttt{\underline{0}} at the first position of the bitmap. Finally, we move again to the left child of the
node and reach the last level where we set the last bit of the code of \texttt{ad}, which is  \texttt{1}.

The counting operation of a symbol  $c$ in the sequence  $S[1,i]$\footnote{For simplicity, we will use the notation $V[i,j]$ to refer to the sequence of elements $\langle V[i],\dots, V[j]\rangle$. } is translated into 
counting operations over the  bitmaps in the path of the symbol $c$.
In order to show how the counting algorithm works, let us use the operation $\rank_b(B,i)$.\footnote{Given a bitmap $B$, $\rank_b(B,i)$ computes the number of occurrences of bit $b$ in $B[1,i]$.}
The algorithm works as follows.
At the root node, if the first bit of symbol $c$ is 0 (1) we descend through the left (right) child of the node.
At the child node, the position $i$ is updated to $\rank_0(B_v,i)$ ($\rank_1(B_v,i)$), if the first bit of the symbol $c$ is 0 (1).
This process is recursively repeated until we reach a leaf node.
At the leaf node, the number of occurrences of the symbol $c$ corresponds to the updated value of $i$.
In total, this counting strategy requires to answer $O(\log n)$ $\rank$ operations over the bitmaps in the path of a symbol.
 Figure~\ref{fig:iwt}.b shows, with a darker background, the bitmaps used to count how many times  the symbol $ad$ appears
until the fifth position of the sequence. 

%


\subsubsection{Strengths and weaknesses of \cet}

One advantage of \cet is its ability to retrieve reverse neighbors with the same time performance of direct neighbors,
due to the bi-dimensional representation used for storing the events of activation/deactivation of edges.
Indeed, we just need to update the retrieval range to $(*,v)$ to obtain the frequency of neighboring changes of the 
edges whose target vertex is $v$.

Another advantage is that the time performance in operations about vertexes and edges is 
independent of the number of contacts per query in the graph. 
This is because
\iwt\ allows the counting of events  in logarithmic time with respect to the number of edges (instead of 
a sequential counting on the history of events). 
Due to the temporal arrangement of events of activation/deactivation of edges, operations regarding events on edges are easily 
obtained by extracting the subsequence related to the time instant of the query.
For example, to obtain the edges that change their state at time instant $t$, we just need to recover the pairs of vertexes 
in the section related to events occurred at time $t$.

Despite the advantages of \cet, its main weakness is related to the counting strategy used to recover the states of edges
when contacts are active for short time intervals.
For example, if we want to retrieve a snapshot at a time instant $t$ in a graph where
all the edges were activated and deactivated before $t$,
we are forced to retrieve the frequency of all the edges (i.e., visiting each node of the \iwt), although only a small fraction 
of them will actually be in the output.
In addition, the frequency counting does not allow the representation of  temporal graphs with overlapping contacts.
This is because a symbol representing an overlapping contact will be interpreted as a symbol denoting the deactivation of the contact.

\subsection{Improved representations of \edgelog and \cet}
In the previous section, the descriptions of \edgelog and \cet are given for temporal graphs  where edges can freely 
appear and disappear along time, with no restrictions on the number of contacts per edge.
The representation of these data structures can be improved by taking into account properties of the graph being represented. In particular, properties such as the duration and the dynamism of contacts~\cite{Holme:2012jo}. 


When all contacts last only one time instant, both \edgelog and \cet can be modified to
only store the event that activates an edge because, by definition, all edges will only remain active for one time instant. This small modification
invalidates the strategy used to check if the edge  is active (i.e., the counting strategy in \cet, 
and the check of the interval in \edgelog). However, it enables a new strategy to check if an edge is active. 
For example, in \edgelog, the list of time intervals per edge is replaced by a list of time instants when an edge was active.
Thus, the updated algorithm for checking the activation state of an edge at time $t$
is replaced by verifying if the new list of time instants contains $t$.
Similarly in \cet, the activation state of an edge is replaced by checking if the edge appears in the subsequence related with the events
occurred at time instant $t$.

 \no { For graphs whose contacts last only one time instant, both \edgelog and \cet 
only store the event that activate an edge, as they know that all edges will only remain active for one time instant. This small modification
invalidates the strategy used to check if the edge  is active (i.e., the counting strategy in \cet, 
and the check of the interval in \edgelog). However, it enables a new strategy to check if an edge is active. 
In \edgelog, the list of time intervals per edge is replaced by a list of time instants when an edge was active.
Thus, the updated algorithm for checking the activation state of an edge at time $t$
is replaced by verifying if the new list of time instants contains $t$.
Similarly, in \cet the activation state of an edge is replaced by checking if the edge appears in the subsequence related with the events
occurred at time instant $t$. }

The data structures were also specialized for temporal graphs where each edge has only one contact, 
and  once activated, this contact
remains active until the end of the lifetime. In the literature, these graphs are called 
incremental graphs~\cite{Demetrescu:2010ti}.
With this kind of temporal graphs, the modification is straightforward. 
As all contacts end at the same time instant (i.e., at the end
of the lifetime), it is not necessary to explicitly store the events that deactivate the edges. 
Caro \textit{et al.}~\cite{DBLP:journals/is/CaroRB15} also used this strategy to improve the space cost of both \edgelog and \cet data structures, 
without the need of updating the query algorithms. 
Nevertheless, its usefulness depends on how many contacts effectively end at the last time instant of the graph.

\section{CSA for Temporal graphs (\tgcsa)} \label{sec:TGCSA}

{\em  The Compressed Suffix Array for
Temporal Graphs} (\tgcsa) is a new data structure adapted from  Sadakane's Compressed
Suffix Array (\csa) \cite{Sad03} to represent temporal graphs. Unlike \edgelog and \cet,
it can  represent  contacts of the same edge that temporally overlap, what makes
\tgcsa a more general representation for temporal graphs.

Below we provide a brief presentation of  the \csa. Then, we
include a detailed description of \tgcsa where we show how to create a \tgcsa and we present a modification of the main
structure ($\Psi$) of \tgcsa (Section~\ref{psiStudy}) that targets at improving its efficiency. 
Finally, we also show how it solves the most relevant temporal queries.

\subsection{Sadakane's Compressed Suffix Array (CSA)} \label{sec:csasad}

\begin{figure}[thbp]
\centering
\includegraphics[width=1.0\textwidth]{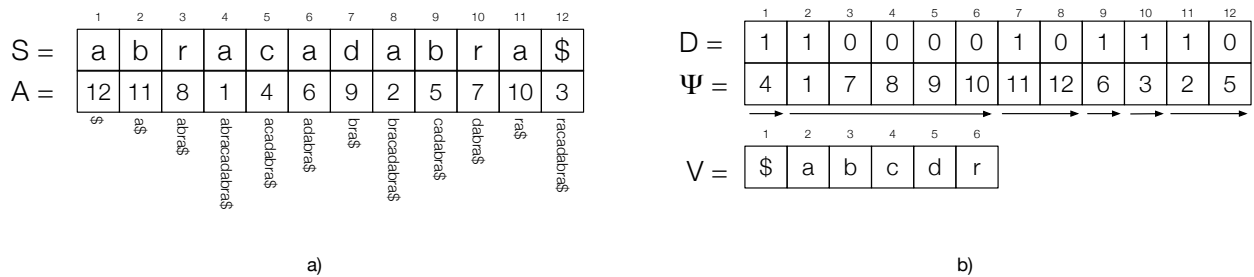}
\caption{The Compressed Suffix Array for the text $S=$\texttt{"abracadabra"}. The left part shows the Suffix Array ($A$). 
The right part depicts the permutation $\Psi$, the bitmap $D$, and the vocabulary $V$. Arrows under the elements of $\Psi$ denote (highly compressible) increasing values. 
In addition, the inverse of the Suffix Array would be $A^{-1}= \langle 4, 8, 12, 5, 9, 6, 10, 3, 7, 11, 2, 1\rangle$.
}
\label{fig:csaexample}
\vspace{-0.3cm}
\end{figure}

Given a sequence $S[1,n]$ built over an alphabet $\Sigma$ of length
$\sigma$, the {\em suffix array} $A[1,n]$ built on $S$ 
is a permutation of $[1,n]$ of all the suffixes $S[i,n]$ such that
$S[A[i],n] \prec S[A[i+1],n]$ for all $1 \le i < n$, being $\prec$
the lexicographic ordering~\cite{MM93}.
In Figure \ref{fig:csaexample}.a, we show the suffix array $A$ for the text $S=$\texttt{"abracadabra"}.%
\footnote{The $\$$ at the end of $S$ is a terminator that must be lexicographically smaller than all the other symbols in $S$.}

Because $A$ contains all the suffixes of $S$ in lexicographic order,
this structure permits to search for any pattern $P[1,m]$ in time
$O(m \log n)$ with a simple binary search of the range $A[l,r]$ (i.e., $[l,r] \leftarrow binSearch(P)$) that
contains pointers to all the positions in $S$ where $P$ occurs. The
 term $m$ of the cost appears because, at each step of the binary search, one could
need to compare up to $m$ symbols from $P$ with those in the suffix
$S[A[i],A[i]+m-1]$. Unfortunately, the space needs of $A$ are high.

To reduce the space needs, \csa \cite{Sad03} uses
another permutation $\Psi[1,n]$ defined in \cite{GV00}. For each
position $j$ in $S$ pointed by $A[i]=j$, 
$\Psi[i]$ gives the
position $z$ such that $A[z]$ points to $j+1 = A[i]+1$. There is a
special case when  $A[i]=n$, in which case $\Psi[i]$ gives the
position $z$ such that $A[z]=1$. In addition, two other structures are
needed, a vocabulary array $V[1,\sigma']$ with all the different
symbols that appear in $S$, and a bitmap $D[1,n]$ aligned to $A$ so
that $D[i] \leftarrow 1$ if $i=1$ or if $S[A[i]] \neq S[A[i-1]]$
($D[i]\leftarrow 0$; otherwise). Basically, a $1$ in $D$ marks the
beginning of a range of suffixes pointed from $A$ such that the
first symbol of these suffixes coincides. Therefore, if the $i^{th}$
and ${(i+1)}^{th}$ ones in $D$ occur in $D[l]$ and
$D[r]$, respectively, that is, if $select_1(D,i)=l$ and
$select_1(D,i+1)=r$, it means that all 
the suffixes 
$S[A[l],n]$, $S[A[l+1],n]$,... $S[A[r-1],n]$ pointed from the 
entries $A[l,r-1]$ start by the same symbol
of the vocabulary.  The bitmap $D$ is used to index the vocabulary
array. Note that $V[rank_1(D,l)] = V[rank_1(D,x)]~ \forall x \in
[l,r-1]$. Recall that $rank_1(D,i)$ returns the number of 1s in
$D[1,i]$ and can be computed in constant time using $o(n)$ extra
bits \cite{Jac89,Mun96}, whereas  $select_1(D,i)$ returns the position
of the $i^{th}$ $1$ in $D$. 
In Figure \ref{fig:csaexample}.b, we show the components of the \csa for the text \texttt{"abracadabra"}.

By using $\Psi$, $D$, and $V$, it is possible to perform binary
search without the need of accessing $A$ or $S$. Note that, the
symbol $S[A[i]]$ pointed by $A[i]$ can be obtained by
$V[rank_1(D,i)]$, symbol $S[A[i]+1]$ can be obtained by $V[rank_1(D, \Psi[i])]$, symbol  $S[A[i]+2]$
can be obtained by $V[rank_1(D, \Psi[\Psi[i]])]$, and so on. Recall
that $\Psi[i]$ basically indicates the position in $A$ that points
to the symbol $S[A[i]+1]$. Therefore, by using $\Psi$, $D$, and $V$
we can obtain the symbols $S[A[i], A[i]+m-1]$ that we could need to
compare with $P[1,m]$ in each step of the binary search.

In principle, $\Psi$ would have the same space requirements
as $A$. Fortunately, $\Psi$ is highly compressible. It was shown to
be formed by $\sigma$ subsequences of increasing values~\cite{GV00} and, therefore,
it can be compressed to around the zero-order entropy of $S$
\cite{Sad03}, and by using $\delta$-codes to represent the
differential values, a space cost of $nH_0+O(n\log\log\sigma)$ bits is obtained.
Note that, in Figure~\ref{fig:csaexample}.b, the arrows under $\Psi$  denote the $\sigma$ subsequences of increasing values in $\Psi$.
In \cite{NM07}, they showed that $\Psi$ can be split into
$nH_k+\sigma^k$ (for any $k$) {\em runs} of consecutive values so
that the differences within those runs are always $1$. This
permitted them to combine $\delta$-coding of gaps with run-length
encoding (of {\em 1-runs}) yielding higher-order compression of $\Psi$.
In addition, to maintain fast random access to $\Psi$, absolute
samples at regular intervals are kept. 

In \cite{FBNCPR12}, authors adapted \csa to deal with
large (integer-based) alphabets and created the {\em integer-based
CSA} (\icsa). They also showed that, in this scenario, the best
compression of $\Psi$ was obtained by combining differential
encoding of runs with Huffman \cite{Huffman1952} and run-length encoding.

As said before, $\Psi$, $D$, and $V$ are enough to simulate the
binary search for the interval $A[l,r]$ where pattern $P$ occurs
without keeping $A$ and $S$ ($[l,r] \leftarrow CSA\_binSearch(P)$). 
Being $r-l+1$ the number of occurrences
of $P$ in $S$, this permits to solve the well-known {\em count}
operation. However, if one is interested in  {\em locating} those
occurrences in $S$, $A$ is still needed. In addition, to be able to
{\em extract} the  subsequence $S[i,j]$, we also need to keep
$A^{-1}$ so that we know the position in $A$ that points to $S[i]$.
In practice, only sampled values of $A$ and $A^{-1}$ are stored.
Non-sampled values $A[i']$ can be retrieved by applying
$i'\leftarrow \Psi[i']$ $k$-times until a sampled position $A[x]$ is
reached (then $A[i'] \leftarrow A[x]-k$). Similarly, sampled
values of $A^{-1}[i]$ can be obtained by applying k-times
$i'\leftarrow \Psi[i']$ from the previous sample $A^{-1}[x]$
(starting with $i' \leftarrow x$). In this case, $A^{-1}[i]
\leftarrow A^{-1}[x] +k$. From this point, the \csa is a 
{\em self-index} built on $S$ that replaces $S$ (as any substring $S[i,j]$
could be extracted) and does not need $A$ anymore to perform
searches.

\subsection{Modifying \csa to represent Temporal Graphs}\label{sec:csa}

Recall that a temporal graph is a set $\mathcal{C}$ of contacts of the form $c=
(u,v,t_s,t_e)$, where $u$ and $v$ are vertexes ($V$) and a link or
edge between them is active during a time interval $[t_s,t_e)$. Also
 $[t_s,t_e) \subset \mathcal{T} \times \mathcal{T}$, with $\mathcal{T}$ being the
time instants representing the \textit{lifetime} of the graph. In 
Example~\ref{ex:tgcsa}, we include a set of five contacts that we will use in our discussion below.

\begin{figure}[t]
\begin{center}
{\includegraphics[width=1.0\textwidth]{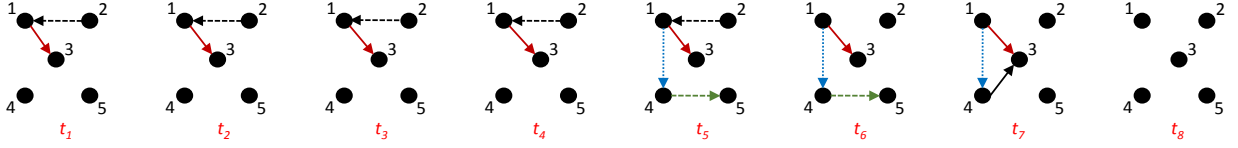}}
\end{center}
\vspace{-0.3cm}
\caption{The temporal graph from Example~\ref{ex:tgcsa}.}
\label{fig:tgraphworking}
\end{figure}

\begin{example} \label{ex:tgcsa} \em
 Let us consider the temporal graph in Figure~\ref{fig:tgraphworking} with $|V|=5$ vertexes numbered $1\dots5$ and $|\mathcal{T}|=8$
 time instants numbered $1\dots8$. 
 This graph contains the following five contacts: $(1,3,1,8)$, $(1,4,5,8)$, $(2,1,1,6)$, $(4,3,7,8)$, and $(4,5,5,7)$.
\qed
\end{example}

Targeting at using a \csa to obtain a self-indexed representation of a set of contacts (i.e. all their terms 
regarded as a unique sequence), we 
discuss in this section two adaptations that we performed. The first one, {\em using-disjoint-alphabets},
consists in assigning {\em ids} from disjoint alphabets to both vertexes and time instants. Then,  when we perform a query for
a given $id$ (or a sequence of $ids$) within the \csa, that $id$ will correspond either to a source vertex, a target vertex, a starting time instant,
or an ending time instant. 
The second modification consists in {\em making $\Psi$ cyclical} on 
the elements of the 4-tuple representing a contact.  
This will permit us to use the regular binary search procedure of the \csa to efficiently search for (and retrieve) 
those contacts matching some constraints on their terms.

\subsubsection{Using disjoint alphabets}

\begin{figure}[thbp]
\centering
\includegraphics[width=0.75\textwidth]{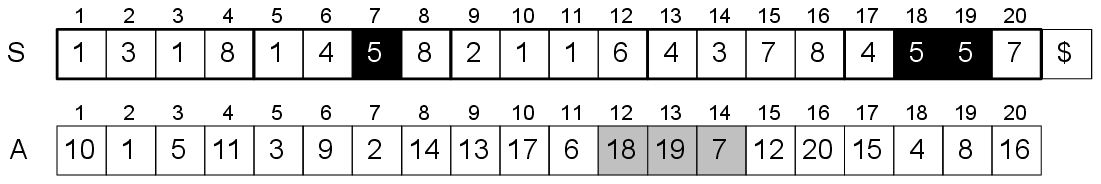}
\caption{Suffix Array for the contacts from Example~\ref{ex:tgcsa} using a unique alphabet  $\Sigma=\{1,2,3,4,5,6,7,8  \}$. }
\label{fig:UniqueAlp}
\vspace{-0.3cm}
\end{figure}

Given a set of $n$ contacts, such as the one in Example~\ref{ex:tgcsa}, our procedure to create \tgcsa starts by 
creating an ordered list of the $n$ contacts, so that they are sorted by their first term, then (if they have
the same first term) by the second term, and so on. After that, these sorted contacts are regarded as a sequence 
with $4n$ elements ($S[1,4n]$), and a suffix array $A[1,4n]$ is built over it. This is depicted in Figure~\ref{fig:UniqueAlp}.

If $S$ were made up of text, $A$ and $S$ (or a \csa built on $S$) would be enough to perform searches for any word or text 
substring $P[1,m]$. In such case, if we looked for the occurrences of symbol $5$ (i.e $P[1,1]=\langle 5\rangle$), 
$A[12,14] = \langle18,19,7\rangle$ would indicate that there are $3$ occurrences of symbol $5$. They occur at $S[18]$, $S[19]$, and $S[7]$ respectively.  
However, in our scenario, when we search for  symbol $5$ (i.e. $P[1,1]=\langle 5 \rangle$) we have to be able to 
distinguish among the source vertex $5$, the target vertex $5$, the starting time instant $5$ and the ending time instant $5$. 
This would require accessing all the entries $A[i]$, $\forall i \in [12,14]$, and checking the positions in $S$ they are pointing to. In practice,
if  $A[i]\mod 4 = 1$ then  $A[i]$ points to a source vertex; otherwise, if $A[i]\mod 4 = 2$ then it points to a target vertex, 
and so on. However, this procedure would ruin the $O( m \log n)$ search time that would now become $O( m\log n + occ)$, 
where $occ$ is the number of occurrences of the query pattern in $S$.

%
%

\begin{figure}[thbp]
\centering
\includegraphics[width=0.75\textwidth]{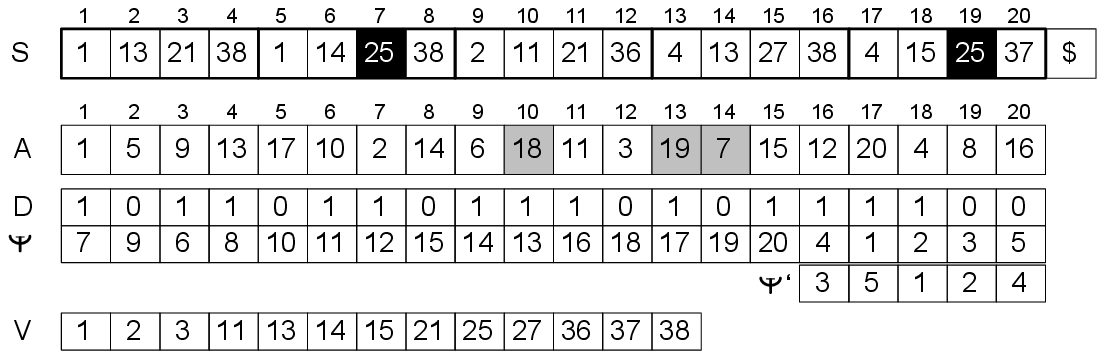}
\caption{Suffix Array for the contacts from Example~\ref{ex:tgcsa} using disjoint alphabets. The structures
$\Psi$, $D$, and $V$ for the corresponding \csa are also depicted. }
\label{fig:disjointAlp}
\vspace{-0.3cm}
\end{figure}

\medskip
A simple workaround to the problem above consists in using disjoint alphabets for the four terms in a contact. In our case,
we use  alphabets $ \Sigma_1, \Sigma_2, \Sigma_3$, and $\Sigma_4$
satisfying that $\Sigma_1 \prec \Sigma_2 \prec \Sigma_3 \prec \Sigma_4$ ($\prec$ indicates lexicographic order).
Note that we can always replace vertexes and time instants in the original set of contacts by new $ids$ satisfying this 
property. For example, in Figure~\ref{fig:disjointAlp}, we have created a new sequence $S$ where: 
(i) the $ids$ of the source vertexes have been kept as they were initially ($\Sigma_1=\{1,2,4\}$); 
(ii) the $ids$ of the target vertexes have been added $+10$ 
  ($\Sigma_2=\{\underline{1}1,\underline{1}3,\underline{1}4,\underline{1}5\}$); 
(iii) the $ids$ of the starting time instants have been added $+20$ 
  ($\Sigma_3=\{\underline{2}1,\underline{2}5,\underline{2}7\}$); and
(iv) the $ids$ of the ending time instants have been added $+30$ 
  ($\Sigma_4=\{\underline{3}6,\underline{3}7,\underline{3}8\}$). 
Now, when we build the suffix array for the new $S$, we can search for either the pattern 
$\langle 5 \rangle$, $\langle 15 \rangle$, $\langle 25 \rangle$, or $\langle 35 \rangle$, 
depending on if we want to find the occurrences of the term $5$ that corresponds  to a source vertex, 
target vertex, starting time, or ending time, respectively. For example, we can see in the figure that when we are searching for
the starting time $5$, we can simply add $+20$ to its $id$ and actually use the suffix array (or the \csa) to look for 
$P=\langle \underline{2}5\rangle$ obtaining its two occurrences pointed  by $A[13]$ and $A[14]$. However, to search for the target vertex $5$ 
we would add $+10$ to its $id$ and found that $A[10]$ points to its unique occurrence in $S$. In any case, we 
retain the original $O(m\log n)$ search time as expected.

\medskip
An interesting by-product that arises from the use of disjoint alphabets is that, since values
from $\Sigma_i$ are smaller than those from  $\Sigma_j$ ($\forall i<j$), the first quarter of
entries in $A$ ($A[1,n]$) will point to the first terms of all the
contacts ($S[1+4k], \forall k \in [0,n)$), the next $n$ entries in $A$ ($A[n+1,2n]$)  to the second terms ($S[2+4k], \forall k \in [0,n)$),
and so on. Consequently, the first quarter of entries of $\Psi$
($\Psi[1,n]$) will point to a position in the range $[n+1, 2n]$,
because in the indexed sequence $S$ each symbol $ u \in \Sigma_1$ is
followed by a symbol $ v \in \Sigma_2$, and so on.
In this way, each entry in the last quarter of $\Psi$ will point to a
position in the range $[1,n]$, corresponding to the first quarter of
entries in $A$.

In our example, recall we have $n=5$ contacts. We can see that the entries in the four
quarters of $A$ discussed above match that: $\forall i \in [1,5], A[i]\mod 4 = 1$;  
$\forall i \in [6,10], A[i]\mod 4 = 2$; $\forall i \in [11,15], A[i]\mod 4 = 3$; and
$\forall i \in [16,20], A[i]\mod 4 = 0$. 
In addition, in Figure~\ref{fig:disjointAlp}, we have also included the $\Psi$ structure that
arises when we build the corresponding \csa. In this case, we can also verify that it holds that:
$\forall i \in [1,5], \Psi[i] \in [6,10]$;  
$\forall i \in [6,10], \Psi[i] \in [11,15]$;   $\forall i \in [11,15], \Psi[i] \in [16,20]$;  and
$\forall i \in [16,20], \Psi[i] \in [1,5]$. This property will be of interest in the following section.

\subsubsection{Modifying $\Psi$ to make it cyclical on the terms of each contact} \label{sec:cyclical}

Recall that in a regular \csa, once we know that the $i^{th}$ entry in the underlying suffix array $A$
points to a position $z= A[i]$ of the source sequence $S$, we can recover the entries $S[z], S[z+1],...$
from the original sequence $S$ 
as $S[z] = S[A[i]] \leftarrow V[rank_1(D,i)]$,
the next symbol as $S[z+1]= S[A[i]+1]  \leftarrow V[rank_1(D,\Psi[i])]$, 
the next symbol as $S[z+2]= S[A[i]+2]  \leftarrow V[rank_1(D,\Psi[\Psi[i]])]$, and so on. 
Therefore, as shown in Section~\ref{sec:csasad}, by using $\Psi$, $D$, and $V$, we can binary search
for any pattern $P$ obtaining the range $[l,r]$ so that $\forall i \in [l,r], A[i]$ points to the positions
in $S$ where $P$ can be found. Then, from those positions on, we could recover the source data of the 
suffixes $S[A[i],...]$ that start with $P$. Unfortunately, this mechanism allows us to recover
the source data only forward-wise (not backwards), and this is not enough in our scenario because
we typically want to search for the contacts that match a given constraint and then we want to retrieve 
all their terms.

To clarify the issue above, consider, for example, when we look for the contacts whose target vertex
 is $v=5$ ($P=\langle\underline{1}5\rangle$), then we
obtain its unique occurrence at the position $10$ ($A[10]$). Consequently, to retrieve the terms of that contact $(u,v,t_s,t_e)$, 
we would compute:
$v\leftarrow V[rank_1(D,10)] = \underline{1}5$;
$t_s\leftarrow V[rank_1(D,\Psi[10])] =\underline{2}5 $;
$t_e\leftarrow V[rank_1(D,\Psi[\Psi[10]])] =\underline{3}7 $. However, 
$u' \leftarrow V[rank_1(D,\Psi[\Psi[\Psi[10]]])]$ would not recover the first term of the current
contact, but the first term of the next contact in $S$. As in a regular \csa, 
to retrieve $u$, we would have to access
$A[10]=18$ to know that the target vertex $v$ occurs at position $S[18]$, and consequently the source
vertex $u$ should be retrieved from $S[18-1]$. Now, because $S$ is not actually kept in the \csa, to extract $S[17]$, we have to
know the entry $x$ in $A$ such that $A[x]=17$. We can use that $x=A^{-1}[17]=\mathbf{5}$.\footnote{Recall 
$A^{-1}[j]=x$ indicates which position $x$ from $A$ points to the $j^{th}$ entry of $S$. That is, such that $A[x]=j$. } Finally, 
by using $u \leftarrow V[rank_1(D,\mathbf{5})] = 4$ we have fully recovered the contact 
$(4,\underline{1}5, \underline{2}5, \underline{3}7 )$ we were searching for. 
To sum up, the previous procedure would make it necessary to use not only $\Psi$, $D$, and $V$,
but also $A$ and $A^{-1}$ as explained in Section~\ref{sec:csasad}. Fortunately, we can modify $\Psi$
in such a way that it allows us to move circularly from one term to the next term within a given 
contact.

Recall that, due to our disjoint alphabets, if $A[i] (i\in[3n+1,4n])$ points
to the last term of the $j^{th}$ contact,
then $\Psi[i]$ would store the position in $A$ pointing to the first term of the following
$(j+1)^{th}$ contact  ($A[i]+1 = A[\Psi[i]]$), which would be in the range $[1,n]$. 
For \tgcsa, we modified these
pointers in the last quarter of $\Psi$ in such a way that, instead
of pointing to the position  $x =A[\Psi [i]] $ corresponding to the
first term of the following contact, they point to the first
term of the same contact; that is, $A[\Psi' [i]] = x-1$ or $A[\Psi'
[i]]=n$ if $x=1$. The modified quarter of $\Psi$ is depicted as $\Psi'$ in Figure~\ref{fig:disjointAlp}.
In this way, starting at any entry $i$ in $\Psi$, and following the
pointers $\Psi[i]$, $\Psi[\Psi[i]]$, and $\Psi[\Psi[\Psi[i]]]$, all the elements of the
current contact can be retrieved, but no entry from any other tuple
will be reached. Due to this modification, in the example above, we can recover 
$u \leftarrow V[rank_1(D,\Psi[\Psi[\Psi[10]]])]$, and $A$ and $A^{-1}$ are no longer needed.

Note that it is not possible now to traverse the whole
\csa by just using $\Psi$ because consecutive applications of the
$\Psi$ function will cyclically obtain the four elements of the
corresponding contact. However, this small change in $\Psi$ to make it cyclical on the
terms of each contact, brings additional interesting searching
capabilities that we will exploit in Section~\ref{sec:tgcsa:search}.

\subsection{Detailed construction of \tgcsa}

\begin{figure}[t]
\begin{center}
{\includegraphics[width=0.95\textwidth]{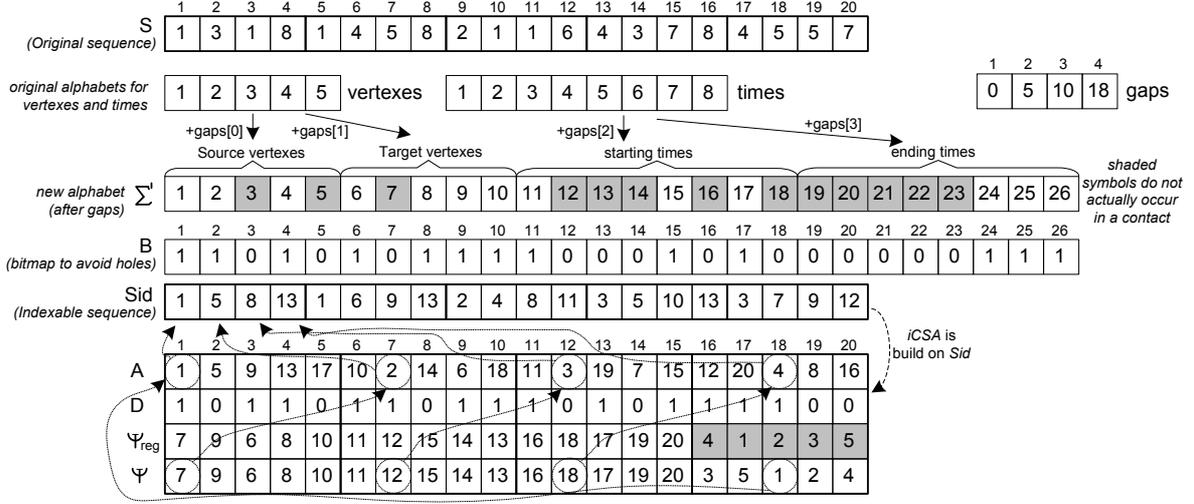}}
\end{center}
\vspace{-0.3cm}
\caption{Structures involved in the creation of a \tgcsa for the temporal graph in Example~\ref{ex:tgcsa}.}
\label{fig:tgcsa}
\end{figure}

Once we have explained the need of using disjoint alphabets and the reason why
we use a modified $\Psi$, in this section we explain the actual procedure to build
our \tgcsa. In Figure \ref{fig:tgcsa}, we depict all the structures involved in the creation of a \tgcsa 
representing the temporal graph in Example~\ref{ex:tgcsa}.

As indicated above, the first step to build a \tgcsa is to create a
sequence $S$ with the ordered $n$ contacts. Hence
we obtain, $S[1,4n] = \langle u^1 ,v^1,t_s^1,t_e^1,u^2
,v^2,t_s^2,t_e^2,\dots, u^n ,v^n,t_s^n,t_e^n\rangle$.\footnote{Note
that the ordering is not relevant because we have a {\em set} of contacts.
Therefore, we will assume that contacts are sorted by the first term, then
by the second one, and so on.}

The second step involves defining a reversible mapping that enables us to use disjoint alphabets.
Let us assume we have $\nu=|V|$ different vertexes and $\tau =
|\mathcal{T}|$ time instants. It is possible to define a
reversible mapping function that maps the terms of any original
contact  $c= (u,v,t_s,t_e)$ to $c'=
(u,v+\nu,t_s+2\nu,t_e+2\nu+\tau)$. To achieve this, we define an
array $ gaps[1,4] \leftarrow \langle 0,\nu,2\nu, 2\nu+\tau \rangle$ and a set with elements
$c'[i]\leftarrow c[i]+gaps[i],~\forall i=1\dots4$. This mapping
defines four ranges of entries in an alphabet $\Sigma'$ for both
vertexes and time instants such that $|\Sigma'|= 2\nu + 2\tau$. Note that
vertex $i$ is mapped to either the integer $i$ or $i+\nu$ depending
on whether it is the source or target vertex of an edge. Similarly,
the time instant $t$ is mapped to either $t+gaps[3]$ or $t+gaps[4]$.
This allows us to distinguish between starting/ending
vertexes/time instants by simply checking the range where their value falls
into. 
\no{In addition, it brings an interesting property during suffix
sorting as we will show below.}

Observe that even though vertex $i$ always exists in the temporal
graph, either source vertex $u' =i+gaps[1] = i$ or target vertex
$v'=i+gaps[2]$ could actually not be used. Similarly, a time instant
$t'$ could not occur either as an initial or as an ending time of a
contact, yet we could be interested in retrieving all the edges that
were active at that time $t'$.

To overcome the existence of holes in the alphabet $\Sigma'$,  a
bitmap $B[1,2\nu+2\tau]$  is used. We set $B[i] \leftarrow 1$ if the
symbol $i$ from $\Sigma'$ occurs in a contact, and $B[i] \leftarrow
0$; otherwise. Therefore, each of the four terms within a contact
$(u,v,t_s,t_e)$ will correspond to a $1$ in $B$. Then an alphabet
$\Sigma$ of size $\sigma = rank_1(B,4n)$\footnote{Recall $rank_1(B,i)$
returns the number of 1s in $B[1,i]$.} is created containing the
positions in $B$ where  $1$ occurs. For each symbol $i \in \Sigma'$,
a {\em mapID($i$)} function 
 assigns an integer
$id \in \Sigma$ to $i$, so that $id \leftarrow$ {\em mapID($i$)} =
$rank_1(B,i)$ if $B[i] = 1$, and $0 \leftarrow${\em mapID}$(i)$ if
$B[i]=0$. The reverse mapping function is provided via {\em unmapID}$(id) =
select_1(B,id)$.\footnote{Recall $select_1(B,i)$ computes the
position of the $i^{th}$ $1$ in B.}

At this point, a sequence of ids $Sid[1,4n]$ can be created by
setting $Sid[i] \leftarrow ${\em mapID}$(S[i] + gaps[i \mod 4+1]),
\forall i=1\dots 4n$.
Indeed, being $type=1,2,3, 4$, respectively, the types of source
vertexes, target vertexes, starting time instants, and ending time instants from the
original sequence $S$, we can map any source symbol $i$ from $S$ into
$Sid$ by  $id = ${\em getmap}$(i,type) \leftarrow rank_1(B,i+gaps[type])$. Similarly, the reverse mapping obtains $i= ${\em getunmap}$(id,type)  \leftarrow  select_1(B,id) -gaps[type]$. 

Once we have made up our indexable sequence $Sid$,  an \icsa\ is built over it.\footnote{We actually added
four integers set to $zero$ that make up a dummy contact
($0$,$0$,$0$,$0$) at the beginning of $Sid$. This is required to
avoid limit-checks at query time.} Then, as discussed in Section~\ref{sec:cyclical}, we modified the array $\Psi$  in our \tgcsa to
allow $\Psi$ to move circularly from one term to the next one within
the same contact. To do this, we simply have to modify the last quarter of the regular
$\Psi$ array so that, $\forall i=3n+1\ldots 4n,~ \Psi[i] \leftarrow
((\Psi[i] -2) \mod n)+1$. This small change brings an interesting
property that allows us  to perform a query for any term of a contact in
the same way.  We use the \icsa\ to binary search for a term of a contact(s), obtaining a
range $A[l,r]$, and then by circularly applying $\Psi$ up to three
times, we can retrieve the other terms of the contact(s).

\medskip
To sum up,  \tgcsa  consists of a bitmap $B$, and the structures $D$ and $\Psi$ of
the \icsa. In practice, $B$ is compressed using  Raman \textit{et al.} strategy\footnote{Raman et al strategy allows
both $select_1$ and $rank_1$ in $O(1)$ time and requires $|B|\mathcal{H}_0(B) + o(|B|)$ bits.}~\cite{Raman:2007:SID:1290672.1290680},
and for $D$ we used a faster bitmap representation~\cite{FBNCPR12} using $1.375 |D|$ bits. For the representation
of $\Psi$ we also used the best option (named $\huffrle$) that samples
$\Psi$ at regular intervals and then differentially encodes the remaining values~\cite{FBNCPR12}. Yet, we also created an
alternative representation for $\Psi$ that is discussed in Section~\ref{psiStudy}.

\subsection{A more suitable representation of $\Psi$ for temporal graphs: $\vbyte$ strategy} \label{psiStudy}
The regular representation of $\Psi$ is based on  sampling the $\Psi$ array at regular intervals (one sample
every $t_{\Psi}$ entries) and then, differentially encoding the remaining values between two samples. 
In~\cite{FBNCPR12}, they studied different alternative encodings for the non-sampled values, and showed
that the best space/time trade-off in a text-indexing scenario was reported by coupling run-length encoding
of {\em 1-runs} (sequences of $+1$ values)  with bit-oriented Huffman ($\huffrle$ approach). In practice, they used $t_{\Psi}$ Huffman codes
to indicate the presence of $1$-runs of length $1\dots t_{\Psi}$. They  also reserved $n_{sv}$ Huffman codes to represent
short gaps (where $n_{sv}$ is a parameter typically set to $2^{14}$). Finally, being $\omega$ the machine word size, 
$2\times \omega$ additional Huffman codes are used as escape codes to mark the number of bits needed to either
represent a large positive gap ($g$) or a negative gap ($-g$). In both cases, such a escape code is followed by
$g$ represented with $\lceil \log_2 g \rceil$ bits. 

\begin{figure}[t]
\begin{center}
{\includegraphics[width=0.95\textwidth]{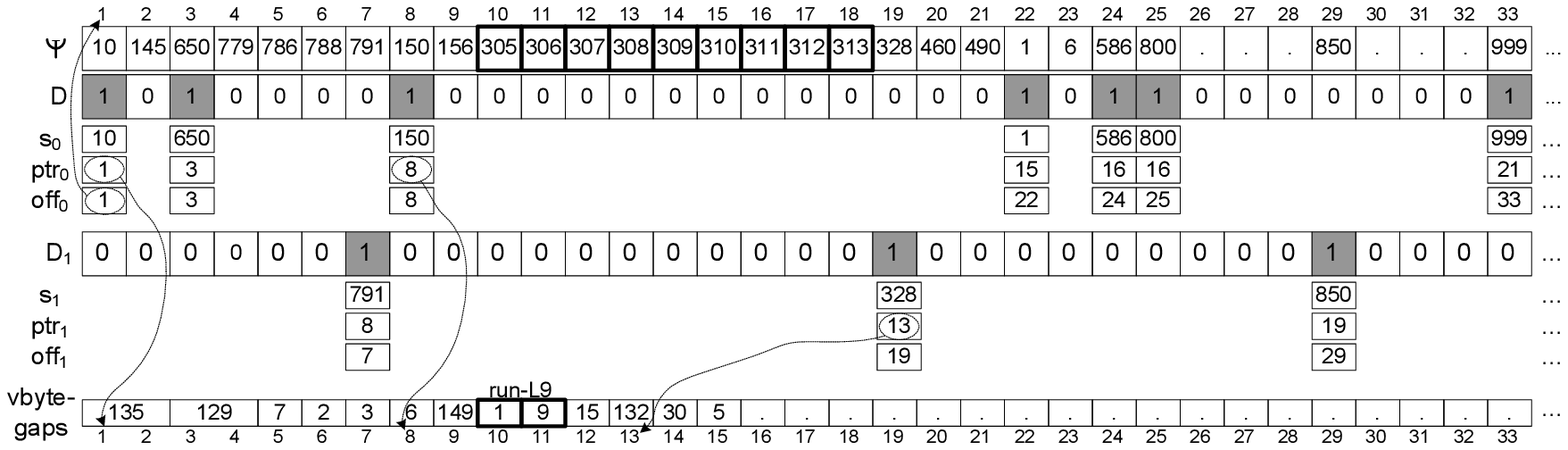}}
\end{center}
\vspace{-0.5cm}
\caption{Example of $\vbyte$ representation of $\Psi$  assuming $t_{\Psi} = 4$.}
\label{fig:psi_structure}
\end{figure}

In this paper, we present a new strategy to represent $\Psi$, that we called $\vbyte$, where we try to 
speed up the $\Psi$ access performance at the cost of using a little more space. An example of the structure for the 
resulting $\Psi$ 
representation is shown in Figure~\ref{fig:psi_structure}.
We also use sampling and differentially encode non-sampled values. Yet, we made some changes with respect to the 
traditional $\Psi$ representations (i.e., $\huffrle$), which are summarized as follows:

\begin{itemize}
 \item We used {\em vbyte} (byte-aligned) codes \cite{WZ99} rather that bit-oriented Huffman codes to differentially
 encode non-sampled values. This should result
 in around one order of magnitude improvement in decoding speed when sequential values of $\Psi$ are to be retrieved.
 Note that in the bottom part of Figure~\ref{fig:psi_structure}, we include a sequence of 
 byte-oriented codewords (either 1 or 2-byte codewords in our example) that are used to represent the gaps from
 the original $\Psi$ structure. It can also contain a pair of codewords for the pair $\langle 1,L \rangle$ to 
 encode a {\em 1-run} of length $L$. Of course, using byte-aligned rather than bit-oriented codes will imply a loss
 in compression effectiveness.
 
 \item We do not sample $\Psi$ at regular intervals. Instead of that, we keep samples aligned with the {\em ones}
 in bitmap $D$, that is, there is a sample at the beginning of the interval in $[l_c,r_c]$ corresponding
 to each symbol $c$.  This modification brings three main advantages: 
 \begin{enumerate}
    \item[(ii)] We ensure that $\Psi[l_c]$ is always sampled, 
 whereas with the traditional  representation  of $\Psi$ the previous sampled position could be in the range 
 $[l_c - t_{\Psi}+1, l_c]$. Therefore, $l_c$ was sampled with probability $1/t_{\Psi}$. 
 Note that, in \tgcsa, a typical access pattern to $\Psi$ during searches (see Section~\ref{sec:tgcsa:search}) 
 consists in traversing all the  values $\Psi[l_c, r_c]$ once we know the interval 
$[l_c,r_c]$ corresponding to a given symbol $c$. This requires decoding gaps from the previous sample to $l_c$ in
$\huffrle$ to obtain synchronization at value $\Psi[l_c]$, and sequentially decoding gaps from there on.
Since $l_c$ is always sampled in $\vbyte$, we avoid that synchronization cost.
    \item[(ii)] While  in
 the traditional representation of $\Psi$, the  differential sequence $\Psi[j] − \Psi[j-1]$ ($j \in [2,4n]$) 
  could contain up to $\sigma / 2$
 negative values (when $i=l_c$ belongs to a symbol $c$ and $j-1 = r_{c-1}$ to symbol $c-1$)\cite{GV00}, the 
$\vbyte$ representation does not deal with negative values because $j=l_c$ is always a sampled  position.
   \item[(iii)] We do not break {\em 1-runs}. Recall  that {\em 1-runs} could occur mainly within the 
   range $[l_c,r_c]$ corresponding to a 
 given symbol $c$. Because our first-level sampling  stores only a sample at position $l_c$, {\em 1-runs} are no longer 
 split. This is interesting  for both  space and access time  because a unique codeword can be used to represent a large 1-run 
 sequence. In our example, we can see that the codewords $\langle 1, 9 \rangle$ in \vbytegapsm$[10,11]$ represent 
 the {\em 1-run} of length $9$ within $\Psi[10,18]$. That is, we do not break the {\em 1-run} every $t_{\Psi}=4$ values.
  
 \end{enumerate}

 In Figure~\ref{fig:psi_structure}, we can see that 
 samples consist of a triple of values $\langle s', \ptrprime, \offprime \rangle$ that are aligned with the {\em ones} in $D$:
 $s'$ indicates the
 absolute value, $\ptrprime$ is a pointer to \vbytegapsm\ sequence, and $\offprime$ indicates the index of 
 the sampled position. In practice, these values are set  in three arrays $s_0[1,\sigma]$, 
 $\ptrzero[1,\sigma]$, and $\offzero[1,\sigma]$, respectively,  such that if $\Psi[j]=s'$ is sampled, we set $s_0[rank_1[D,j]] = s'$,
 $\offzero[rank_1[D,j]] = j$, and $\ptrzero[rank_1[D,j]] = x$. 

Note that the absolute values  $s'$ are kept explicitly in $s_0$ and are not 
 represented within the sequence \vbytegapsm\ (exactly as in  $\huffrle$). 
 For example, $\Psi[1]=10$ is stored at the first
 entry of $s_0$, and the first codeword in \vbytegapsm\ represents value $135$, which corresponds to the gap $\Psi[2]-\Psi[1]$.
 Hence, no codeword in \vbytegapsm\ is associated with the sampled value $\Psi[1]$.
 Note also that
 $x$ is the position in \vbytegapsm\ that we have to access to recover values $\Psi[j+1,...]$.
 In our example, we can see that $\Psi[9]$ can be recovered by accessing the previous sampled value
 $s_0[3]=150 = \Psi[8]$, then accessing sequence \vbytegapsm\ at position $x=\ptrzero[3]=8$ to obtain the gap $\Psi[9]-\Psi[8]$
by  $gap=decode\_vbyte(x)=6$. Finally, we recover $\Psi[9]=150 + 6 = 156$. 
 As an important remark, observe that given a symbol $c$, we will use $\offzero[c]$ to obtain the starting sampled position $l_c$ 
 for the range $\Psi[l_c,r_c]$. We could skip storing array $\offzero$ as we can compute $l_c= select_1(D,c)$. This introduces
 a space/time trade-off that we discuss in the next section. 

\medskip

Despite the advantages of the sampling structures described above, our representation has also a main drawback: we cannot 
parameterize the number of samples we want to use. Thus, we can be using a rather too dense sampling for infrequent 
symbols (consequently, we expect that compression will suffer in datasets with very large vocabularies ($\sigma \approx n$)),  
or  we can 
be using a very sparse sampling for frequent symbols $c$, as they will have only one sample  at the beginning of 
the corresponding interval $[l_c,r_c]$. This  fact could slow down the access to an individual position $\Psi[j]$, 
with  $j\in[l_c+1, r_c]$.
To overcome this, we added a second-level sampling where we sample the positions
$l_c + t_{\Psi}, l_c + 2\times t_{\Psi}, \dots$ ($t_{\Psi}$ is again the sampling interval). We use
a bitmap $D_1$ (see Figure~\ref{fig:psi_structure}) to mark the positions of these samples in $\Psi$, and, 
aligned with the {\em ones} in $D_1$, arrays  $s_1[1,n_1]$, 
$\ptrone[1,n_1]$, and $\offone[1,n_1]$ keep the sampling data ($n_1$ is the number of {\em ones} in $D_1$).
This second-level sampling works exactly like the first-level one with the exception that sampled values
are also retained in the \vbytegapsm\ sequence. This redundant data is kept to allow us to 
sequentially decode the whole values $\Psi[l_c +1, r_c]$ belonging to a given symbol $c$ without the need to
access the second-level sampling data. This is of interest when we want to retrieve a range of consecutive values from
$\Psi$ instead of simply recovering an individual value.

 \end{itemize}

\subsubsection{Comparing the Space/time trade-off of $\vbyte$ with $\huffrle$.}~  


We run experiments to compare the space/time trade-off obtained by $\huffrle$ against 
 $\vbyte$ and $\vbyteselect$ (the latter is the variant of $\vbyte$
where arrays $\offzero$ and $\offone$ are not  stored). We tuned these representations using  four different sampling 
values for $\Psi$. In particular, we used values $t_{\Psi} \in \{256, 64, 16, 8\}$ (from sparser to denser sampling, respectively).
In addition, we include in the comparison a non-compressed baseline representation for $\Psi[1,4n]$ (we refer to it as $\mathsf{plain}$) that represents
each entry of $\Psi$ with $\lceil \log 4n \rceil$ bits and provides direct access to any position.

\begin{figure}[htpb]
  \begin{center}

  \includegraphics[angle=-0,width=0.42\textwidth]{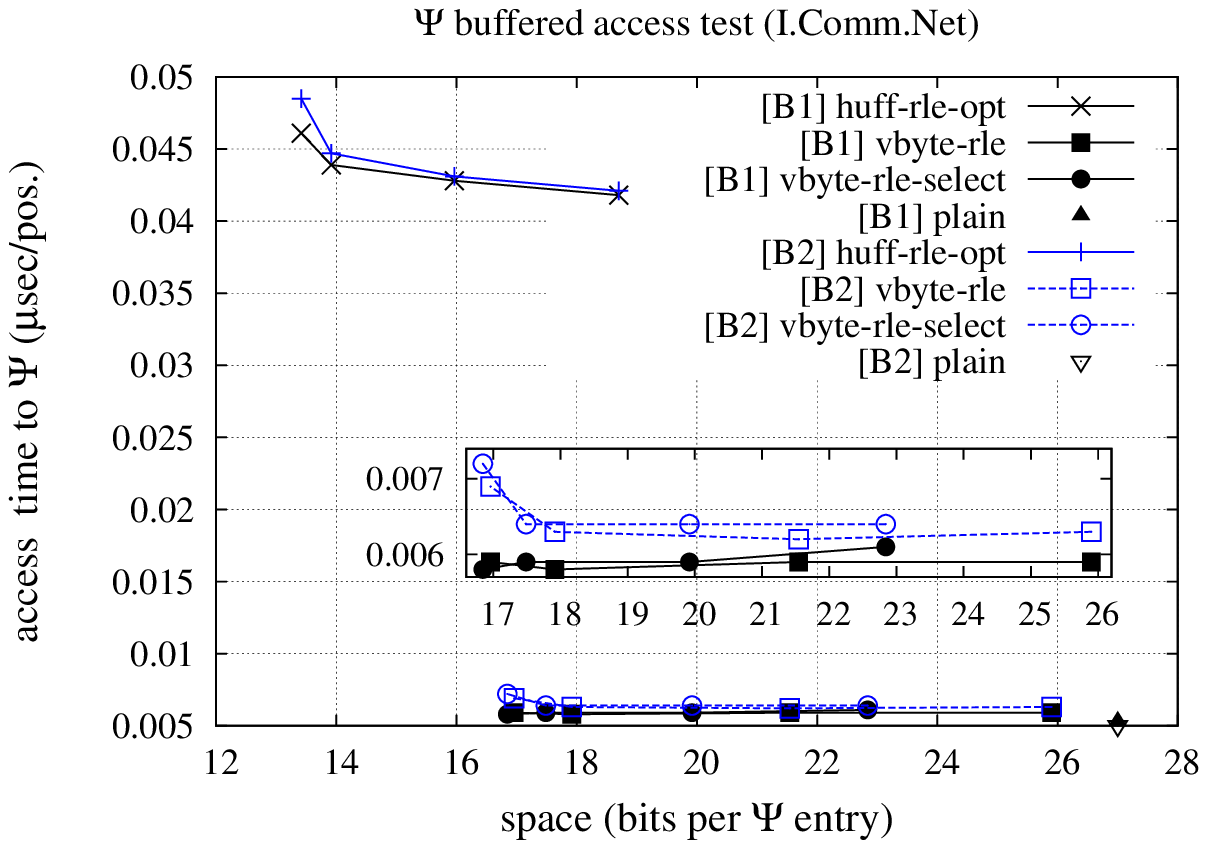}
  \includegraphics[angle=-0,width=0.42\textwidth]{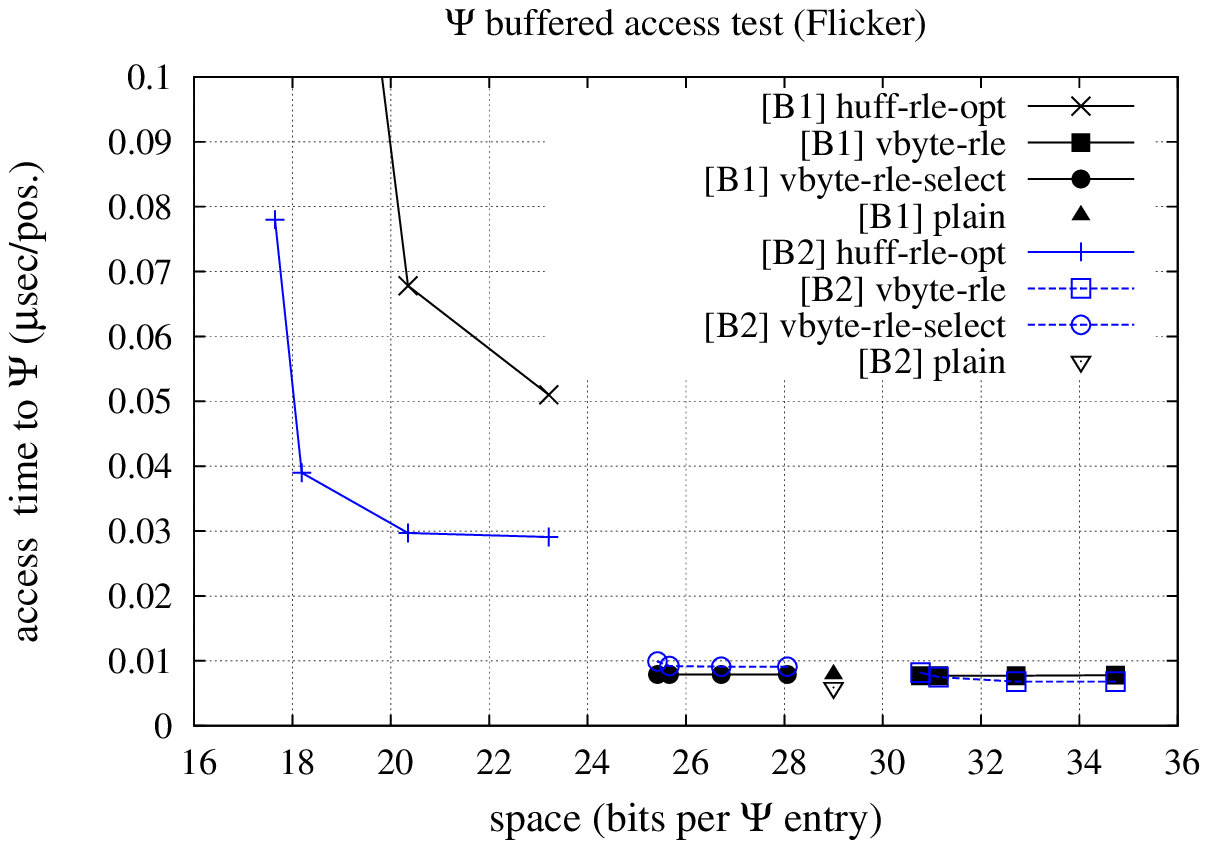}
  \includegraphics[angle=-0,width=0.42\textwidth]{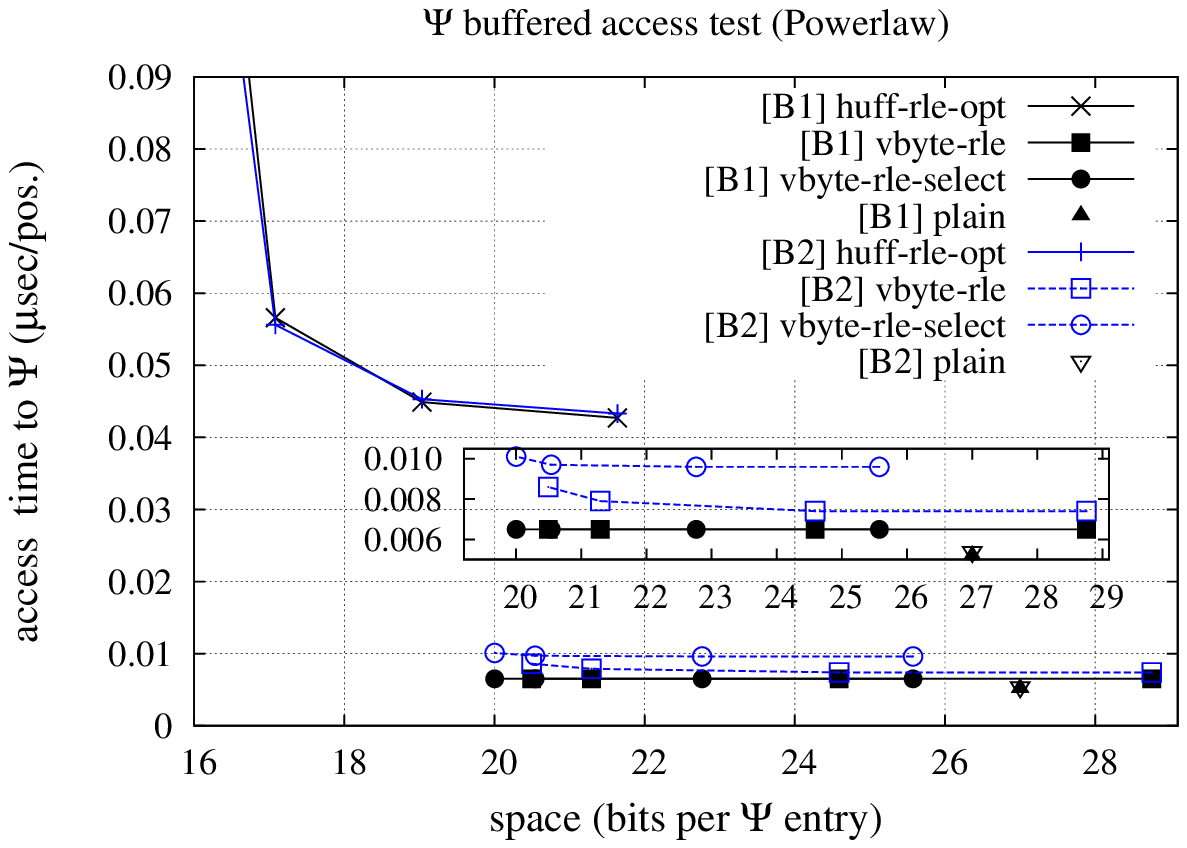}
  \includegraphics[angle=-0,width=0.42\textwidth]{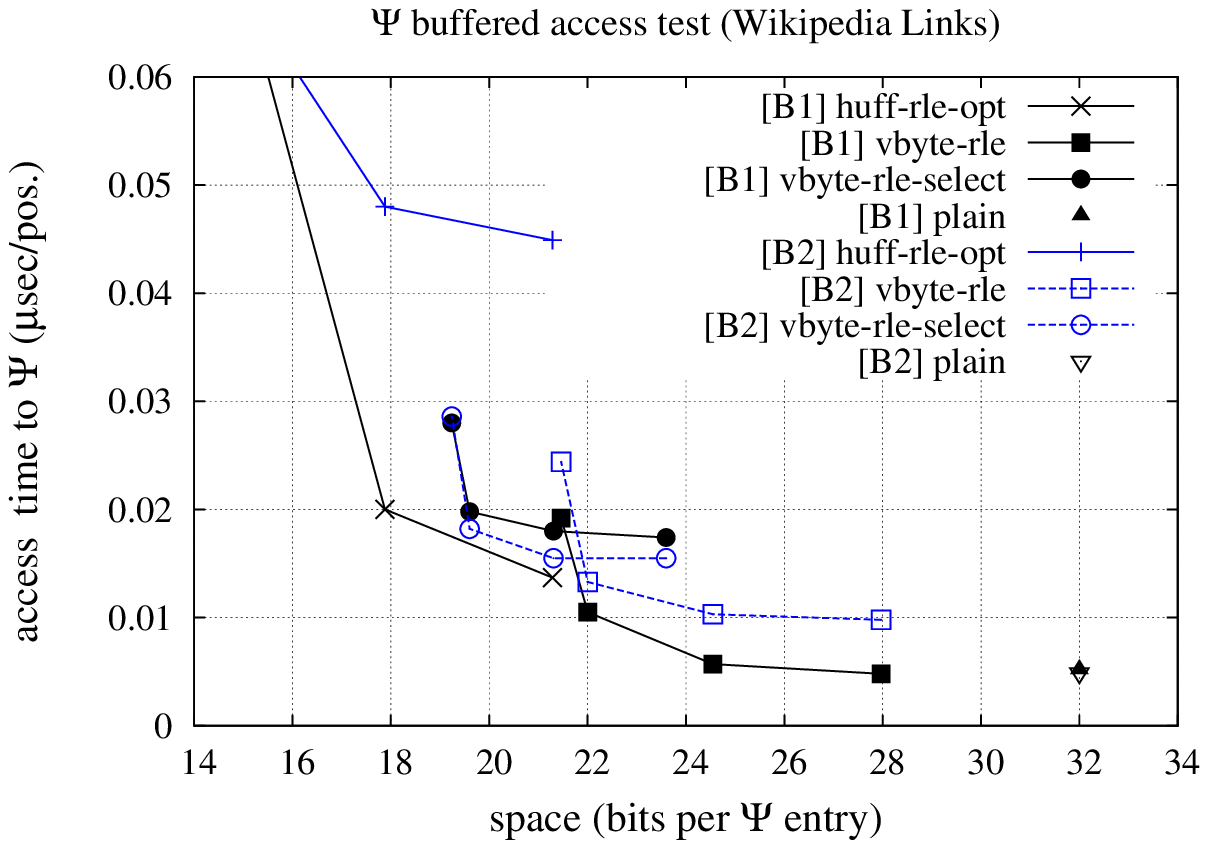}
  \end{center}
  \vspace{-0.3cm}
  \caption{Space/time trade-off for buffered access to $\Psi$.}
  \label{fig:psiBuffered}
  \end{figure}

\begin{figure}[htbp]
  \begin{center}  
  \includegraphics[angle=-0,width=0.42\textwidth]{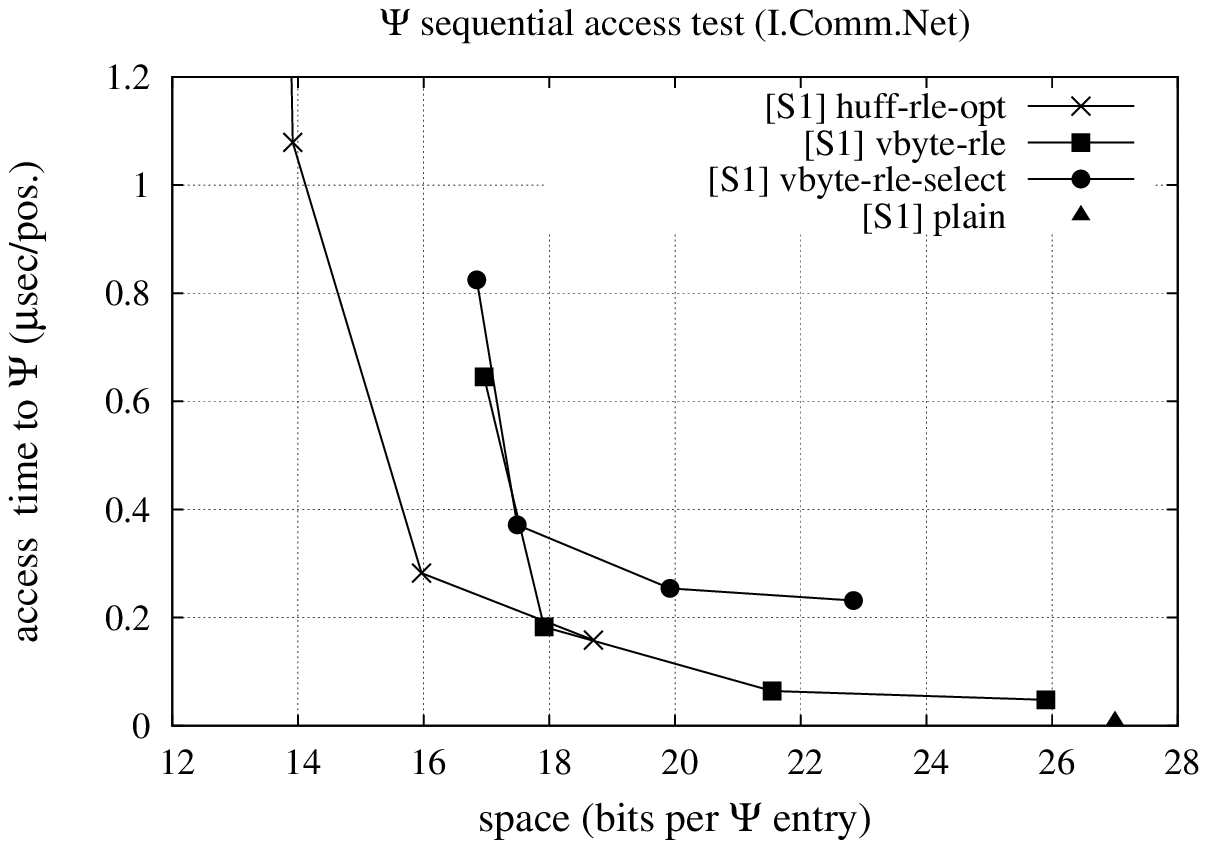}
  \includegraphics[angle=-0,width=0.42\textwidth]{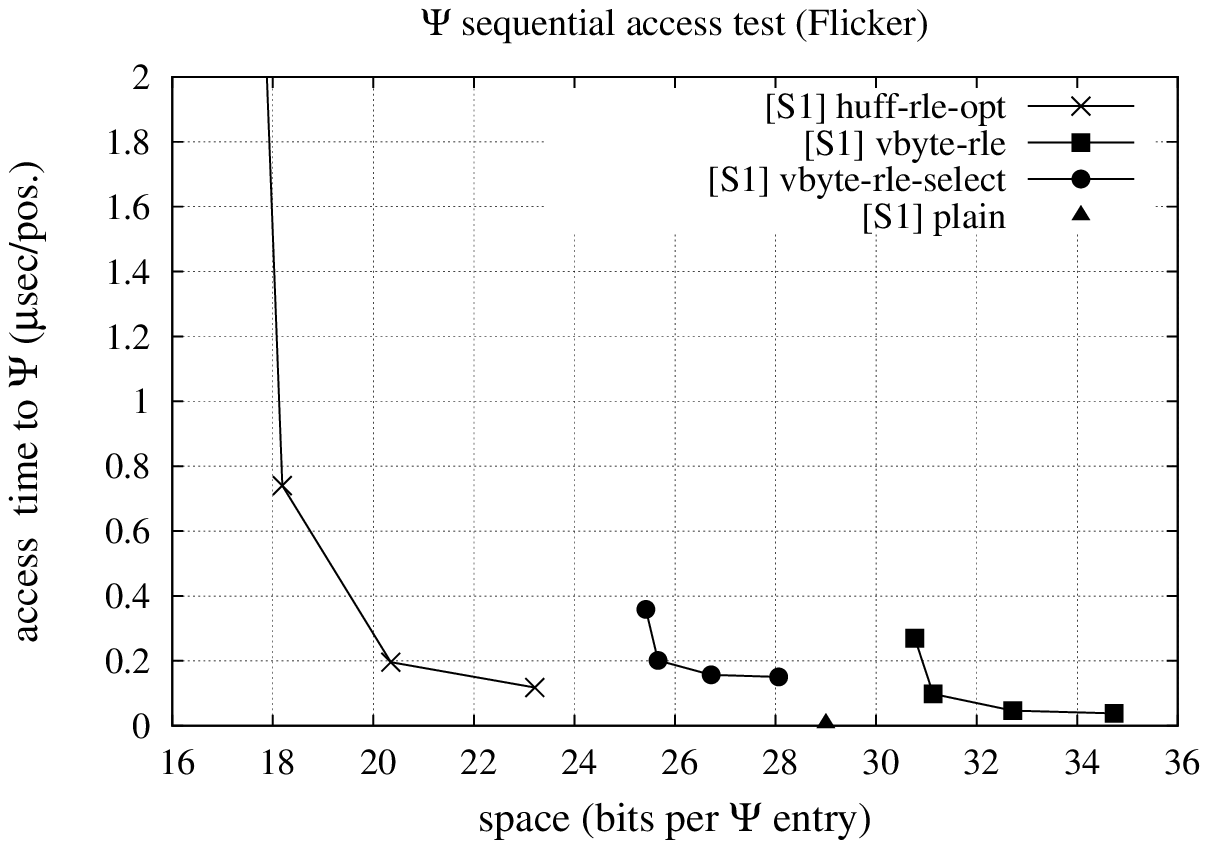}
  \includegraphics[angle=-0,width=0.42\textwidth]{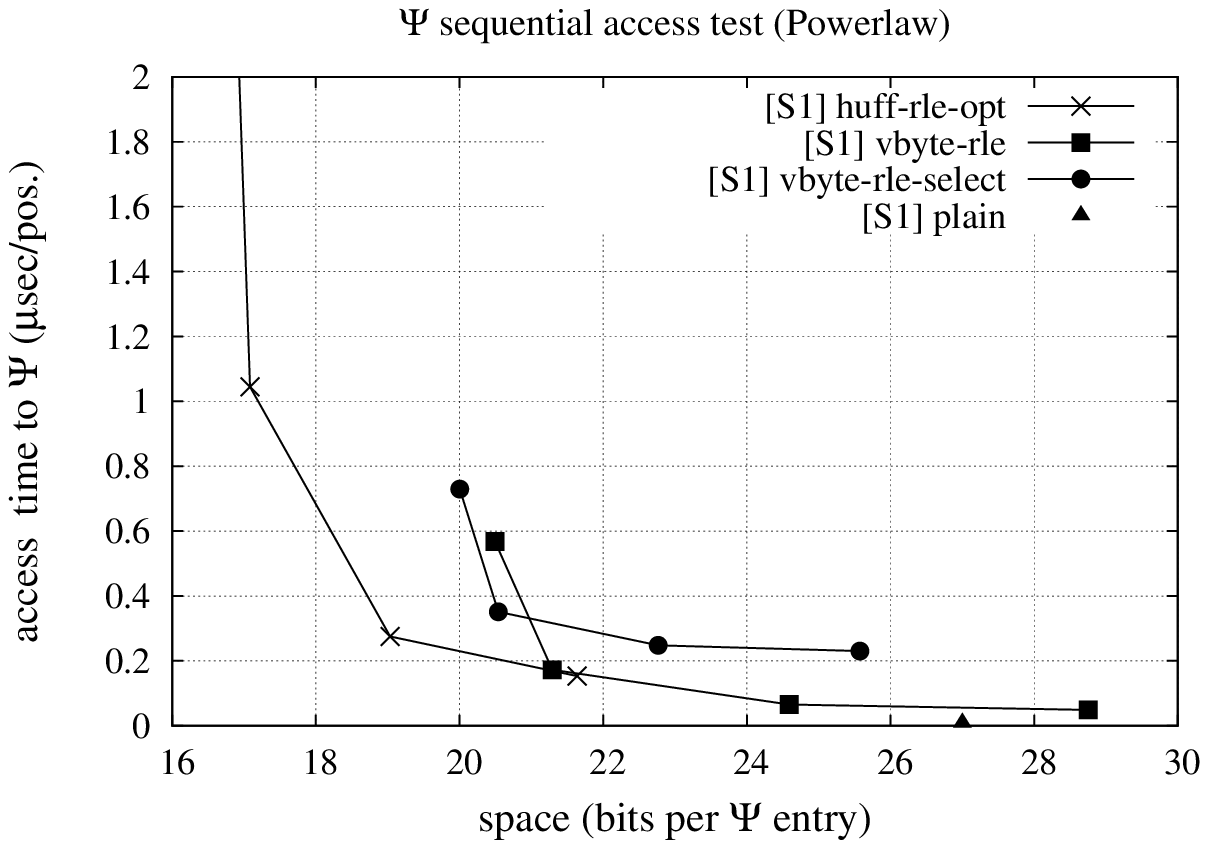}
  \includegraphics[angle=-0,width=0.42\textwidth]{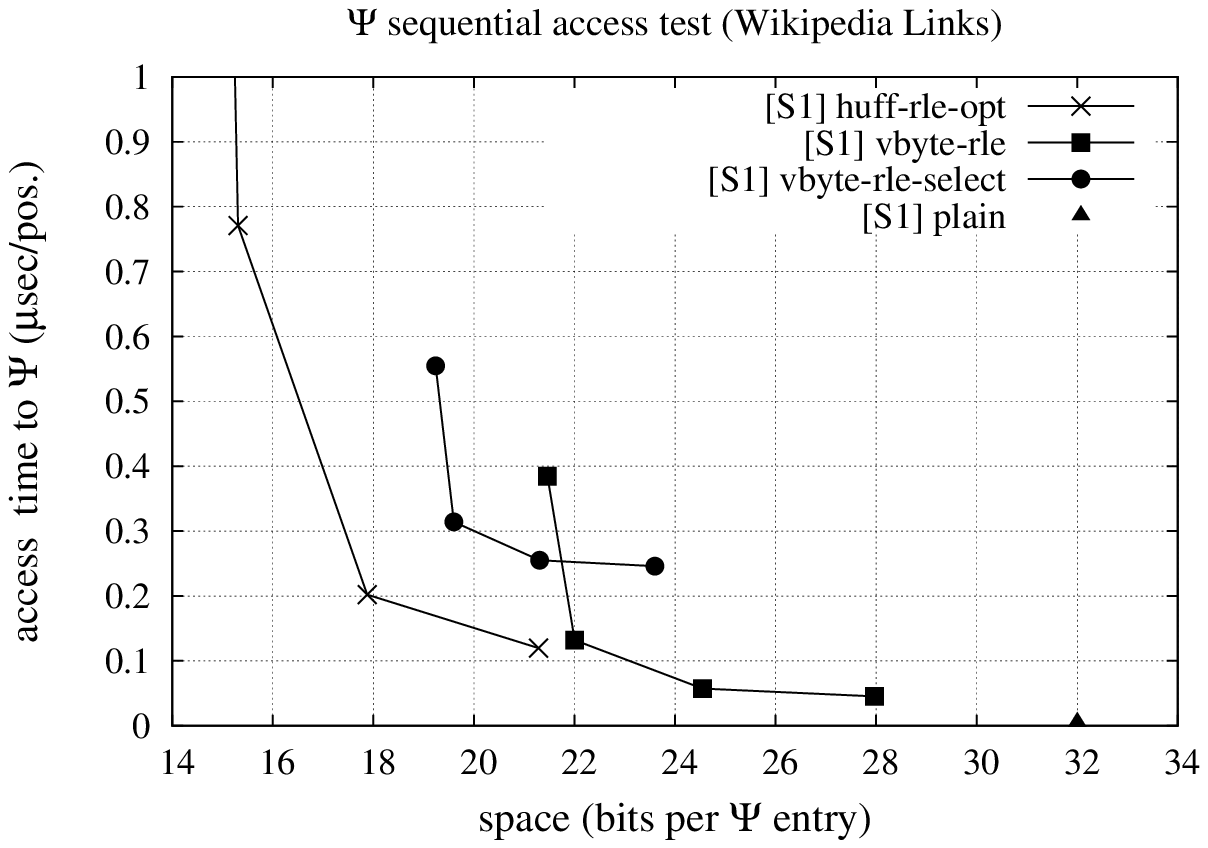}
  \end{center}                    
  \vspace{-0.3cm}
    \caption{Space/time trade-off for sequential access to $\Psi$.}
  \label{fig:psiSequential}

  \end{figure}

In Figures~\ref{fig:psiBuffered} and \ref{fig:psiSequential}, we compare the space (shown as the number of bits needed to represent
each entry in $\Psi$) and time (in $\mu s$ per entry reported) 
required to access all the values in $\Psi$ for three
different scenarios. In the plots labeled by [B1] and [B2], we assume that the ranges $[l_c,r_c]$
for all the symbols $c \in [1,\sigma]$ are known and we perform a buffered access to retrieve the values
$\Psi[l_c,r_c]$ for all these symbols. In scenario [B2], we only retrieve those values $\Psi[l_c,r_c]$ for symbols
occurring at least $8$ times (hence $r_c -l_c -1 \geq 8$).  In these {\em buffered} scenarios, synchronization
is done once to obtain $\Psi[l_c]$ (except in $\mathsf{plain}$ that has direct access and does not require synchronization at all) 
and from there on, we apply sequential decoding of subsequent values.
In the last scenario (plot labeled [S1]), we show the cost of accessing $\Psi$ at 
individual positions (hence synchronization, for the compressed variants, is required for each access to $\Psi$). 
We access sequentially all the positions in $\Psi$,
$\forall j \in [1..4n]$.

We have run tests for all the datasets in Table~\ref{tab:datasets} (described in Section~\ref{sec:experiments})
and show results here
for datasets:  \texttt{I.Comm.Net},  \texttt{Powerlaw}, \texttt{Flickr-Data}, and \texttt{Wikipedia-Links}.
We do not show plots for \texttt{ba*} datasets because they obtain as fairly identical shapes as those for 
\texttt{I.Comm.Net} (yet with slightly different x-axis).

We can see that the cost of the synchronization required by $\huffrle$ and the
slower decoding of bit-Huffman in comparison with vbyte make $\huffrle$ more than 5 times slower than $\vbyte$ when decoding all the entries of $\Psi$ corresponding
to a given symbol $c$.  In Section~\ref{sec:tgcsa:search}, we will see that this particular operation  appears in most \tgcsa\ query
algorithms (a {\em for loop} after a binary search that returns the range of $\Psi$ values for a given symbol).
The  shortcoming of this speed up at recovering $\Psi$ values is that the overall size of $\Psi$ 
increases by around $20$-$25$\%.
As we expected, it can be seen that in the \texttt{Flickr-Data} dataset, due to the large
vocabulary size of this dataset in comparison with the number of contacts, 
the $\vbyte$ representation becomes unsuccessful
because a plain representation of $\Psi$ would even be smaller. We also include results for the $\vbyteselect$ counterpart.
In this case, we do not explicitly store arrays $\offzero$ and $\offone$, and we require $select_1$ operations to know
the position $j$ in $\Psi$ corresponding to the $i$-th sample.  
In general, when the number of synchronization operations is small (this occurs when $\sigma$ is small), 
$\vbyteselect$ offers an interesting space/time trade-off. In particular, we can see that it typically yields the same performance
 of $\mathsf{plain}$ baseline representation while requiring  $5$-$40$\% less space.

%

Unfortunately, not all the accesses to $\Psi$ performed at query time will follow 
a sequential pattern in \tgcsa. In that case, the previous buffered retrieval of $\Psi$ values is not applicable,
and we need to perform many random accesses to positions within $\Psi$. Accessing random positions
implies that each access to $\Psi[j]$ must initially check if $j$ is a sampled position. This is accomplished
by checking if $j\mod t_{\Psi} = 0$ in $\huffrle$ or if $access(D,j)=1$ in $\vbyte$.\footnote{$access(D,j)$ 
returns the value of the bit at position $j$ in the bitmap $D$.} In that
case $\Psi[j] = s_0[\lfloor j/  t_{\Psi} \rfloor]$ or $\Psi[j] = s_0[rank_1(D,j)]$, respectively. Yet, in $\vbyte$
we could still have a sampled value if $access(D_1,j)=1$ and we would obtain the sampled value 
by $\Psi[j] = s_1[rank_1(D_1,j)]$.


In Figure~\ref{fig:psiSequential}, we can see that when we access individual positions of $\Psi$, $\vbyte$
and its two-level sampling approach is still able to improve  the $\Psi$ access time of $\huffrle$. In general,
$\huffrle$ using $t_{\Psi} = 8$ (very dense setup) obtains similar values than $\vbyte$ with $t_{\Psi} = 64$
(a relatively sparse setup). Yet, in $\vbyte$ we still have room to decrease access time at the cost of 
using a denser tuning. As expected, in this scenario, $\vbyteselect$ becomes unsuccessful, and $\mathsf{plain}$ is
unbeatable due to its direct access capabilities.

\subsection{Performing queries in \tgcsa} \label{sec:tgcsa:search}

We can take advantage of the \icsa\ capabilities at search time to solve all the typical queries in a temporal
graph regarding  {\em direct} and {\em reverse} vertexes from contacts that are active at a given time instant $t$
($\directNeighbor$ and $\reverseNeighbor$ queries, respectively).
Basically, we binary search the range in $A[l,r]$ for the given source or target vertex, and for each
position $i \in [l,r]$, we apply $\Psi$ circularly up to the third or four ranges where we can check whether or not the
starting-time and ending-time constrains hold. In Figure~\ref{fig_alg_direct}, we include the 
pseudocode of the algorithms to answer both $\directNeighbor$ and  
$\reverseNeighbor$ queries. Note that they are almost identical with the difference that, in the former, 
the search begins in the range $A[lu,ru]$ corresponding to the source vertex, whereas in the latter the starting range
$A[lv,rv]$ corresponds to the target vertex being searched for.

\begin{figure}[htpb]
\begin{center}
\begin{minipage}{1.0\textwidth}
\begin{code}
\textbf{DirectNeighbors} $(vrtx,t)$~//neighbors ($v$) of $vrtx$ in contact ($vrtx$,$v$,$t_1$,$t_2$) s.t. $t_1\leq t < t_2$ \\
( 1) \> $ u  ~\leftarrow$ \textbf{getmap($vrtx,~  typeVertex=1$)};  ~~ // maps into the final alphabet without holes\\
( 2) \> \textbf{if} $u = 0$ \textbf{then} \textbf{return} $ \emptyset $;  ~~ // vertex does not appear as source vertex \\
( 3) \> $ neighbors  ~\leftarrow~ \emptyset $;  \\
( 4) \> $ t_s ~\leftarrow$ \textbf{getmap$(t,typeStartTime=3)$};  $ t_e ~\leftarrow$ \textbf{getmap$(t,typeEndTime=4)$}; \\
( 5) \> $ [lu,ru]  ~\leftarrow~ $\textbf{CSA\_binSearch($u$)};  ~~ // range $A[lu,ru]$ for vertex $u$ \\
( 6) \> $ [lt_s,rt_s]  ~\leftarrow~ $\textbf{CSA\_binSearch($t_s$)};  ~~ // range $A[lt_s,rt_s]$ for starting time $t_s$ \\
( 7) \> $ [lt_e,rt_e]  ~\leftarrow~ $\textbf{CSA\_binSearch($t_e$)};  ~~ // range $A[lt_e,rt_e]$ for ending time $t_e$ \\

( 8) \> \textbf{for} $i\leftarrow lu$ \textbf{to} $ru$ ~~ // checks time intervals for each occurrence of $u$\\
( 9) \>   \> $ x  ~\leftarrow$ \textbf{$\Psi$}$[i]$;  ~~~~ // $x=$ position of target vertex\\
(10) \>   \> $ y  ~\leftarrow$ \textbf{$\Psi$}$[x]$;  ~~~ // $y=$ position of starting time \\
(11) \>   \> \textbf{if} ($y \leq rt_s$) \textbf{then}    ~~ \\
(12) \>   \>   \> $ z  ~\leftarrow$ \textbf{$\Psi$}$[y]$;   ~~ // $z=$ position of ending time\\
(13) \>   \>   \> \textbf{if} ($z > rt_e$) \textbf{then}  ~~ \\
(14) \>   \>   \>   \> $neighbors  ~\leftarrow~ neighbors \cup \{$\textbf{getunmap$(x,typeRevVertex=2)$}$\}$; ~~ \\

(15) \>    \textbf{return} $ neighbors $; \\
\end{code}
\end{minipage}
\end{center}
\vspace{-0.8cm}
\begin{center}
\begin{minipage}{1.0\textwidth}
\begin{codebottom}
\textbf{ReverseNeighbors} $(vrtx,t)$~//reverse neighbors ($u$) of $vrtx$ in contact ($u$,$vrtx$,$t_1$,$t_2$) s.t. $t_1\leq t < t_2$ \\
( 1) \> $ v  ~\leftarrow$ \textbf{getmap($vrtx,~  typeRevVertex=2$)};  ~~ // maps into the final alphabet without holes\\
( 2) \> \textbf{if} $v = 0$ \textbf{then} \textbf{return} $ \emptyset $;  ~~ // vertex does not appear as target vertex \\
( 3) \> $ rev\_neighbors  ~\leftarrow~ \emptyset $;  \\
( 4) \> $ t_s ~\leftarrow$ \textbf{getmap$(t,typeStartTime=3)$};  $ t_e ~\leftarrow$ \textbf{getmap$(t,typeEndTime=4)$}; \\
( 5) \> $ [lv,rv]  ~\leftarrow~ $\textbf{CSA\_binSearch($v$)};  ~~ // range $A[lv,rv]$ for vertex $v$ \\
( 6) \> $ [lt_s,rt_s]  ~\leftarrow~ $\textbf{CSA\_binSearch($t_s$)};  ~~ // range $A[lt_s,rt_s]$ for starting time $t_s$ \\
( 7) \> $ [lt_e,rt_e]  ~\leftarrow~ $\textbf{CSA\_binSearch($t_e$)};  ~~ // range $A[lt_e,rt_e]$ for ending time $t_e$ \\

( 8) \> \textbf{for} $i\leftarrow lv$ \textbf{to} $rv$ ~~ // checks time intervals for each occurrence of $v$\\
( 9) \>   \> $ y  ~\leftarrow$ \textbf{$\Psi$}$[i]$;  ~~ \\
(10) \>   \> \textbf{if} ($y \leq rt_s$) \textbf{then}    ~~ \\
(11) \>   \>   \> $ z  ~\leftarrow$ \textbf{$\Psi$}$[y]$;   ~~ \\
(12) \>   \>   \> \textbf{if} ($z > rt_e$) \textbf{then}  ~~ \\
(13) \>   \>   \>   \> $ u  ~\leftarrow$ \textbf{$\Psi$}$[z]$; ~~ \\
(14) \>   \>   \>   \> $rev\_neighbors  ~\leftarrow~ rev\_neighbors \cup \{$\textbf{getunmap$(u,typeVertex=1)$}$\}$; ~~ \\
(15) \>    \textbf{return} $ rev\_neighbors $; \\
\end{codebottom}
\end{minipage}
\end{center}
\vspace{-0.3cm}
\caption{Obtaining the direct neighbors ($\directNeighbor$) and the reverse neighbors ($\reverseNeighbor$) 
of a vertex in a contact that is active at time $t$.}
\label{fig_alg_direct}\label{fig_alg_reverse}
\end{figure}

Note that the accesses to $\Psi$ in the \textit{for} loop in line 8 traverse consecutive positions 
$i \in [lu,ru]$ (or $i \in [lv,rv]$ for reverse neighbors). Recall that we do not have {\em direct access} to all the 
values of $\Psi$, but only to sampled positions and the remaining values require accessing 
the previous sample (to gain synchronization on either the Huffman-compressed or Vbyte-compressed stream of gaps) and sequentially decoding gaps from there on up to the desired position (see Section~\ref{psiStudy} for more details). Therefore,
although it is not stated in the pseudocode, we have boosted the access to consecutive positions in $\Psi$ (i.e. $\Psi[lu,ru]$)
by implementing a {\em buffered access} method to $\Psi$. 
%
%
By using this buffered access method to recover $\Psi[lu,ru]$, we only access the sample
before position $lu$, then we synchronize at value $\Psi[lu]$,\footnote{Recall $\Psi[lu]$ 
is always sampled  in $\vbyte$ and no synchronization costs are involved.} 
 and from there on, we sequentially decompress
the remaining values in $\Psi[lu+1,ru]$. The other accesses to $\Psi$ (i.e., $\Psi[x]$ and $\Psi[y]$ 
in $\directNeighbor$)
are completely random and there is no room for optimization there. We will also apply
this buffered access to $\Psi$ in the loops on the following algorithms.

When comparing queries, $\activeEdge$ 
is expected to be faster than  $\directNeighbor$  because we can binary search for a 
phrase $u\cdotp v$ rather than by a unique vertex $u$, hence returning a much shorter initial range. 
The pseudocode for solving the $\activeEdge$ operation at a given time instant is included in Figure~\ref{fig_alg_point_active_Edge}.

\begin{figure}[htb]
\begin{center}
\begin{minipage}{1.0\textwidth}
\begin{code}
\textbf{activeEdge} $(vrtx_u, vrxt_v,t)$~//checks if exists ($vrtx_u$,$vrtx_v$,$t_1$,$t_2$) s.t. $t_1\leq t < t_2$ \\
( 1) \> $ u  ~\leftarrow$ \textbf{getmap($vrtx_u,~  typeVertex=1$)};  ~~ // maps into final alphabet without holes\\
( 2) \> $ v  ~\leftarrow$ \textbf{getmap($vrtx_v,~  typeRevVertex=2$)};  ~~ \\
( 3) \> \textbf{if} $u = 0$ \textbf{or} $v = 0$ \textbf{then} \textbf{return} $ false $;  ~~ // edge does not exist \\
( 4) \> $ t_s ~\leftarrow$ \textbf{getmap$(t,typeStartTime=3)$};  $ t_e ~\leftarrow$ \textbf{getmap$(t,typeEndTime=4)$}; \\
( 5) \> $ [l_{uv},r_{uv}]  ~\leftarrow~ $\textbf{CSA\_binSearch($uv$)};  ~~ // range $A[l_{uv},r_{uv}]$ for edge $uv$ \\
( 6) \> $ [lt_s,rt_s]  ~\leftarrow~ $\textbf{CSA\_binSearch($t_s$)};  ~~ // range $A[lt_s,rt_s]$ for starting time $t_s$ \\
( 7) \> $ [lt_e,rt_e]  ~\leftarrow~ $\textbf{CSA\_binSearch($t_e$)};  ~~ // range $A[lt_e,rt_e]$ for ending time $t_e$ \\

( 8) \> \textbf{for} $i\leftarrow l_{uv}$ \textbf{to} $r_{uv}$ ~~ // checks time intervals for each occurrence of $uv$\\
( 9) \>   \> $ x  ~\leftarrow$ \textbf{$\Psi$}$[i]$;  ~~ \\
(10) \>   \> $ y  ~\leftarrow$ \textbf{$\Psi$}$[x]$;  ~~ \\
(11) \>   \> \textbf{if} ($y \leq rt_s$) \textbf{then}    ~~ \\
(12) \>   \>   \> $ z  ~\leftarrow$ \textbf{$\Psi$}$[y]$;   ~~ \\
(13) \>   \>   \> \textbf{if} ($z > rt_e$) \textbf{then}  ~~ \\
(14) \>   \>   \>   \> \textbf{return} $ true$;  ~~ \\

(15) \>    \textbf{return} $ false $; \\
\end{code}
\end{minipage}
\end{center}
\vspace{-0.3cm}
\caption{Checking if an edge is active at time instant $t$ ($\activeEdge$ operation).}
\label{fig_alg_point_active_Edge}
\end{figure}

To solve $\snapshot$ queries given a time instant $t$, which return the set of active contacts $(u,v,t_1,t_2)$ 
such that $t_1\leq t < t_2$, we can binary search the starting and ending-time intervals: 
$[lt_s,rt_s]\leftarrow$ 
{\em CSA\_binSearch}$(${\em getmap}$(t,3))$ and $[lt_e,rt_e]\leftarrow$ 
{\em CSA\_binSearch}$(${\em getmap}$(t,4))$. All the contacts pointed by $A[2n+1, rt_s]$ 
hold $t_1 \leq t$ and those in $A[rt_e+1, 4n]$ hold $t_2 > t$. 
Therefore, $\forall i \in [2n+1,rt_s]$, if $\Psi[i] >rt_e$, we recover the source and 
target vertexes by $\Psi[\Psi[i]]$ and $\Psi[\Psi[\Psi[i]]]$, respectively. 
The original values are obtained via {\em getunmap()}. Figure~\ref{fig_snapshot_pseudo} includes the
pseudocode to solve  $\snapshot$ queries.

\begin{figure}[htb]
\begin{center}
\begin{minipage}{1.0\textwidth}
\begin{code}
\textbf{snapshot} $(t)$~//returns all the edges ($u$,$v$) s.t. $\exists$ contact ($u$,$v$,$t_1$,$t_2$) where $t_1\leq t < t_2$ \\
( 1) \> $ t_s ~\leftarrow$ \textbf{getmap$(t,typeStartTime=3)$};\\
( 2) \> $ t_e ~\leftarrow$ \textbf{getmap$(t,typeEndTime=4)$};\\ 
( 3) \> $ [lt_s,rt_s]  ~\leftarrow~ $\textbf{CSA\_binSearch($t_s$)};  ~~ // range $A[lt_s,rt_s]$ for starting time $t_s$ \\
( 4) \> $ [lt_e,rt_e]  ~\leftarrow~ $\textbf{CSA\_binSearch($t_e$)};  ~~ // range $A[lt_e,rt_e]$ for ending time $t_e$ \\
( 5) \> $ snap  ~\leftarrow~ \emptyset $;  \\

( 6) \> \textbf{for} $i\leftarrow 2n+1$ \textbf{to} $rt_{s}$ ~~ \\
( 7) \>   \> $ z  ~\leftarrow$ \textbf{$\Psi$}$[i]$;   ~~ \\
( 8) \>   \> \textbf{if} ($z > rt_e$) \textbf{then}  ~~ \\
( 9) \>   \>   \> $ x  ~\leftarrow$ \textbf{$\Psi$}$[z]$;   ~~ \\
(10) \>   \>   \> $ y  ~\leftarrow$ \textbf{$\Psi$}$[x]$;   ~~ \\
(11) \>   \>   \> $ u  ~\leftarrow$ \textbf{getunmap($x,~  typeVertex=1$)};  ~~ \\
(12) \>   \>   \> $ v  ~\leftarrow$ \textbf{getunmap($y,~  typeRevVertex=2$)}; ~~\\
(13) \>   \>   \> $snap  ~\leftarrow~ snap \cup \{(u,v)\}$; ~~ \\
(14) \>    \textbf{return} $ snap $; \\

\end{code}
\end{minipage}
\end{center}
\vspace{-0.3cm}
\caption{$\snapshot$ operation returns the edges that are active at time instant $t$.}
\label{fig_snapshot_pseudo}
\end{figure}

Queries regarding activation/deactivation events at a given time instant $t$ in the graph can be solved very efficiently. 
A unique binary search allows \tgcsa to find all the contacts that have an event at time $t$. In the case of the 
 $\deactivedEdge$ operation, the binary search looks for the range 
$[lt_e,rt_e] \subseteq [3n+1,4n]$ corresponding to contacts ($u,v,t_1, t_2$) where
$t_2= t$, whereas for the $\activedEdge$ operation we obtain an interval 
$[lt_s,rt_s] \subseteq [2n+1,3n] $ corresponding to those contacts where
$t_1 = t$. From these intervals, we apply $\Psi$ circularly (twice or three times, respectively) 
up to reaching the values $u$ and $v$ corresponding to 
the source and target vertex of these contacts. 
In Figure~\ref{fig_deactived_pseudo}, we include the
pseudocode for the $\deactivedEdge$ operation. Note that the  $\activedEdge$ operation would be similar but the loop would traverse positions $i \in [lt_s,rt_s]$ with $x \leftarrow \Psi[\Psi[i]]$
in line 5.

\begin{figure}[htb]
\begin{center}
\begin{minipage}{1.0\textwidth}
\begin{code}
\textbf{DeactivatedEdges} $(t)$~//returns all the edges ($u$,$v$) s.t. $\exists$ contact ($u$,$v$,$t_1$,$t_2$) where $t_2 = t $ \\
( 1) \> $ t_e ~\leftarrow$ \textbf{getmap$(t,typeEndTime=4)$};\\ 
( 2) \> $ [lt_e,rt_e]  ~\leftarrow~ $\textbf{CSA\_binSearch($t_e$)};  ~~ // range $A[lt_e,rt_e]$ for ending time $t_e$ \\
( 3) \> $ edges  ~\leftarrow~ \emptyset $;  \\

( 4) \> \textbf{for} $i\leftarrow lt_e$ \textbf{to} $rt_e$ ~~ \\
( 5) \>   \> $ x  ~\leftarrow$ \textbf{$\Psi$}$[i]$;   ~~ \\
( 6) \>   \> $ y  ~\leftarrow$ \textbf{$\Psi$}$[x]$;   ~~ \\
( 7) \>   \> $ u  ~\leftarrow$ \textbf{getunmap($x,~  typeVertex=1$)};  ~~ \\
( 8) \>   \> $ v  ~\leftarrow$ \textbf{getunmap($y,~  typeRevVertex=2$)}; ~~\\
( 9) \>   \> $ edges  ~\leftarrow~ edges \cup \{(u,v)\}$; ~~ \\
(10) \>    \textbf{return} $ edges $; \\

\end{code}
\end{minipage}
\end{center}
\vspace{-0.3cm}
\caption{$\deactivedEdge$ operation returns the edges that were deactivated at time $t$.}
\label{fig_deactived_pseudo}
\end{figure}

Taking a look at the pseudocodes presented for \tgcsa query operations, we can see that we are using the following operations during searches: (i) {\em getmap} and {\em getunmap} calls that imply performing {\em rank} and {\em select} over $B$ and can be solved in $O(1)$ time. (ii) A call to {\em CSA\_binSearch}($p$)  that requires $O(\log n)$ time and returns a range $A[l,r]$ containing the occurrences of a pattern $p$. Up to two additional calls to {\em CSA\_binSearch}($p'$) could be needed depending on the query, also requiring $O(\log n)$ time.  (iii) A loop traversing the $L=r-l+1$ entries in $A[l,r]$ that involves only $O(1)$ operations, typically {\em getunmap} and accesses to $\Psi$. The exception is the $\snapshot$ operation that traverses always $L \leftarrow rt_s -2n$ entries. To sum up, the temporal queries in \tgcsa can be solved in time $O(\log n + L)$.

\paragraph{\textbf{Dealing with interval queries}}
As indicated in Section~\ref{sec:preliminary}, we have shown how \tgcsa handles  $\directNeighbor$, $\reverseNeighbor$, $\activeEdge$, $\snapshot$, $\activedEdge$, and $\deactivedEdge$ queries at a given time instant $t$. Yet, these operations could be easily extended in \tgcsa to time intervals. In queries that refer to  checking 
the connectivity between vertexes (the first three ones), one would be interested in contacts $(u,v, t_1, t_2)$ occurring not only at a given time instant $t$, but during a whole time interval $[t,t')$; that is, $[t,t')\subseteq [t_1,t_2)$ (this is called {\em strong semantics} for intervals in the literature). A different option (referred to as {\em weak semantics}) consists in reporting those contacts occurring at least at some point of $[t,t')$; that is, such that it holds $[t_1,t_2) \cap [t,t') \neq \emptyset$. Note that for queries retrieving the changes on connectivity
($\activedEdge$ and $\deactivedEdge$), it makes no sense to distinguish between {\em weak} and {\em strong semantics}, 
and we would be interested in simply checking if the connectivity changed at some point of the interval $[t,t')$.

\medskip

If we focus on queries constrained to an interval $[t,t')$ under {\em strong semantics}, 
to solve $\directNeighbor$ queries, we should only adapt the temporal constraint so that contacts 
match $(y \leq rt_s) \textrm{~AND~} (z> rt_e)$. Yet, in this case, $rt_s$ and $rt_e$ must be the right hand of the
ranges $[lt_s, rt_s]$ and $[lt_e, rt_e]$ corresponding to $t$ and $t'$, respectively. 
Therefore, we should modify  line 4 in the pseudocode of Figure~\ref{fig_alg_direct} to set $ t_s\leftarrow$ 
{\em getmap}$(t,3)$ and  $ t_e \leftarrow$ {\em getmap}$(t',4)$; instead of
$ t_s \leftarrow$ {\em getmap}$(t,3)$  and $t_e \leftarrow$ {\em getmap}$(t,4)$.
Algorithms $\reverseNeighbor$ (in Figure~\ref{fig_alg_reverse}) and $\activeEdge$ (in Figure~\ref{fig_alg_point_active_Edge}) could be adapted by simply modifying their line 4 in the same way.

Although not considered in previous works, we could also think of defining a $\snapshot$ operation 
to recover the contacts that were active during the interval $[t,t')$. Under {\em strong semantics}, this interval-wise $\snapshot$ could be defined such that it
would retrieve the contacts that were activated before $t$ and deactivated after $t'$. Therefore, we could see this operation
as the {\em union} of the results of $\snapshot$ at a given time $t_x$, 
$\forall t \leq t_x < t'$. This case would only require
modifying line 2 from Figure~\ref{fig_snapshot_pseudo}, to again set $ t_e ~\leftarrow$ {\em getmap}$(t',4)$.

For $\deactivedEdge$ queries at time interval $[t,t')$ (see Figure~\ref{fig_deactived_pseudo}), 
we would have to replace lines $1-4$ by the following:
First, we map both $t$ and $t'$ values to the ending times $t_s$ and $t_e$; that is, 
$ t_s \leftarrow$ {\em getmap}$(t,4)$ and $ t_e \leftarrow$ {\em getmap}$(t',4)$.
Then, we binary search for the corresponding intervals in \tgcsa: 
$ [lt_s,rt_s]  \leftarrow ${\em CSA\_binSearch($t_s$)} and 
$ [lt_e,rt_e]  \leftarrow ${\em CSA\_binSearch($t_e$)}. And finally, all the ending time instants between $lt_s$
and $lt_e -1$ correspond to contacts deactivated within $[t,t')$. 
Therefore, we have to traverse the entries in that range, that is, we
would iterate (line 4) \textbf{for} $i\leftarrow lt_s$ \textbf{to} $lt_e-1$. A similar adaptation is possible for $\activedEdge$ queries.

\medskip


We can also deal with {\em weak semantics} in \tgcsa. As an example, we show how to adapt 
$\directNeighbor$ queries to this scenario.
The rest of operations can be adapted similarly. Now, a $\directNeighbor$ query for a given 
vertex $u$ constrained to an interval $[t,t')$ must 
retrieve any vertex $v$ from a  contact $(u,v,t_1,t_2)$ 
that were active at some time instant within $[t,t')$. 
Therefore, these contacts must match the time constraint  $(t_1 < t') \textrm{~AND~} (t_2 >t)$.
Focusing on Figure~\ref{fig_alg_direct}, because we need to compare the starting time instant of the contacts ($t_1$)
with $t'$, and their ending time instant ($t_2$) with $t$, we would have to replace line $4$ to set 
$ t_s \leftarrow$ {\em getmap}$(t',3)$  and   $ t_e \leftarrow$ {\em getmap}$(t,4)$.
Finally, the sentences in lines $11-14$ in the for-loop must be changed to modify the temporal condition.
In practice, we replace them by:

\begin{center}
\begin{minipage}{0.8\textwidth}
\begin{codeLarge}

(11) \>   \textbf{if } (($y < lt_s$) \textbf{then}~~\\
(12) \>   \>$z \leftarrow \Psi[y]$;~~ \\
(13) \>   \>\textbf{if } ($z> rt_e$) \textbf{then}~~ \\
(14) \>   \>   \>    $neighbors  ~\leftarrow~ neighbors \cup \{$\textbf{getunmap$(x,typeRevVertex=2)$}\};~~ \\

\end{codeLarge}
\end{minipage}
\end{center}

\subsection {Strengths and weaknesses of  \tgcsa}

The strong expressive power of \tgcsa is probably its main advantage with respect to other state-of-the-art representations such as \edgelog and \cet (\cite{Bernardo:2013iz,DBLP:journals/is/CaroRB15}). Recall \tgcsa  can really represent any set of contacts, including contacts of a given edge that temporally overlap.


Another important property is that it can answer queries over any
term of a contact in the same way; that is, searching for all
the contacts of a source node $u$ is performed exactly with the
same algorithm as searching for all the contacts starting in a
specific time instant $t$: first a binary search is performed over one of
the four sectors of the array $\Psi$, depending on the term of the
contact that is searched for (i.e., bounded in the query), to locate the area
devoted to that value, and then, for each of the entries in that
area,  $\Psi$  is applied three times to recover the other
components of each contact. The overall search time is
$O(\log n + L)$, where $L$ is the length of the range reported by the initial
binary search (with the exception of the $\snapshot$ operation). 
Although other data structures are more efficient 
for some types of queries,  \tgcsa has a more regular behavior over all types of queries. 
Table~\ref{tab:theoAnalysis} compares the cost of the query operations in \tgcsa with 
those of the most representative state-of-the-art counterparts: \cet and \edgelog.
Furthermore, for graphs whose contacts last for only one time instant 
(\textit{Point-contact Temporal Graphs}),  the
behavior of  \tgcsa improves  because  the
suffix array only has three sections and $\Psi$ has only to be applied
twice to recover each contact.

\begin{table}[tbp]
	\scriptsize
	\centering
	\setlength\tabcolsep{4pt} 
	\begin{tabular}{l|c|c|c}
	  \hline
	            Operation         &   \cet     &   \edgelog  &     \tgcsa  \\
	  \hline
	  \hline
			$\directNeighbor$     &    $O(d \log \nu)$   &   $O(d + c)$    & $O(\log n + d ~t_{\Psi})$   \\
			$\reverseNeighbor$    &    $O(d \log \nu)$   &   $O(d^2 + c)$  & $O(\log n + d ~t_{\Psi})$   \\
			$\activeEdge$         &    $O(\log \nu)$     &   $O(d+c')$     & $O(\log n + c' ~t_{\Psi})$  \\
			$\activedEdge$        &    $O( k\log \nu)$   &     $O(n)$      & $O(\log n + k~ t_{\Psi})$   \\
			$\deactivedEdge$      &    $O( k\log \nu)$   &     $O(n)$      & $O(\log n + k ~t_{\Psi})$   \\
			$\snapshot$           &    $O(e \log \nu)$   &     $O(n)$      & $O(\log n + n ~t_{\Psi})$   \\
		\hline
	\end{tabular}
	\caption{Comparison of the costs of the search operations in \tgcsa, \cet, and \edgelog~\cite{DBLP:journals/is/CaroRB15}. The term $d$ denotes the degree of the vertex of the query in the aggregated graph. The term $c$ ($c'$) is the number of contacts related to the vertex (edge) in the query. The term $k$ is the number of contacts starting or ending at the time instant of the queries $\activedEdge$ and $\deactivedEdge$. Finally, $e$ is the number of different edges in the aggregated graph.}
\label{tab:theoAnalysis}
\end{table}

%
  
Observe that within the section devoted to any symbol, in each of
the four quarters of $\Psi$, all the pointers are always growing,
which is a property that allows good compression. However, this
property is also the main drawback of this representation. When
there are few occurrences of the symbols in the vocabulary; that
is, when the vocabulary is huge and there are few occurrences of
each symbol, $\Psi$ will not be very
compressible. As shown  in the experimental results, the
compression in some synthetic collections  is poor
when the relative number of contacts per time instant is low or when the
number of edges per node is low. In these cases,  the
increasing areas of $\Psi$ are  small. Therefore, the differences between
pointer values are rather big, and consequently, not very compressible.





\section{Experimental results} \label{sec:experiments}

We ran several experiments with real and synthetic temporal graphs. Table~\ref{tab:datasets} gives
the main characteristics
of these graphs including: the name of each dataset, the numbers of their vertexes, edges, and contacts,
and the length of the graphs' lifetime. In addition, we show the numbers of contacts per vertex, edges per
vertex, and contacts per edge, respectively. Finally, we show  the
space of a plain representation of the original datasets (in MiB) assuming that each contact was represented 
with four 32-bit integers ($Size^{u32}$), or with $2\lceil \log \nu\rceil + 2 \lceil\log {\tau}\rceil$ bits ($Size^{b}$).

	\begin{table}[hbp]
	\scriptsize
	\centering
	\setlength\tabcolsep{4pt} 
	\begin{tabular}{l|r|r|r|r|r|r|r|r|r}
	  \hline
	  Dataset &   Vertexes &   Edges  &     Lifetime  &   Contacts &  c/$\nu$ &  e/$\nu$ & c/e & Size$^{u32}$ & Size$^b$ \\
	          &    ($\nu$) $\times 10^3$     &   (e) $\times 10^3$   &     ($\tau$) $\times 10^3$ & 
	               (c)  $\times 10^3$ &      &      &     &   (MiB)&   (MiB)\\
	   \hline\hline
	  I.Comm.Net      &     10 &  15,940 &      10 &  19,061 &    1.2 & 1594.1 &    1.2 &    291 &  127 \\
	  Flickr-Data     &  6,204 &  71,345 & 167,943 &  71,345 &    1.0 &   11.5 &    1.0 &  1,089 &  868 \\
	  Powerlaw        &  1,000 &  31,979 &       1 &  32,280 &    1.0 &   32.0 &    1.0 &    493 &  231 \\
	  Wikipedia-Links & 22,608 & 564,224 & 414,347 & 731,468 &    1.3 &   25.0 &    1.3 & 11,161 & 9,417 \\
	  ba100k10u1000   &    100 &     941 &     100 & 941,408 & 1000.0 &    9.4 & 1000.0 & 14,365 & 7,631 \\
	  ba1M10p12       &  1,000 &   9,735 &   1,000 &  50,177 &    5.2 &    9.7 &    5.2 &    766 &  479 \\
	  ba1M10u5        &  1,000 &   9,735 &   1,000 &  48,679 &    5.0 &    9.7 &    5.0 &    743 &  464 \\
	  ba1M10u50       &  1,000 &   9,735 &   1,000 & 486,792 &   50.0 &    9.7 &   50.0 &  7,428 & 4,642 \\
	\hline
	\end{tabular}
	\caption{Description of temporal graphs used in our experiments.}
	\label{tab:datasets}
	\end{table}

The dataset \texttt{I.Comm.Net} is a synthetic dataset where short communications between random
vertexes are simulated.  The dataset  \texttt{Powerlaw}  is also synthetic; it simulates a
power-law degree graph, where few vertexes have many more connections than the
other vertexes (following a power-law distribution), but with a short lifetime.
\texttt{Flickr-Data}  is a real dataset that  
consists in an incremental temporal graph that indicates the time instant in which two people became
friends in the Flickr social network, with a temporal granularity given in seconds, and a lifetime 
that starts with the creation of Flickr and ends in April 2008.
The  dataset \texttt{Wikipedia-Links} contains the history of
links between articles from the English version of the Wikipedia with a time granularity given also in 
seconds. This dataset corresponds to a history dump of the
Wikipedia\footnote{Downloaded from \url{http://dumps.wikimedia.org/enwiki/}.}
downloaded on 2014-03-04.
Other synthetic datasets were built by first setting a given
degree distribution on the aggregated
graph, and then assigning a number of contacts to each
edge that follows a given distribution.
The time interval of each edge was selected uniformly over the lifetime.
We used the Barab\'asi-Albert model~\cite{RevModPhys.74.47} (see datasets ba* below) to
generate a powerlaw degree distribution. Then we used a uniform ($U$) and a pareto ($P$)
distribution to assign the number of contacts per edge.
Pareto distributions were generated with $\alpha =1.2$, whereas
for the uniform distributions, we created graphs 
with $5, 50$, and $1000$ contacts per edge.

Even though \tgcsa allows us to deal with datasets where contacts could have
overlapping times, in order to allow the comparison with \edgelog\ and \cet, the datasets above
have contacts with no time  overlapping. Yet, these datasets still allow us to show the behavior of \tgcsa.

Our tests were run on a machine with two Intel(R) Xeon(R) Intel(R) E5620 CPUs @ 2.40GHz. They sum
 eight-cores (sixteen siblings), yet our experiments run in a single core. The system has  64GB DDR3 RAM @ 1066Mhz. The operating system
was Ubuntu 12.04 (kernel Linux version 3.2.0-79-generic), and the compiler used was gcc 4.6.3 (option -O3).
Time measures refer to CPU user-time.


In the following sections, we include experiments to compare both the space  and time performance of \cet, \edgelog, and
\tgcsa. In particular, we compare the time performance for the following queries: $\directNeighbor$, $\reverseNeighbor$,
$\activedEdge$, $\deactivedEdge$, and $\snapshot$ at a given time instant.

For \edgelog\ and \cet\ we used the same source code as in~\cite{DBLP:journals/is/CaroRB15}.
Therefore, \edgelog\ uses an implementation in {\em C} of $\pfordelta$ from the PolyIRTK
project,\footnote{Available at \url{http://code.google.com/p/poly-ir-toolkit/}.} and the best space
was obtained by tuning $\pfordelta$ block-size to $32$ (rather than the usual $128$ value). In
addition, when the number of elements to compress is smaller than the block size, $\pfordelta$ is
replaced either by the word-wise $\simple$ coding \cite{DBLP:conf/www/ZhangLS08}, when $\tau < 2^{28}$, or 
by $\rice$ \cite{WMB99} when $\tau \geq 2^{28}$ (both are also available in the PolyIRTK project).

The Interleaved Wavelet Tree in \cet\ is implemented as a Wavelet Matrix \cite{CNOis14}, which
keeps a good space/time trade-off for sequences with large
alphabets. Compressed bitmaps \cite{Raman:2007:SID:1290672.1290680,Claude:2008:PRQ:1483948.1483966}
included in \cet\ can be found in the Compact Data Structures
Library ({\em libcds}\footnote{Available at https://github.com/fclaude/libcds}).

The implementation of  \tgcsa is an adaptation of the implementation of \icsa
\footnote{Available at \url{http://vios.dc.fi.udc.es/indexing/wsi}}~\cite{FBNCPR12}. The bitmap
representation used by $D$ is exactly the same  than in  \icsa, whereas bitmap  $B$ uses
the same {\em libcds} implementation of Raman \textit{et al.}~\cite{Raman:2007:SID:1290672.1290680} in
\cet. In addition, \tgcsa uses $\huffrle$ strategy to represent $\Psi$.
We will show results including three different
configurations by setting the sampling parameter on $\Psi$ to values $t_{\Psi} \in \{16, 64, 256\}$. 
Note that  $t_{\Psi}= 16$ ($\Psi_{16}$ in advance)
corresponds to the densest sampling and $\Psi_{256}$ to the most sparse one. 
We have also included
results for \tgcsavb, the variant of \tgcsa that uses the $\vbyte$ strategy to represent $\Psi$.
Again, we set $t_{\Psi} \in \{16, 64, 256\}$ for the second-level sampling in  \tgcsavb.

A further detail is related to the \texttt{Flickr-Data} dataset. In this case, the {\em ending time} of
all the contacts is set to the same value (the last time instant in the timeline). Therefore, we could
avoid representing this value explicitly. We have adapted \tgcsa, and also used adapted versions of
\cet and \edgelog \cite{DBLP:journals/is/CaroRB15}, in order to index only the first three elements of the contacts. 
This reduces (rather slightly) the size of
the resulting structures, and also improves their overall performance. We will include both the regular
\tgcsa and the \tgcsa built over 3-element contacts (\tgcsavvt) when showing time performance
 on the \texttt{Flickr-Data} dataset. 

\subsection{Space comparison}\label{sec:spacecomparison}


Table~\ref{exp:tab:space} shows the comparison   of \tgcsa and \tgcsavb against 
\cet, \edgelog, and a {\em plain} baseline representation using $2 \lceil\log \nu\rceil + 2 \lceil\log \tau\rceil$ bits.
Finally, we also include {\em gzip} in that table (run over the source plain-text-wise
datasets) because this will allow us to compare the compressibility obtained by \icsa\ in our datasets
with that originally obtained when dealing with text \cite{FBNCPR12}. 
Note that for the \texttt{Flickr-Data} dataset we include two rows. The first one refers
to the space obtained by the structures when we assume contacts containing only three elements
, hence excluding the final time instant (the plain baseline uses only $2 \lceil\log \nu\rceil +  \lceil\log \tau\rceil$ $bpc$ = $79 bpc$). 
In the case of \tgcsa, this  corresponds to the variant \tgcsavvt.
The space needs are shown as the number of bits needed to represent each contact ($bpc$).

Even
tough an \icsa-based self-index built on English text typically
reached the compression of {\em gzip} \cite{FBNCPR12}, the
compressibility of temporal graphs is not so good. Actually, the
large number of $1$-runs that appear in $\Psi$ when dealing with
text is now much smaller in the \tgcsa, and we are not able to
reach the compression levels of {\em gzip} in most cases. As expected, taking into account the 
experiments regarding the $\vbyte$ representation of $\Psi$ that we showed in 
Section~\ref{psiStudy}, we typically obtain that \tgcsavb requires around $20$-$30$\% more space 
than \tgcsa. With the \texttt{Flickr-Data} dataset, the space usage of \tgcsavb is huge due to the 
non-parameterizable first-level sampling and the large vocabulary in such dataset.

Focusing on \edgelog, we see that it is also
unsuccessful when the number of contacts per edge is very small.
However, when there are few edges and the number of contacts per
edge grows, it becomes very interesting because its inverted lists become highly
compressible. \tgcsa shows a more stable behavior, with reasonable
space needs in most cases. It does not require as much space as
\edgelog\ when the number of contacts per edge is small, but it
cannot cope with many contacts per edge because  $\Psi$ is irregular, as discussed above.

With respect to \cet, we can see that \cet\  obtains  always a more compact representation
than  \tgcsa, and becomes the best overall alternative if one aims at
obtaining little space cost (with the exception of \texttt{ba100k10u1000} and
\texttt{ba1M10u50} datasets). Yet, in the following sections we will show that \tgcsa typically
performs faster.

\begin{table}[tbp]
\scriptsize
\centering
\setlength\tabcolsep{4pt} 
\begin{tabular}{l||r|r|r||r|r|r||r||r||r|r}
  \hline
  Dataset            & \multicolumn{3}{|c|}{\tgcsa} & \multicolumn{3}{|c|}{\tgcsavb}     & ~\cet\ & \edgelog\ &{\em plain} &{\em gzip} \\
  \cline{2-7}
  & $\Psi_{16}$ & $\Psi_{64}$ & $\Psi_{256}$ &$\Psi_{16}$ & $\Psi_{64}$ & $\Psi_{256}$ &  & ~ &{\em bit-wise}  & {\em def} \\
  \hline\hline
  %
  I.Comm.Net           & 69.36 & 61.17 & 59.17 & 91.68  &  77.17 &  73.34 & 52.28 &  82.48 &  56.00 & 66.13 \\ 
  Flickr-Data          & 82.90 & 77.60 & 76.29 & 139.12 & 132.80 & 131.34 & 49.71 & 187.39 &  79.00 &    -- \\ 
  ~                    & 89.65 & 81.01 & 78.84 & --     &     -- &     -- &    -- &     -- & 102.00 & 97.89 \\ 
  Powerlaw             & 81.66 & 73.85 & 71.92 & 103.88 &  90.69 &  87.50 & 67.97 & 129.88 &  60.00 & 70.01 \\ 
  Wikipedia-Links      & 78.02 & 67.73 & 65.14 & 104.69 &  94.51 &  92.34 & 57.75 & 137.08 & 108.00 & 50.67 \\ 
  ba100k10u1000        & 74.62 & 64.47 & 61.93 & 96.52  &  79.65 &  75.18 & 43.63 &  18.22 &  68.00 & 49.88 \\ 
  ba1M10p12            & 87.42 & 79.51 & 77.54 & 109.03 &  93.96 &  92.04 & 56.35 &  65.77 &  80.00 & 69.63 \\ 
  ba1M10u5             & 92.74 & 85.10 & 83.22 & 115.70 & 100.68 &  98.82 & 61.37 &  67.98 &  80.00 & 72.34 \\ 
  ba1M10u50            & 89.20 & 80.18 & 77.98 & 112.41 &  95.97 &  91.51 & 56.56 &  37.26 &  80.00 & 68.24 \\ 
 
  \hline
\end{tabular}
\caption{Space comparison shown as number of bits per contact ($bpc$). 
For \texttt{Flick-Data} the first row assumes contacts with no ending time.}
\label{exp:tab:space}
\end{table}

\subsection{Time comparison: Direct and Reverse neighbors operations}

This section presents the evaluation of the time performance to retrieve the set of direct and
reverse neighbors that were active at a given time instant. To evaluate these operations, 
we generated $2,000$ queries by randomly choosing $2,000$ contacts from each graph dataset.
For each selected contact $(u,v,t_s,t_e)$, we took the pairs $(u,t_s)$ and ($v,t_s$) to create
the query patterns to use for $\directNeighbor$ and $\reverseNeighbor$, respectively.
The time performance is measured in $\mu s$ per contact reported and the space usage 
in bits per contact (as in Table~\ref{exp:tab:space}).

  \begin{figure}[tpb]
  \begin{center}

  \includegraphics[angle=-90,width=0.32\textwidth]{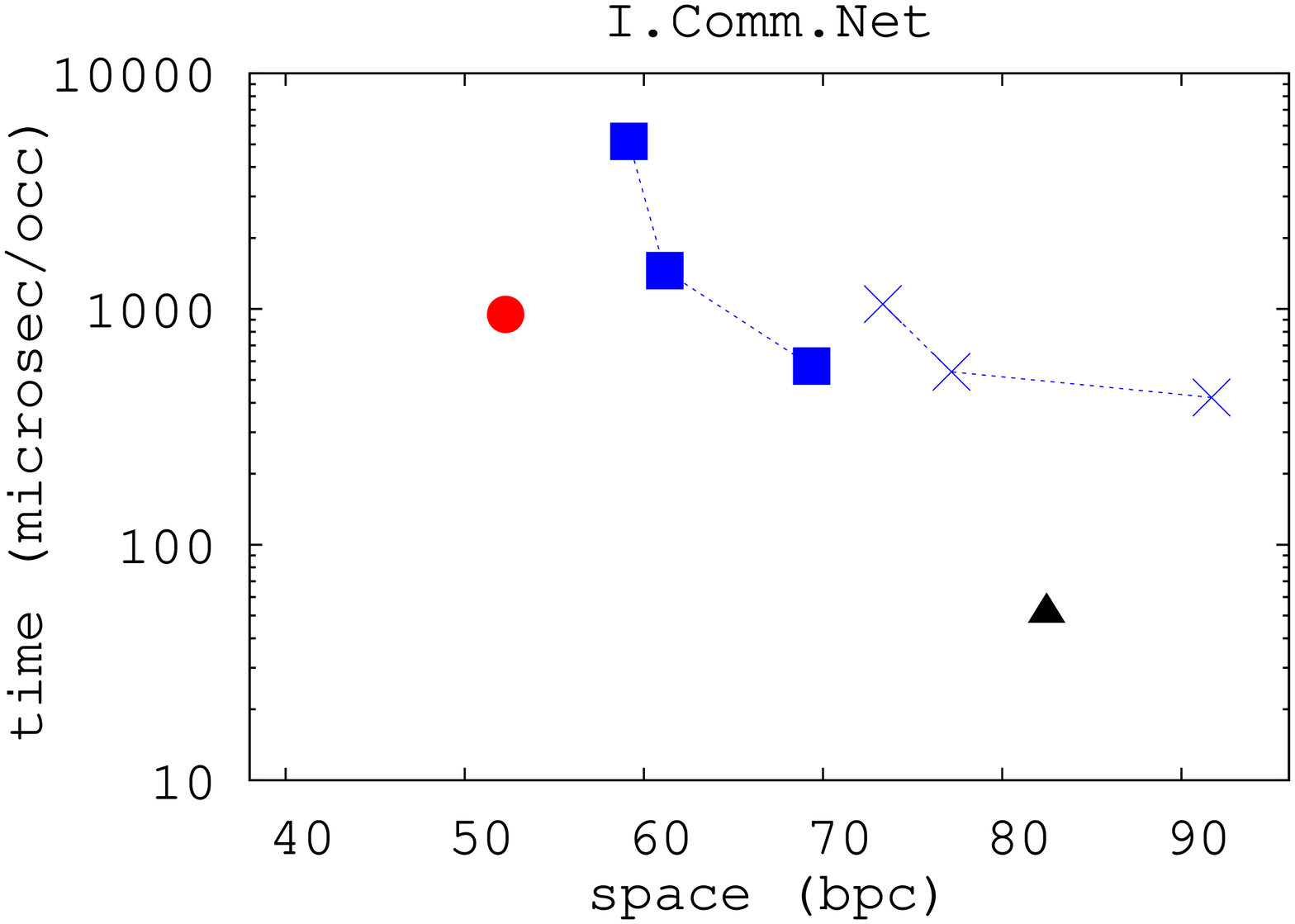}
  \includegraphics[angle=-90,width=0.32\textwidth]{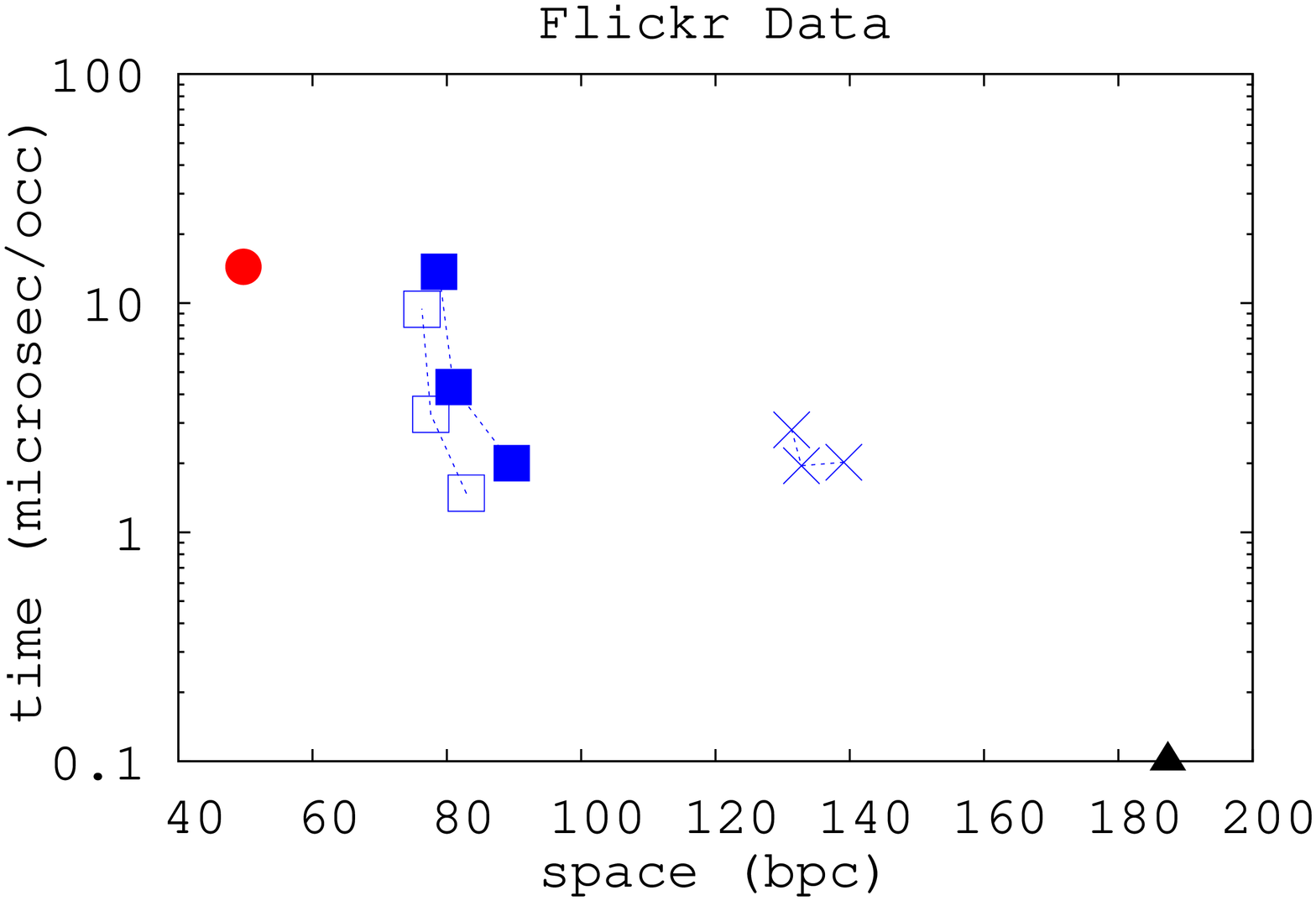}
  \includegraphics[angle=-90,width=0.32\textwidth]{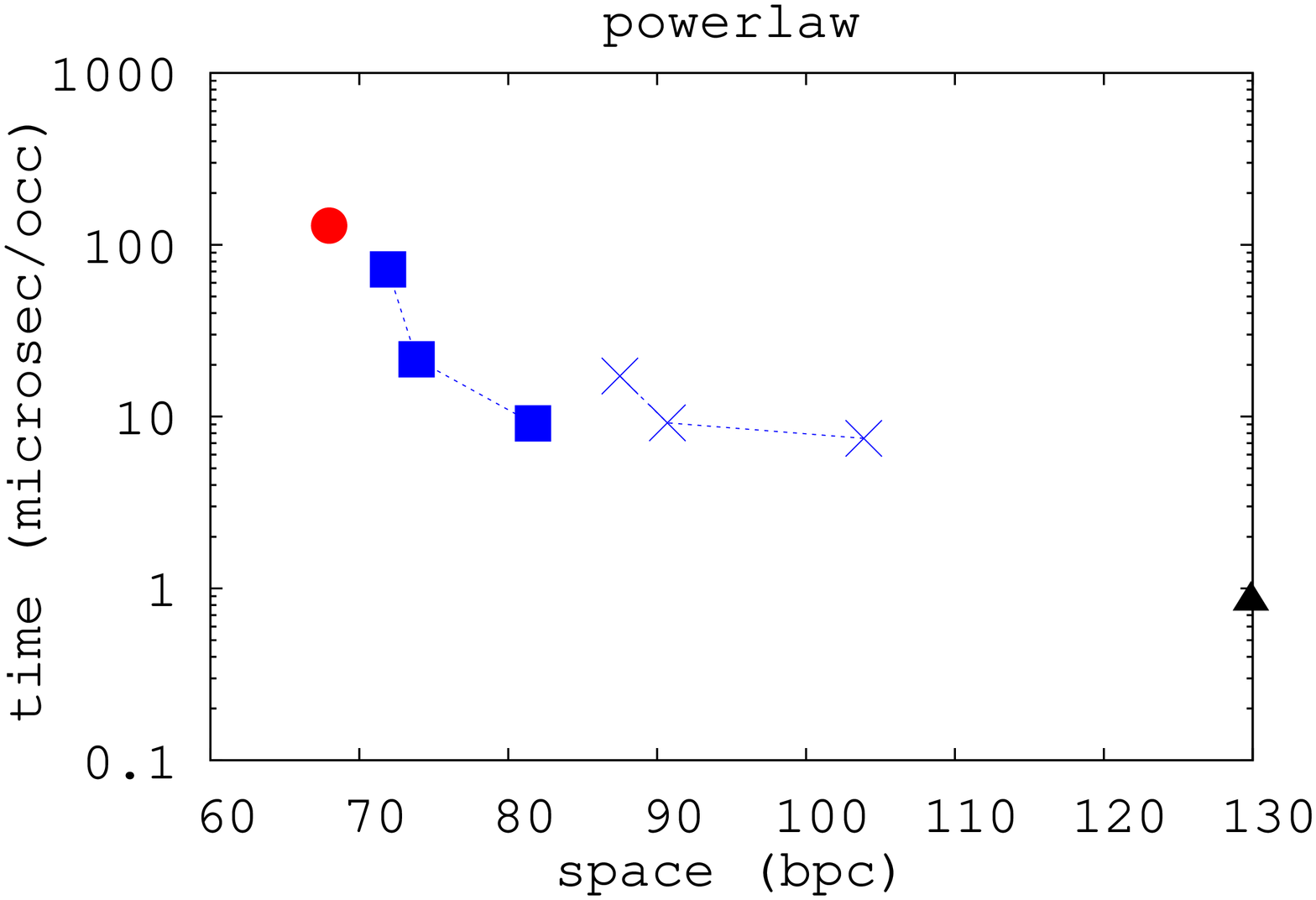}
  \includegraphics[angle=-90,width=0.32\textwidth]{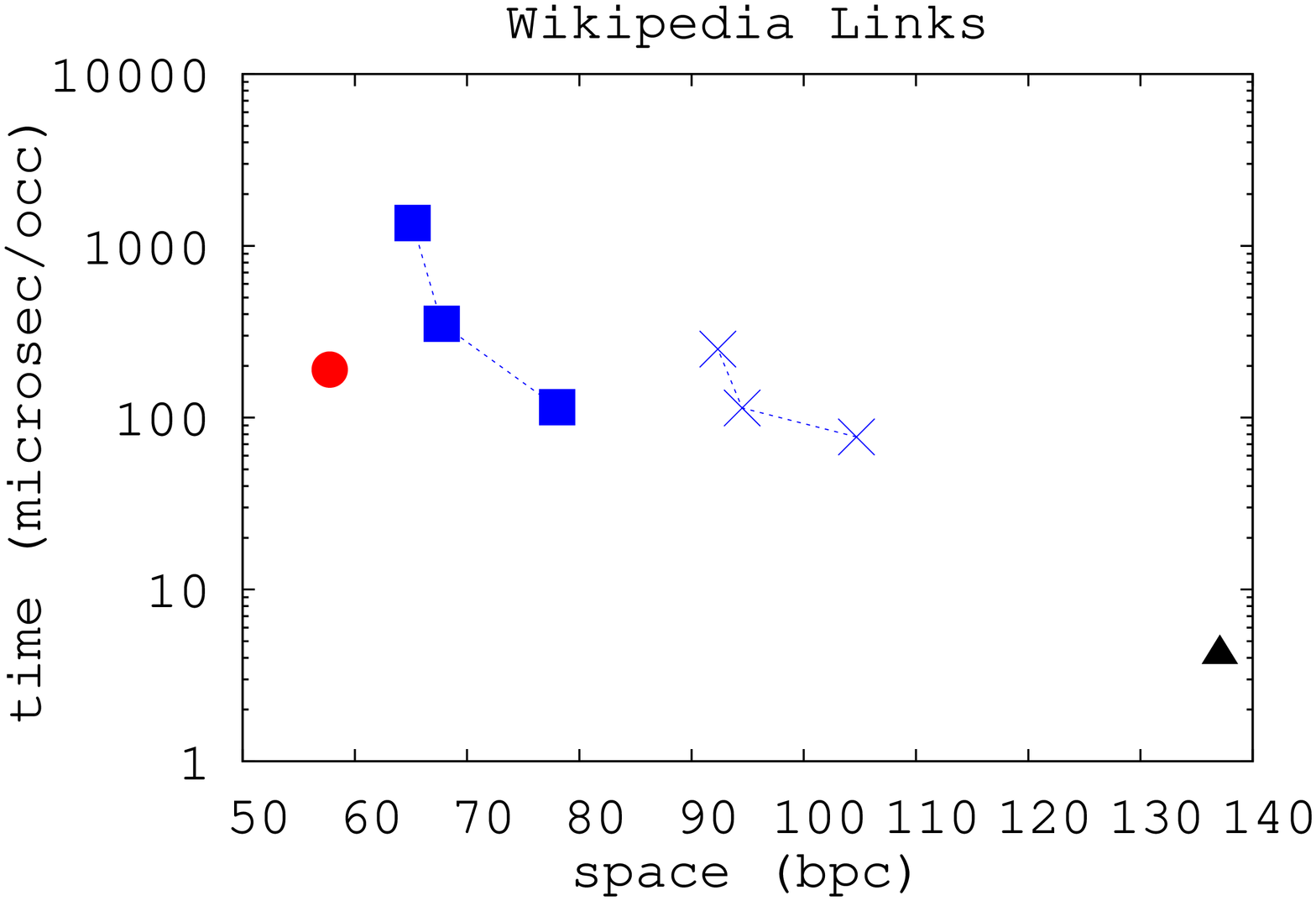}
  \includegraphics[angle=-90,width=0.32\textwidth]{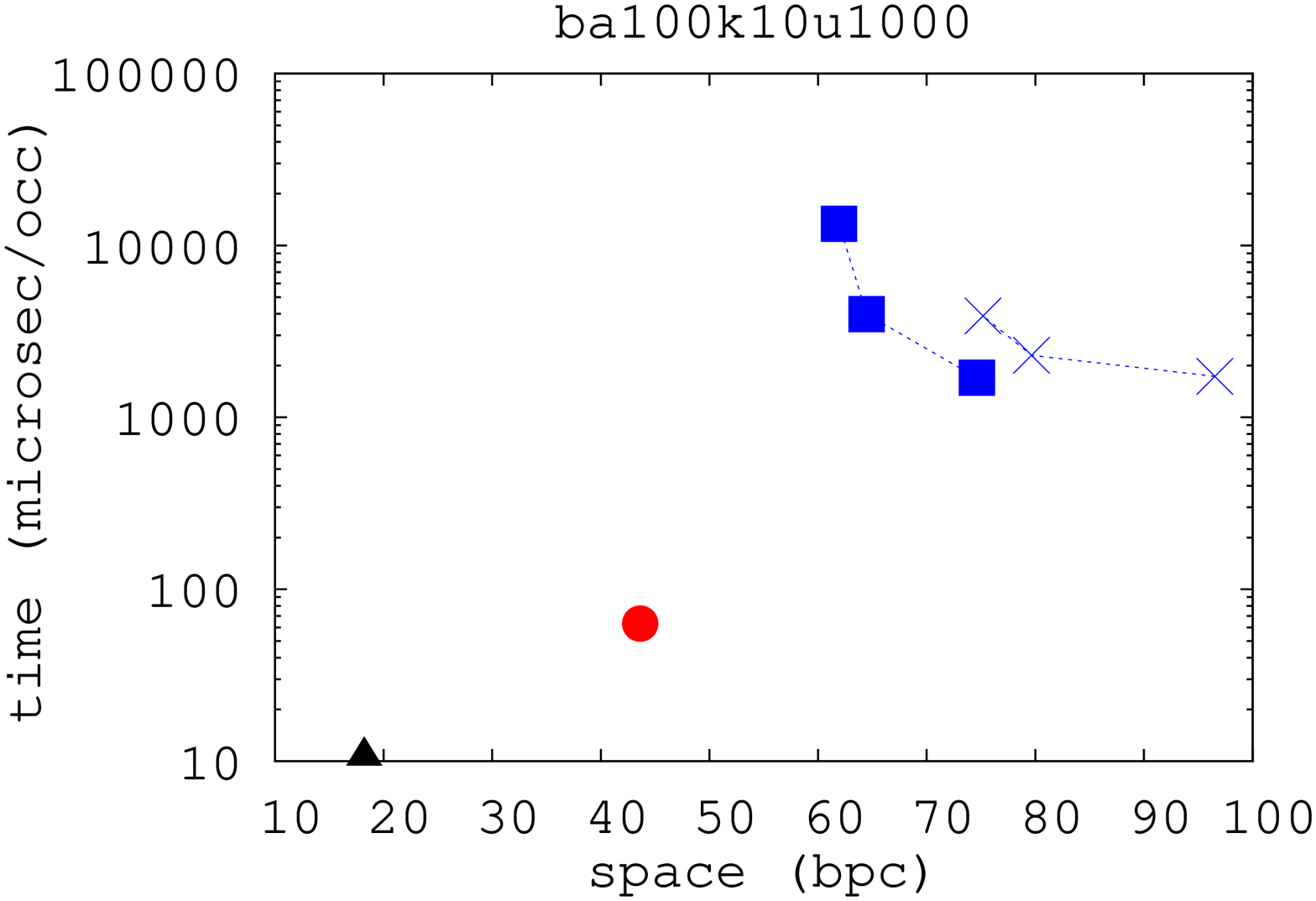}
  \includegraphics[angle=-90,width=0.32\textwidth]{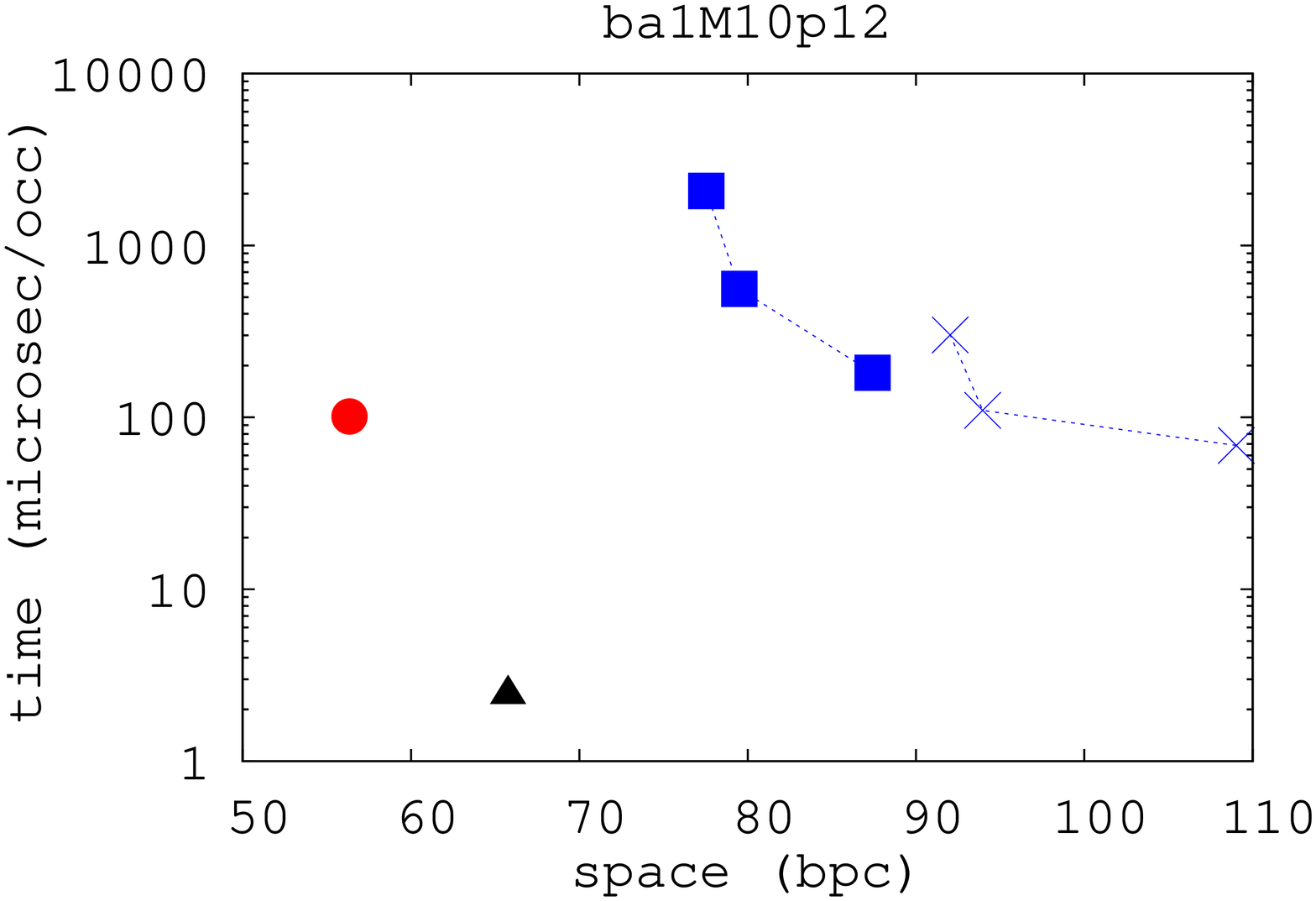}
  \includegraphics[angle=-90,width=0.32\textwidth]{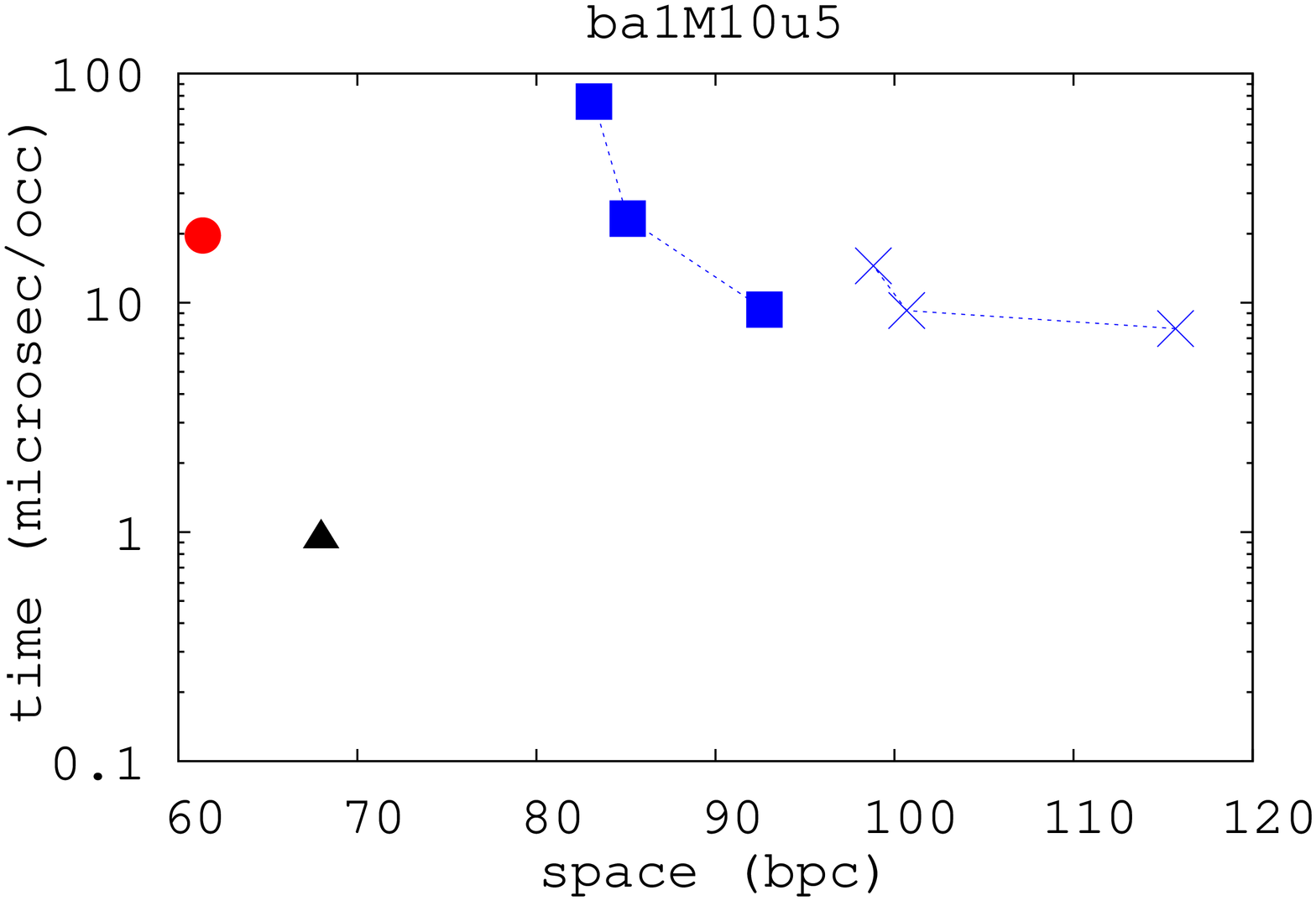}
  \includegraphics[angle=-90,width=0.32\textwidth]{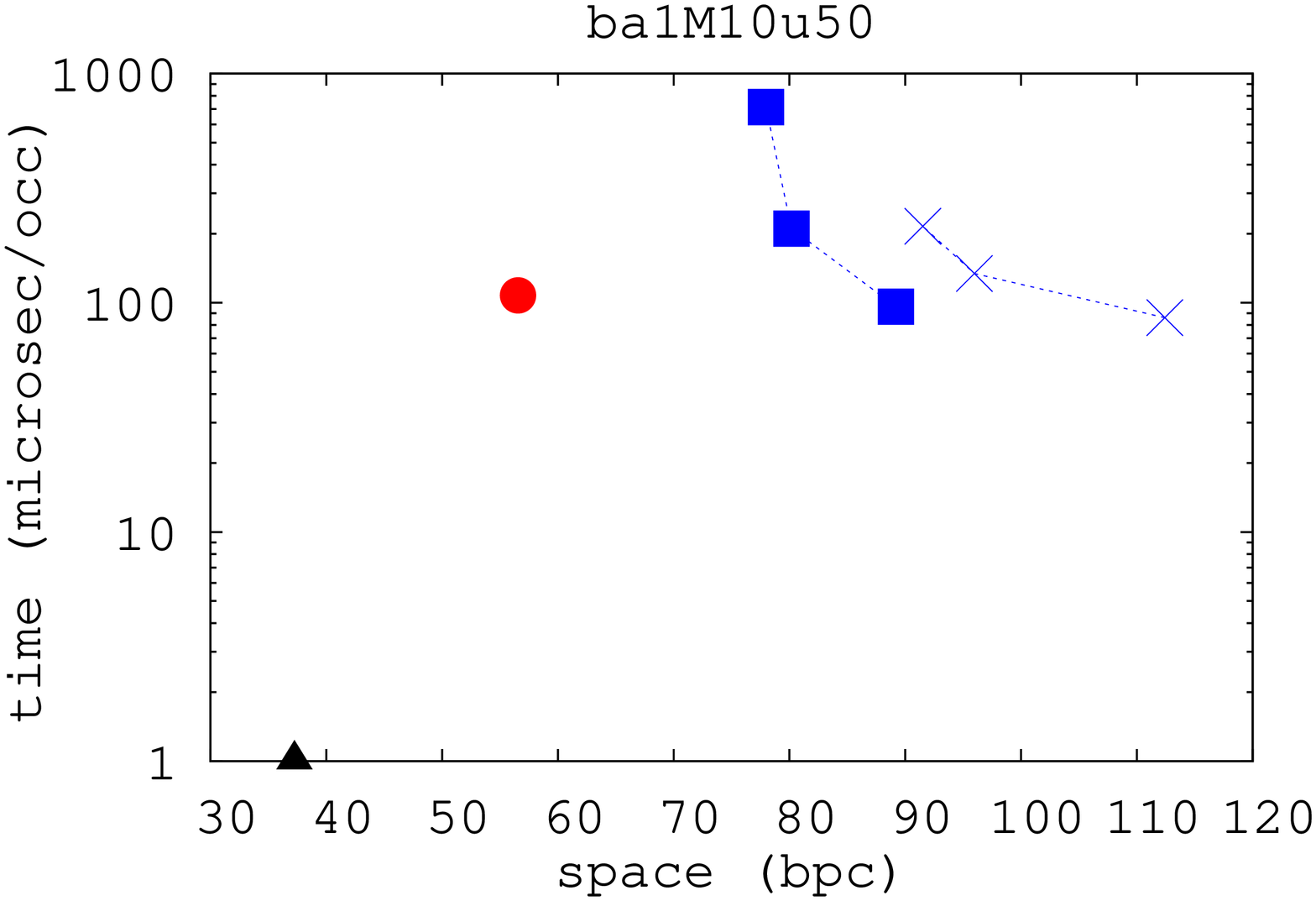}
  \includegraphics[angle=-90,width=0.32\textwidth]{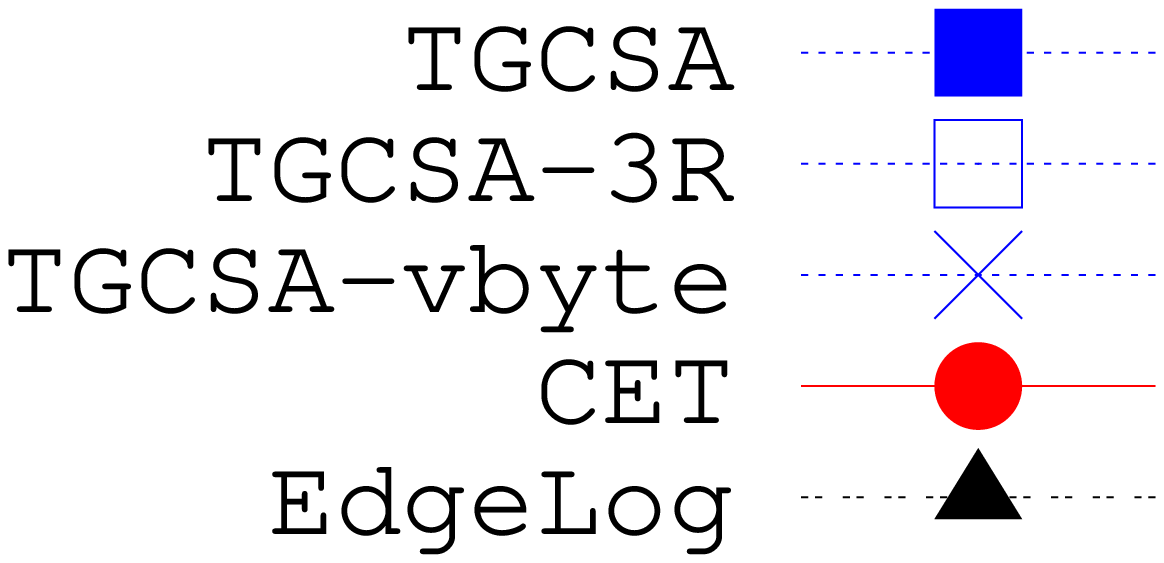}
  \end{center}
  \vspace{-0.3cm}
  \caption{Space/time trade-off for $\directNeighbor$ queries.}

  \label{fig:dirnei}
  \end{figure}

  \begin{figure}[tpb]
  \begin{center}

  \includegraphics[angle=-90,width=0.32\textwidth]{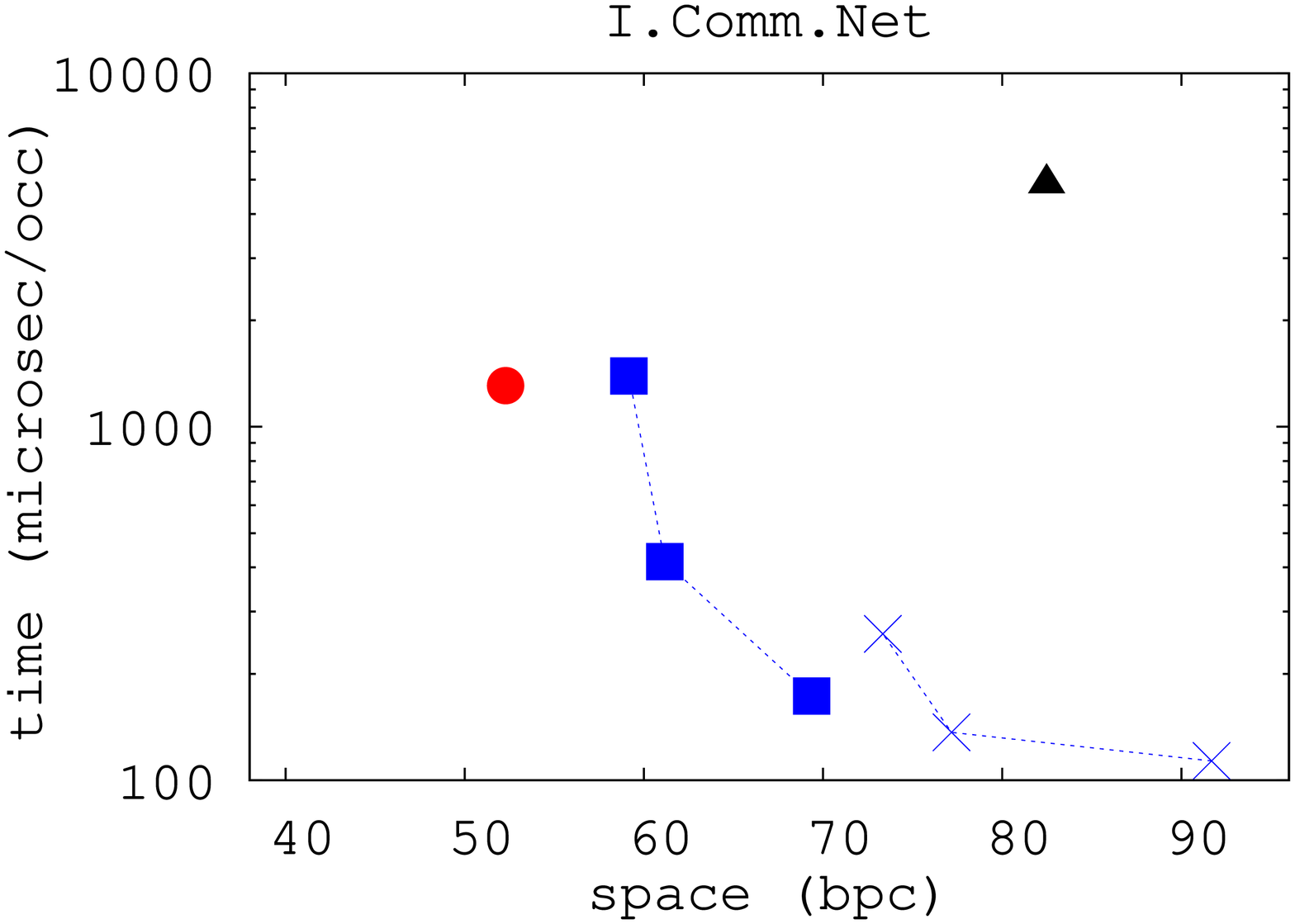}
  \includegraphics[angle=-90,width=0.32\textwidth]{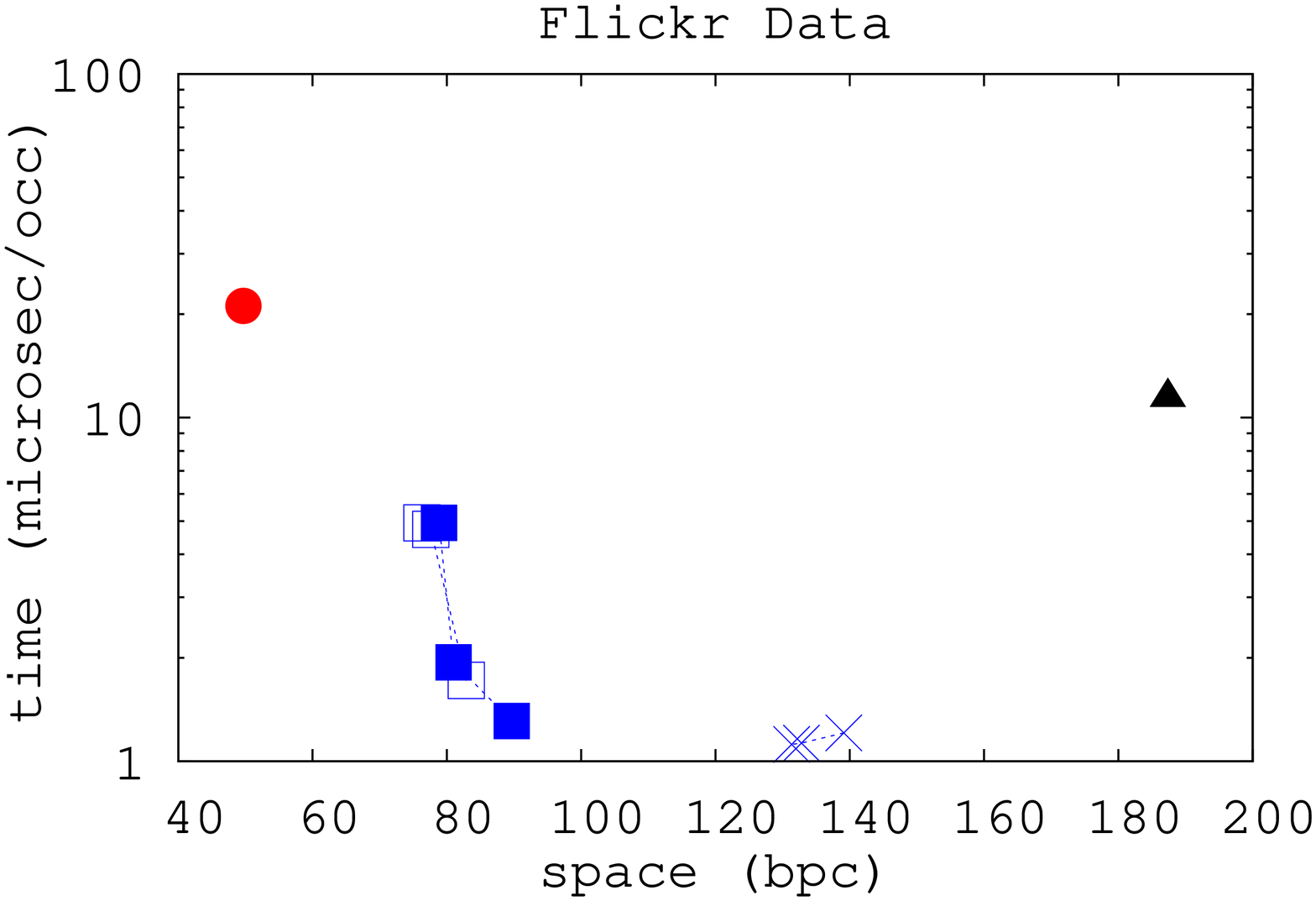}
  \includegraphics[angle=-90,width=0.32\textwidth]{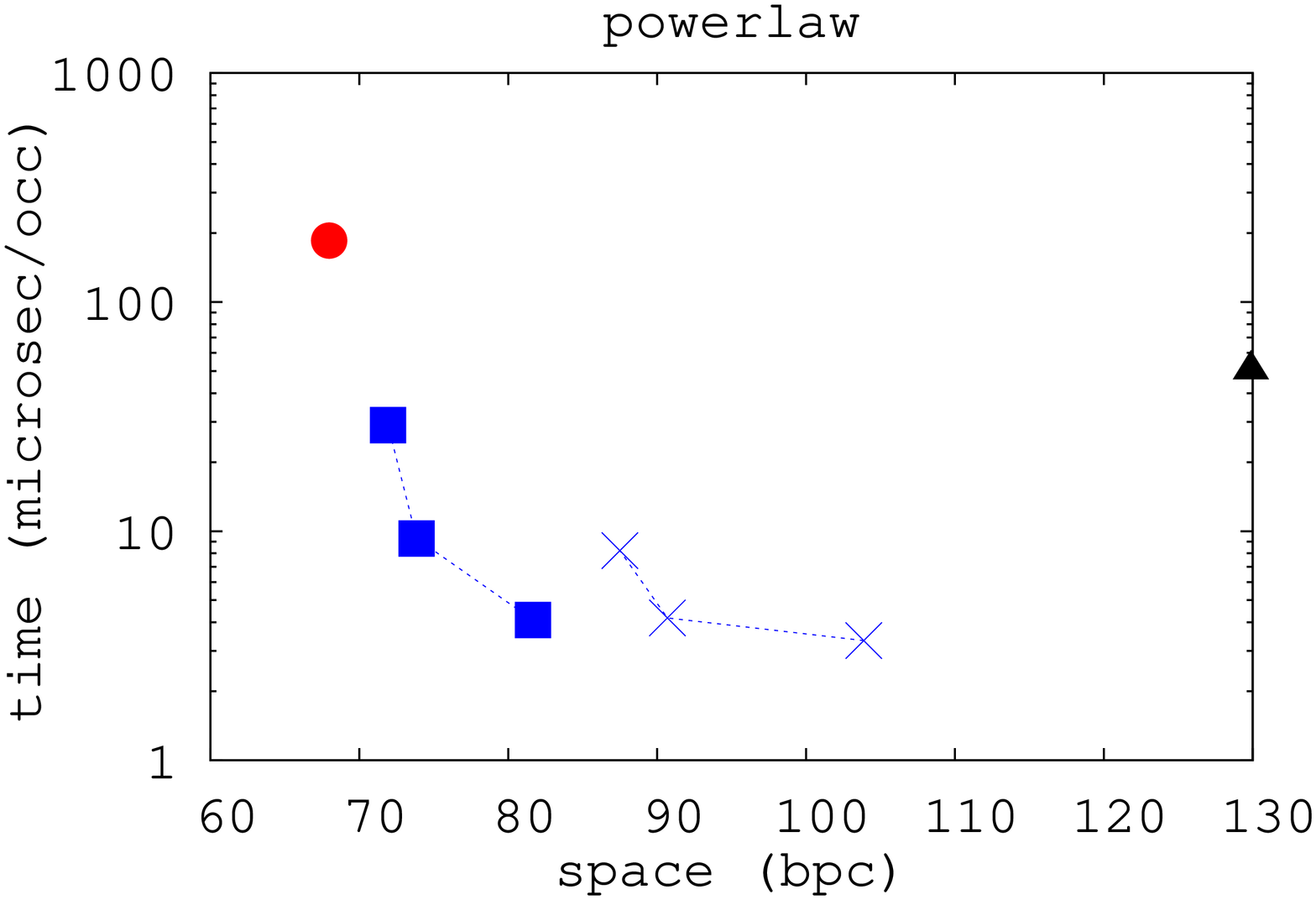}
  \includegraphics[angle=-90,width=0.32\textwidth]{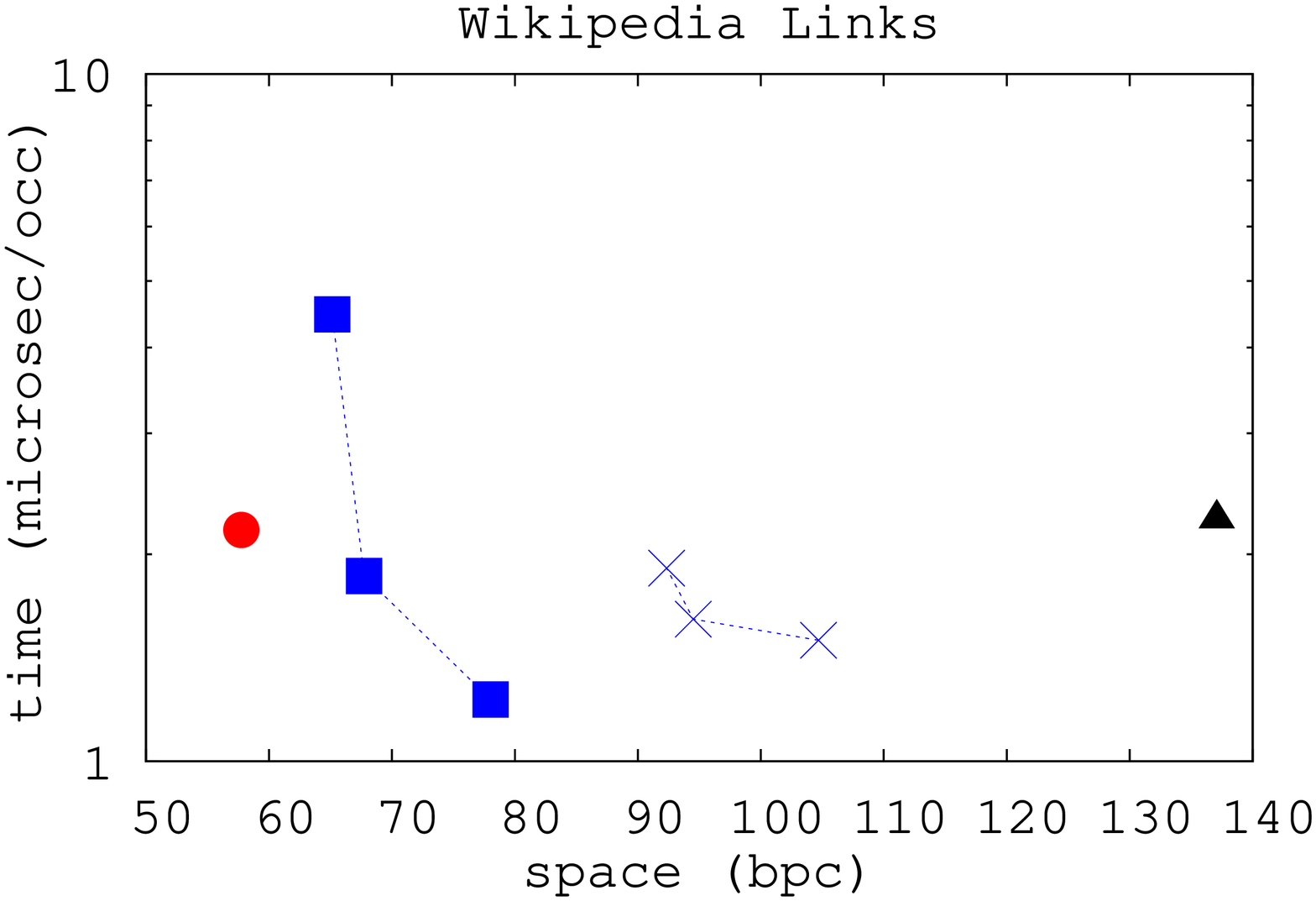}
  \includegraphics[angle=-90,width=0.32\textwidth]{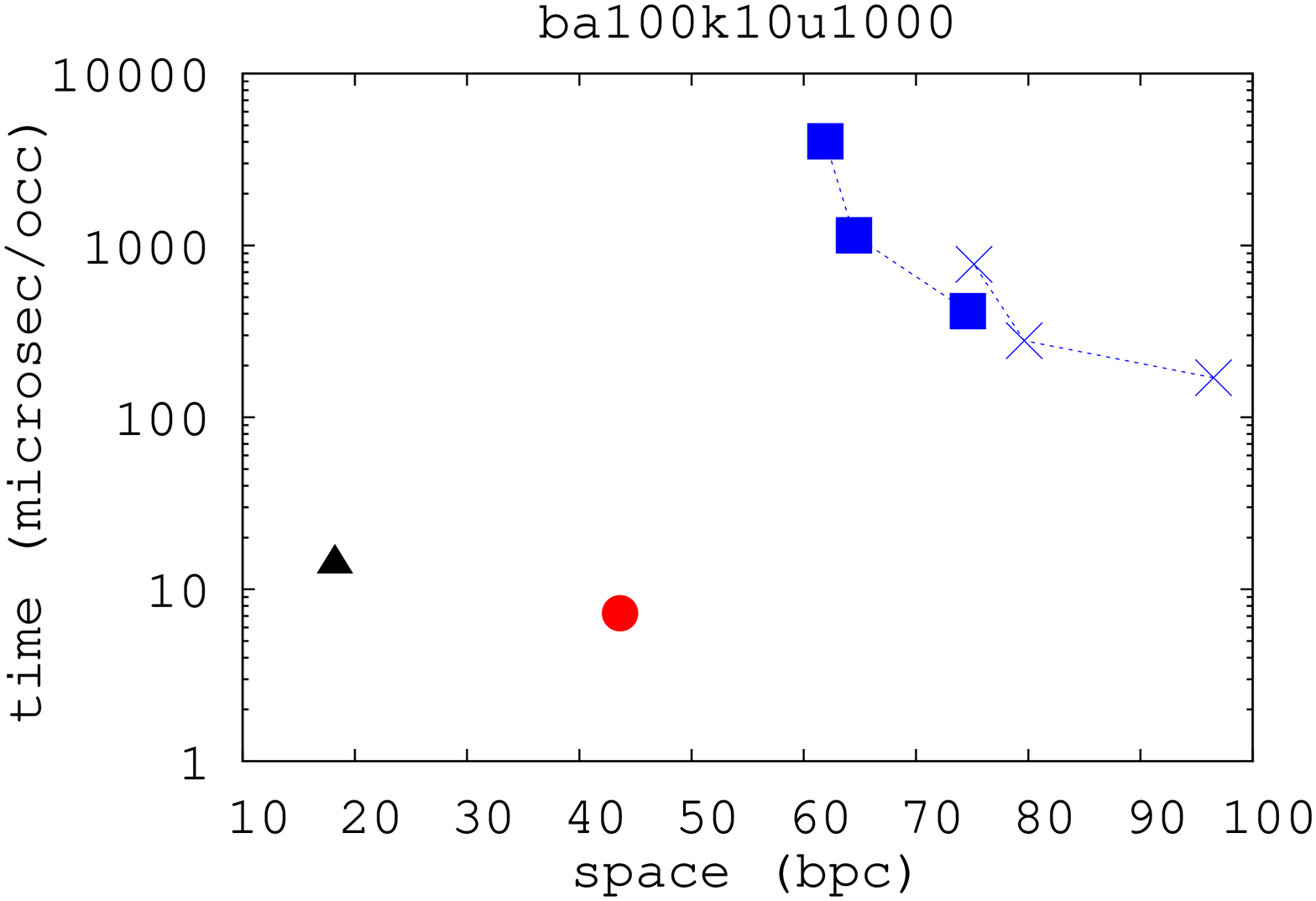}
  \includegraphics[angle=-90,width=0.32\textwidth]{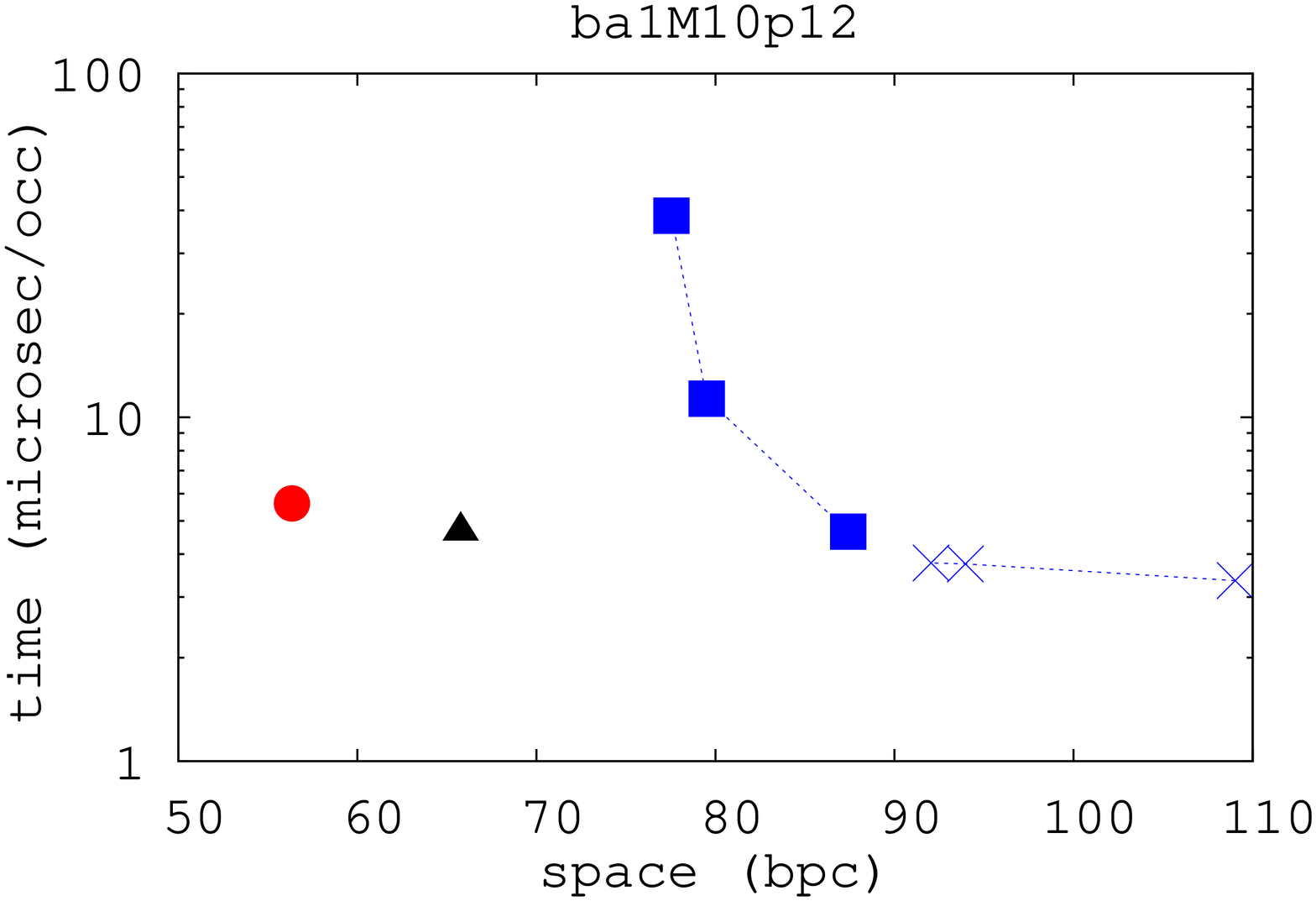}
  \includegraphics[angle=-90,width=0.32\textwidth]{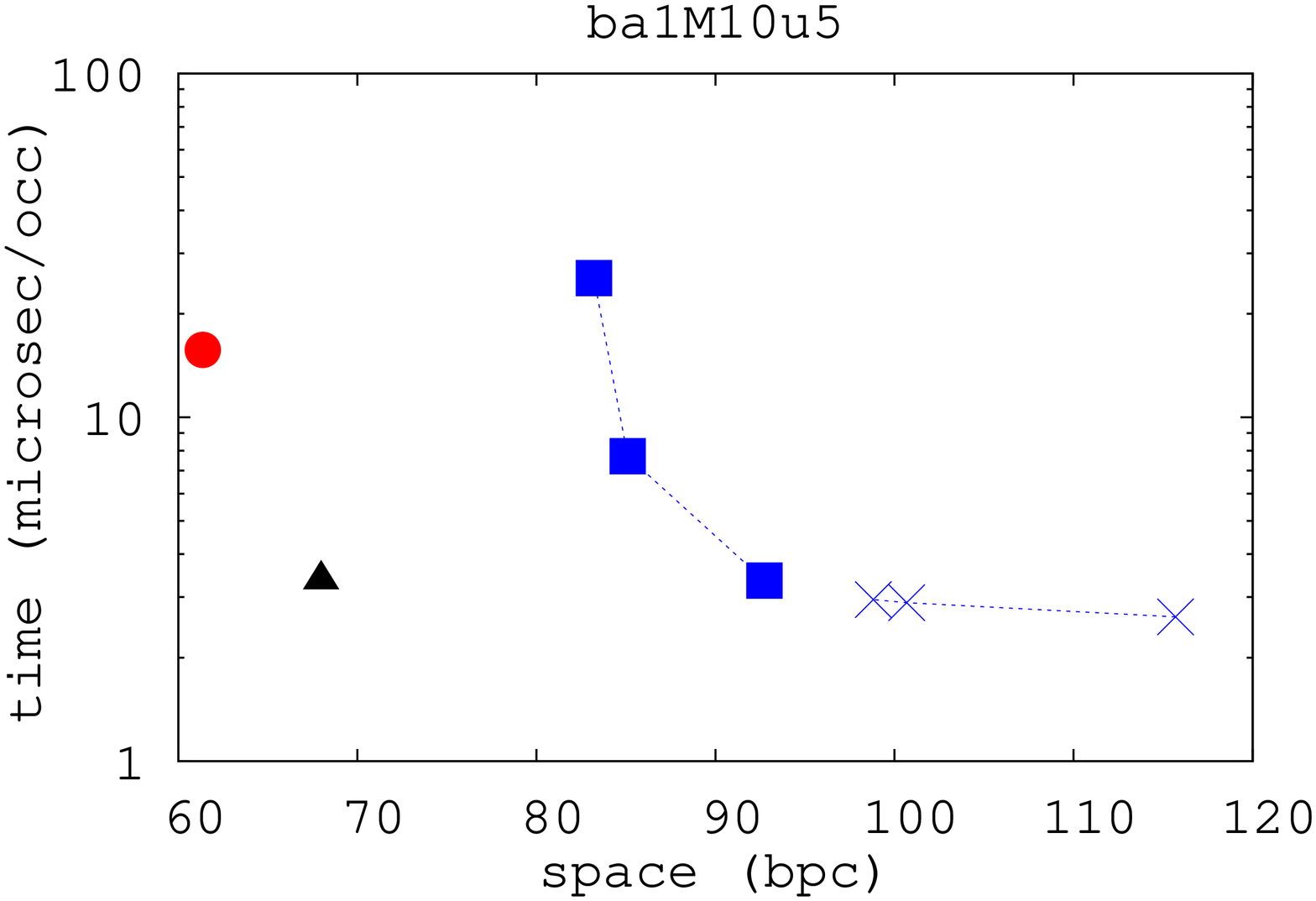}
  \includegraphics[angle=-90,width=0.32\textwidth]{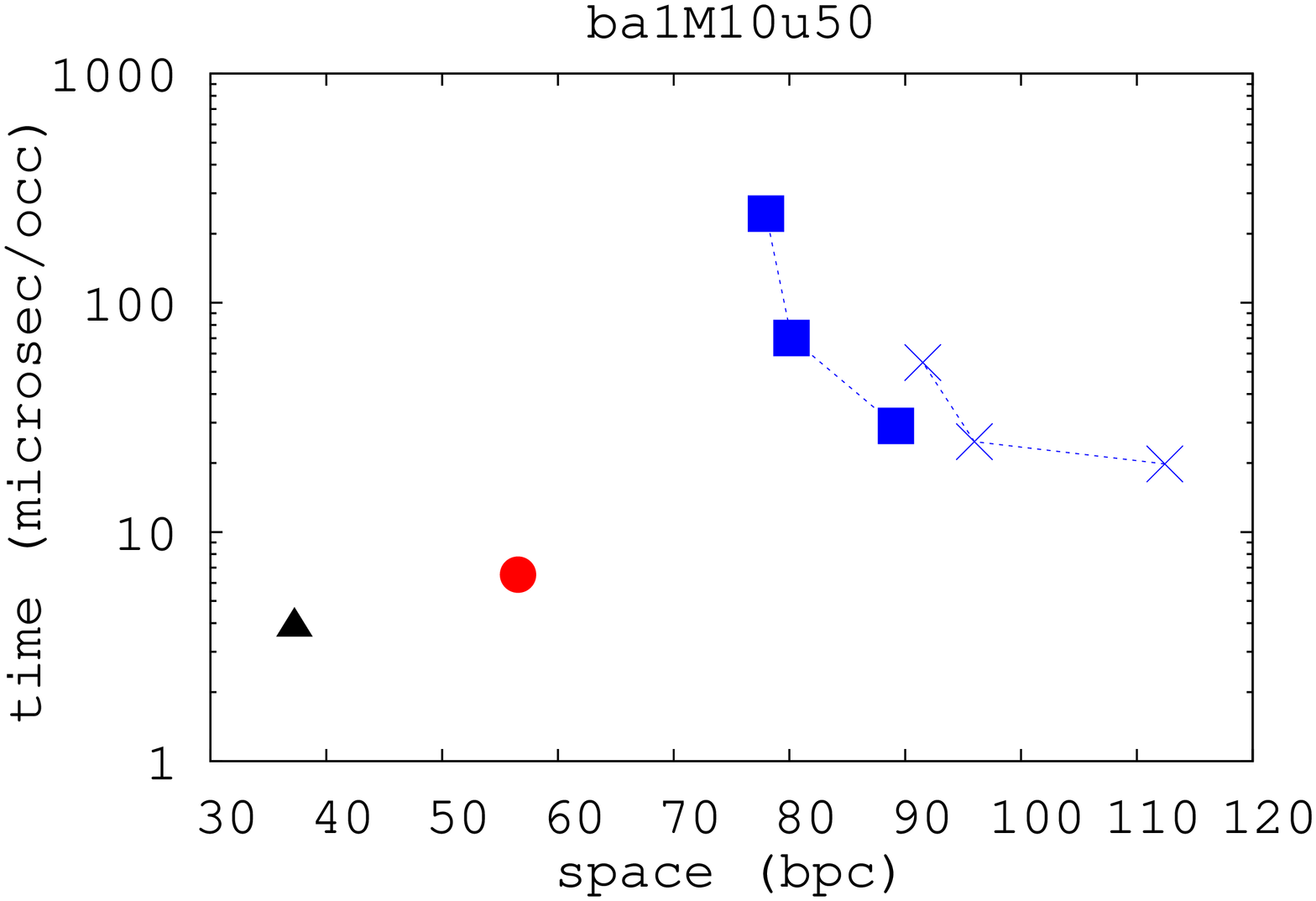}
  \includegraphics[angle=-90,width=0.32\textwidth]{dirnei-9legend.eps}
  \end{center}
  \vspace{-0.3cm}  
  \caption{ Space/time trade-off for $\reverseNeighbor$ queries.}

  \label{fig:revnei}
  \end{figure}

Figures \ref{fig:dirnei} and  \ref{fig:revnei} show the results.
  Despite the fact that \tgcsa uses always more space than \cet\ to represent our temporal graphs,
  we can see that both techniques have similar performance at solving $\directNeighbor$ queries when the
  number of contacts per vertex is small. The only exception is the synthetic
  dataset \texttt{ba100k10u1000} where there are $1,000$ direct neighbors for each vertex, which
  forces \tgcsa to sequentially check a lot of probably unsuccessful direct neighbors. We
  can see that in the \texttt{Powerlaw} and \texttt{Flickr-Data} datasets, \tgcsa clearly overcomes
  \cet. Considering \tgcsavb, it is typically faster (around 3-5 times) than \tgcsa when using the densest
  sampling setup. Yet, assuming that we could tune \tgcsavb and \tgcsa to use similar space, \tgcsavb would
  always be slower than \tgcsa because it would use a very sparse sampling.
  
  Finally, in the plot corresponding to the \texttt{Flickr-Data} dataset,  we show the gain in both
  space and time that \tgcsavvt obtains with respect to \tgcsa. As shown, it is worth not to
  explicitly represent the fourth component (ending-time) of the contacts for incremental graphs.
  When comparing \tgcsavvt  with \edgelog, results show that solving $\directNeighbor$ queries
  is indeed one of the main strengths of \edgelog, because  \edgelog\ only needs to  traverse the corresponding adjacency list.

  \medskip

  With respect to $\reverseNeighbor$ queries, we can see similar results as for $\directNeighbor$ queries
  when comparing \cet\ with \tgcsa. Yet, now we can see that  \tgcsa (and \tgcsavb) are clearly
  faster to solve reverse- instead of direct-neighbors operations, whereas the results of \cet\ are very similar for both types of operations.

  It is easy to understand why \tgcsa is faster at $\reverseNeighbor$ queries than
  at $\directNeighbor$ operations. Note that the time instants are the third and forth elements of
  the contacts, and the source vertex and target vertex are, respectively, the first and second elements.
  Therefore, in the case of $\directNeighbor$ operations \tgcsa must traverse a range $[l,r]$ of
  source vertexes $i\in [l,r]$ and it has to  apply $\Psi\Psi[i]$ and $\Psi^3[i]$, respectively, to
  reach the starting and ending time instants (in order to either accept or discard the contact due to the time constraints).
  In the case of $\reverseNeighbor$ operations, the  traversal starts in the range of the target vertexes,  
  and we save one application of $\Psi$ to reach the time components of the contact (we apply $\Psi[i]$
  and $\Psi[\Psi[i]]$, respectively, to reach the starting and ending time instants of the contact). Recall that in
  these operations, the first application of $\Psi$ to obtain $\Psi[i]$ is performed over a range 
  of consecutive positions $i \in [l,r]$, which benefits from the buffered access to $\Psi$. From there on,
  obtaining $\Psi\Psi[i]$ or $\Psi\Psi\Psi[i]$ requires, respectively, one or two (slower) 
  additional random accesses to $\Psi$.

  As expected, \edgelog performance drastically worsens in $\reverseNeighbor$ queries.
  Yet, the use of the reverse aggregated graph still allows a good performance in most cases.
  The exception is  in the \texttt{I.Comm.Net} graph, where the number of edges per vertex is high. In
  the other cases, the number of edges per vertex is relatively small (from $10$ to $30$) and
  the time performance does not degrade in excess.

\subsection{Time comparison: Activation and deactivation at a given time instant}

This section shows the performance of $\activedEdge$ and $\deactivedEdge$ queries;
that is, retrieving the set of edges that have been either activated or deactivated at a given time instant.
For the evaluation, we generated $2,000$ random time instants, uniformly distributed over
the lifetime of the corresponding graph.
Again, time measures are shown as the average time in $\mu s$ per contact reported.

\begin{figure}[tbp]
\begin{center}

\includegraphics[angle=-90,width=0.32\textwidth]{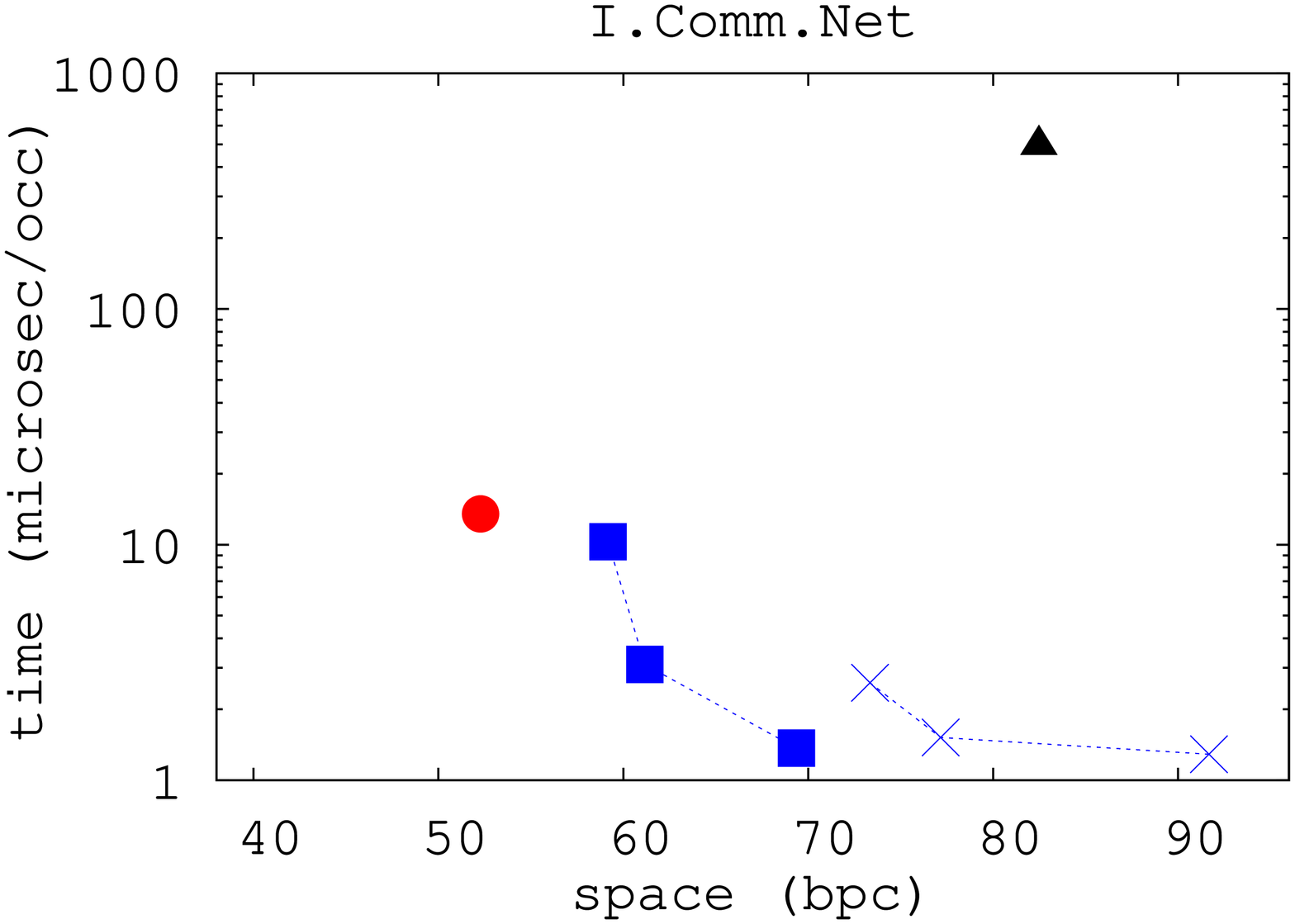}
\includegraphics[angle=-90,width=0.32\textwidth]{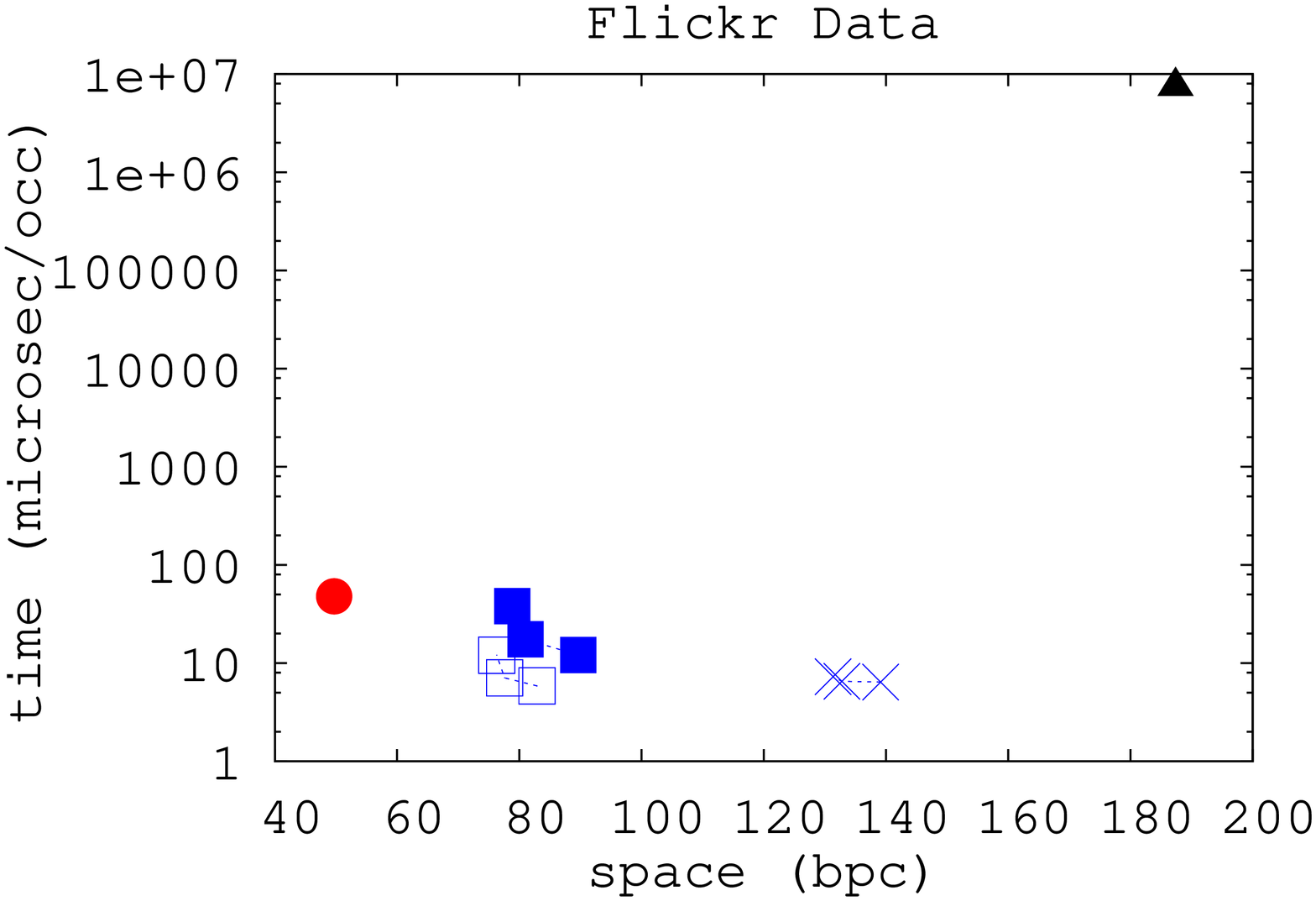}
\includegraphics[angle=-90,width=0.32\textwidth]{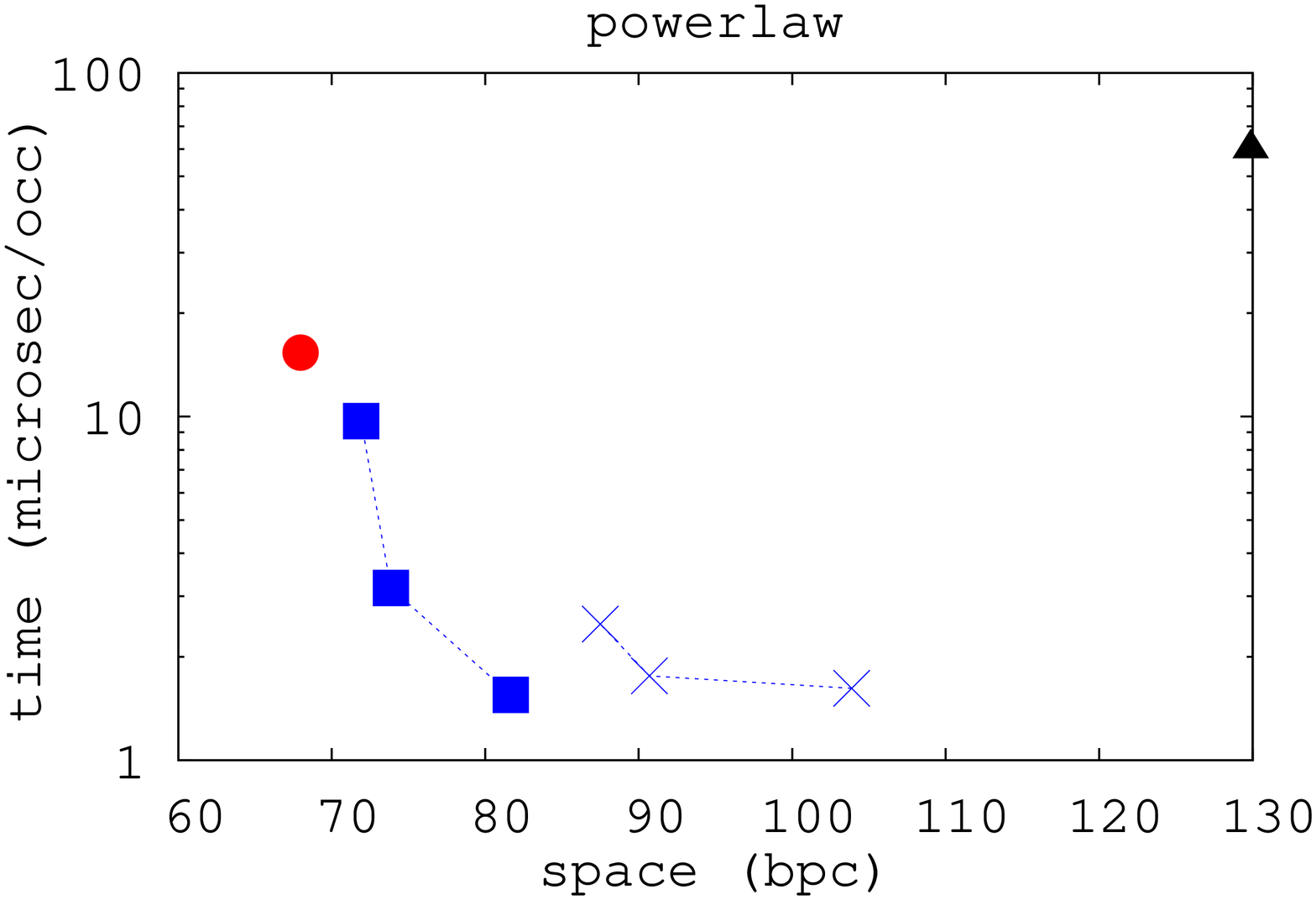}
\includegraphics[angle=-90,width=0.32\textwidth]{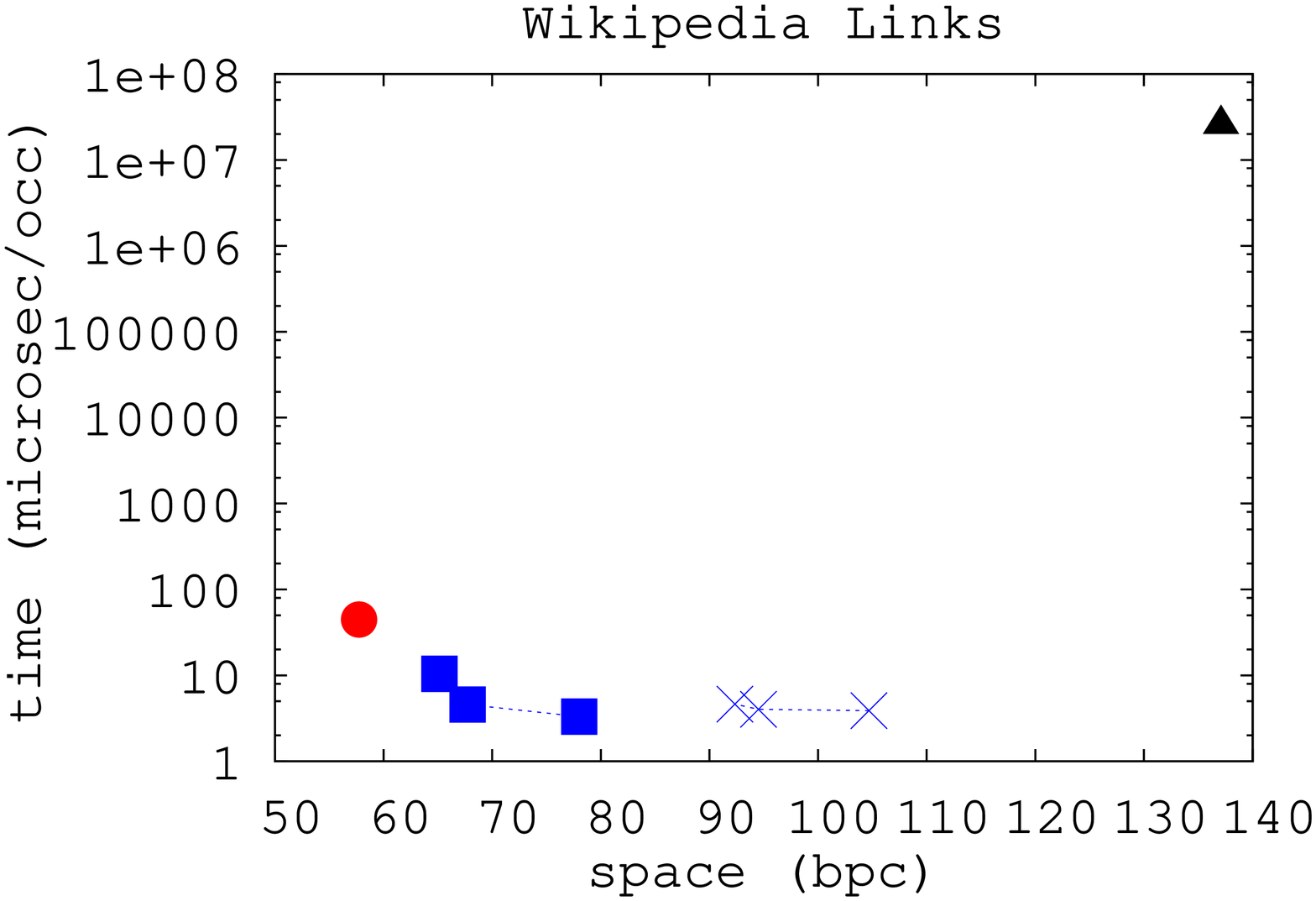}
\includegraphics[angle=-90,width=0.32\textwidth]{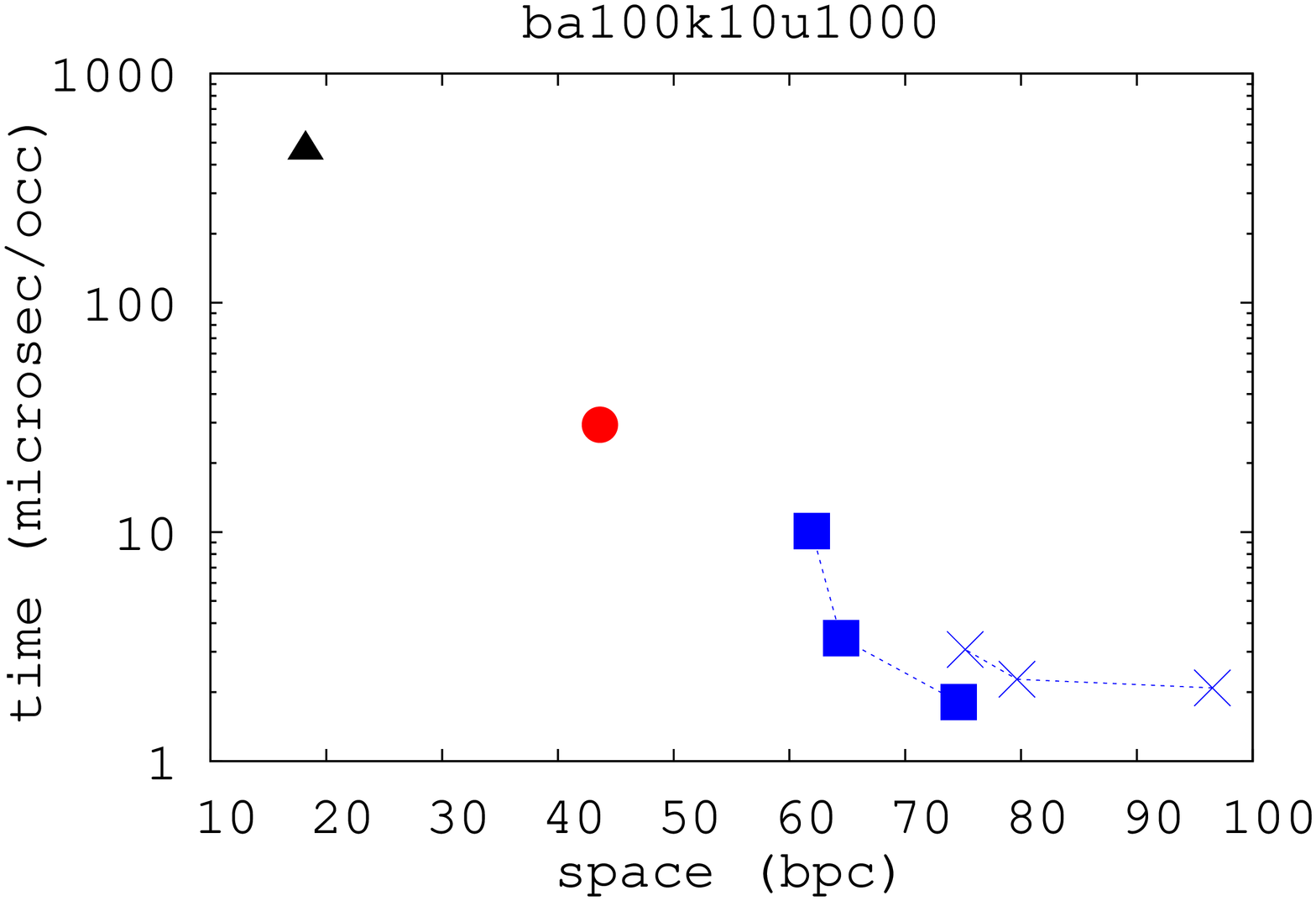}
\includegraphics[angle=-90,width=0.32\textwidth]{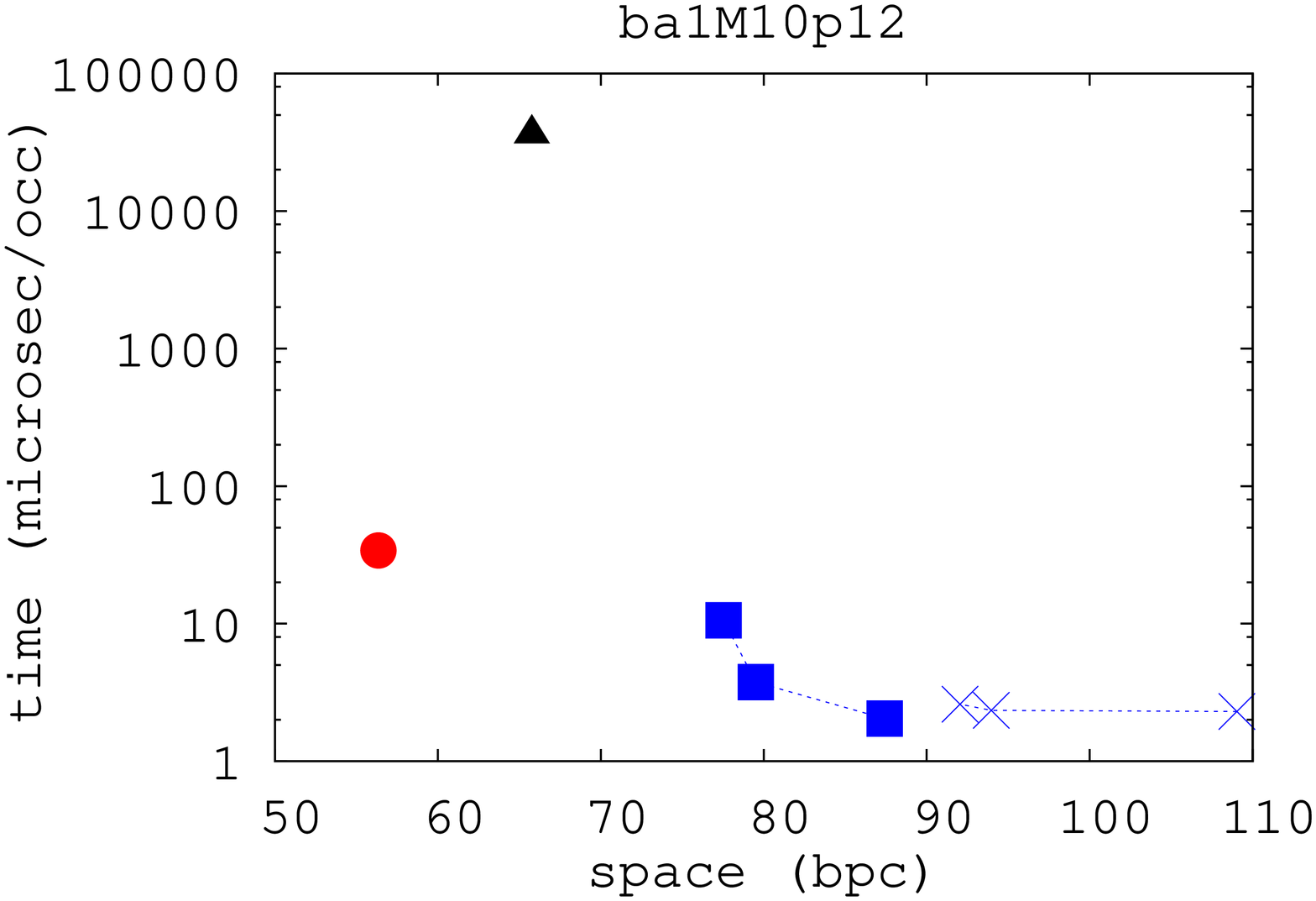}
\includegraphics[angle=-90,width=0.32\textwidth]{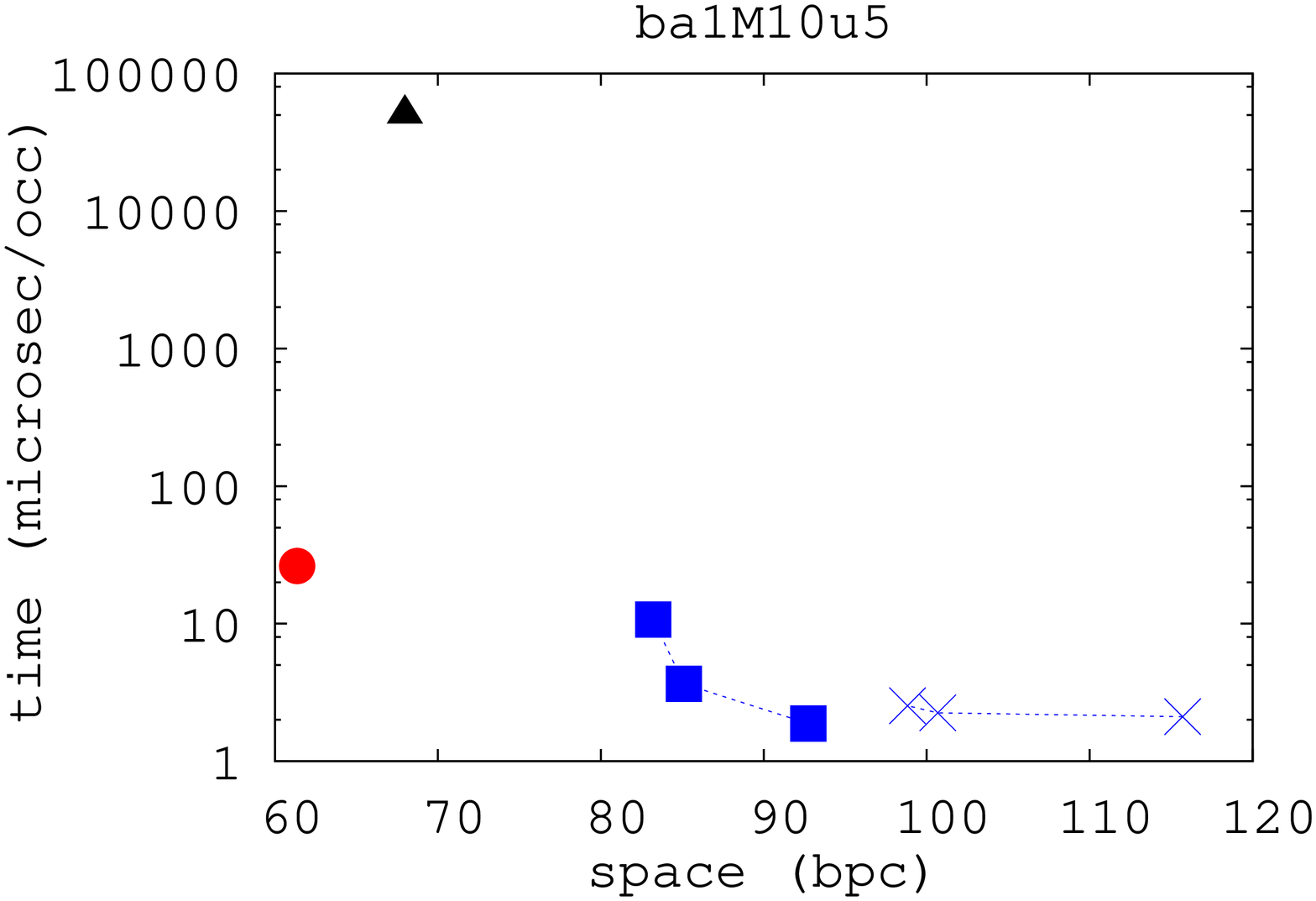}
\includegraphics[angle=-90,width=0.32\textwidth]{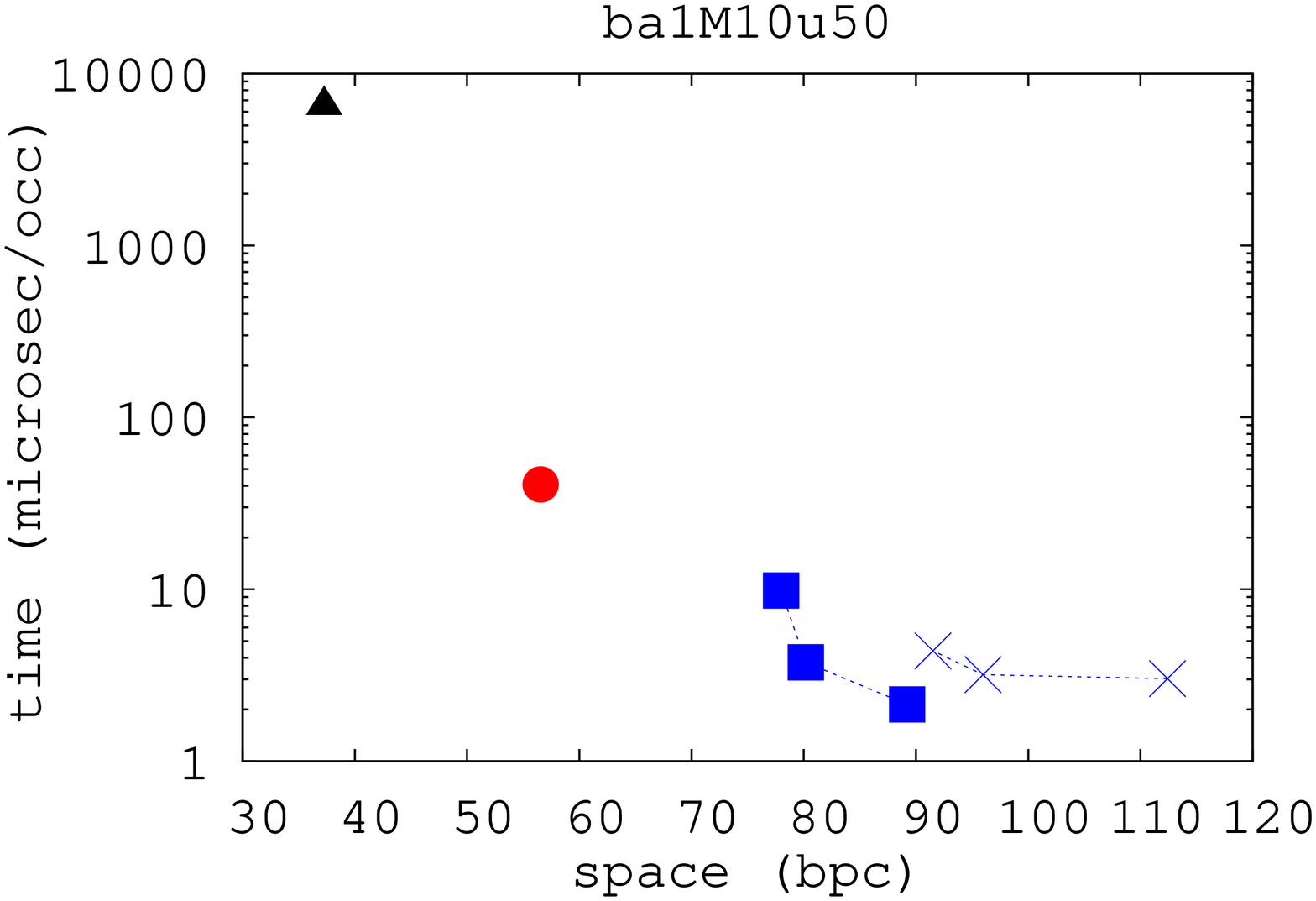}
\includegraphics[angle=-90,width=0.32\textwidth]{dirnei-9legend.eps}
  \end{center}
  \vspace{-0.3cm}
\caption{ Space/time trade-off for $\activedEdge$ operations.}

\label{fig:active}
\end{figure}

\begin{figure}[tbp]
\begin{center}

\includegraphics[angle=-90,width=0.32\textwidth]{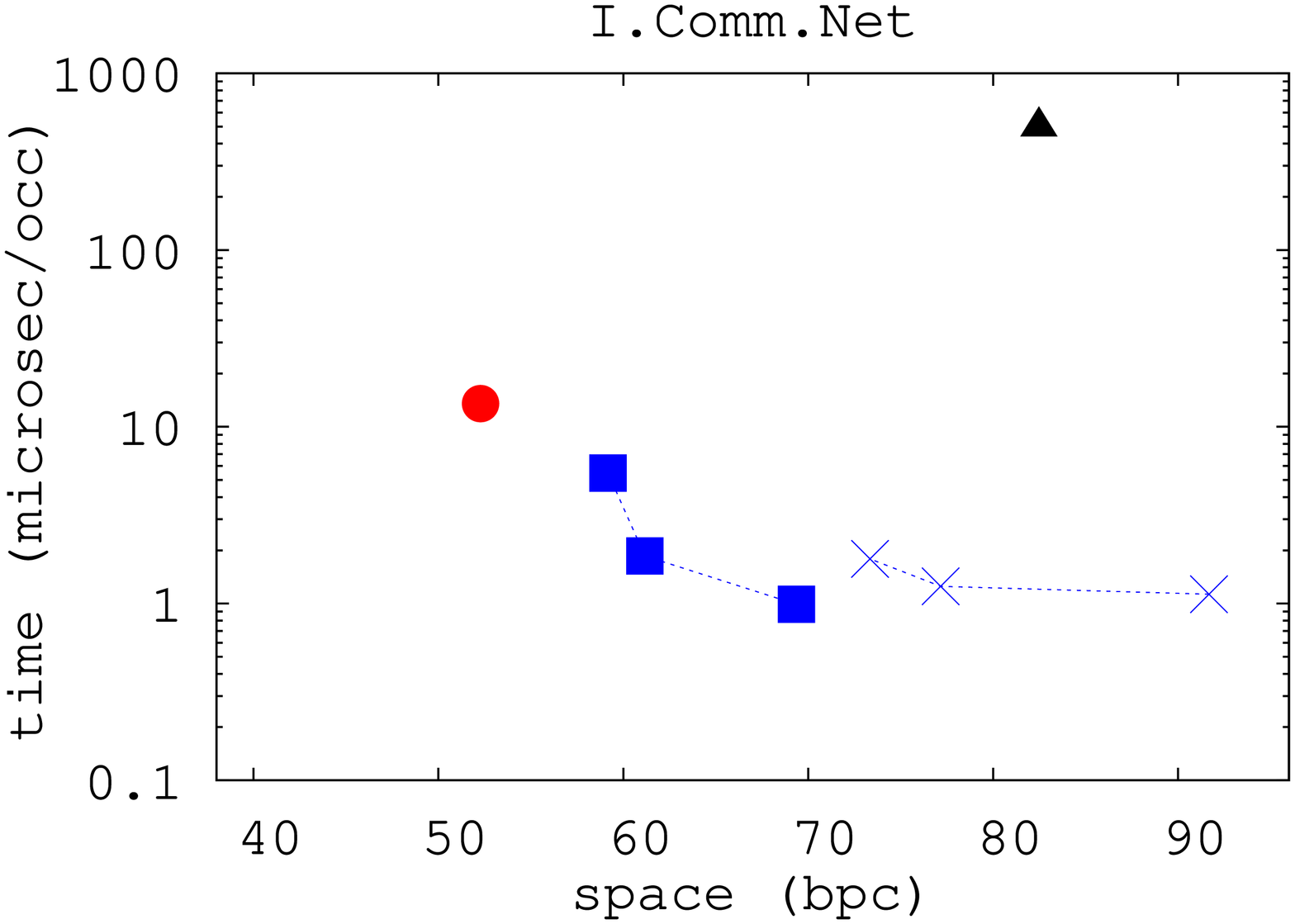}
\includegraphics[angle=-90,width=0.32\textwidth]{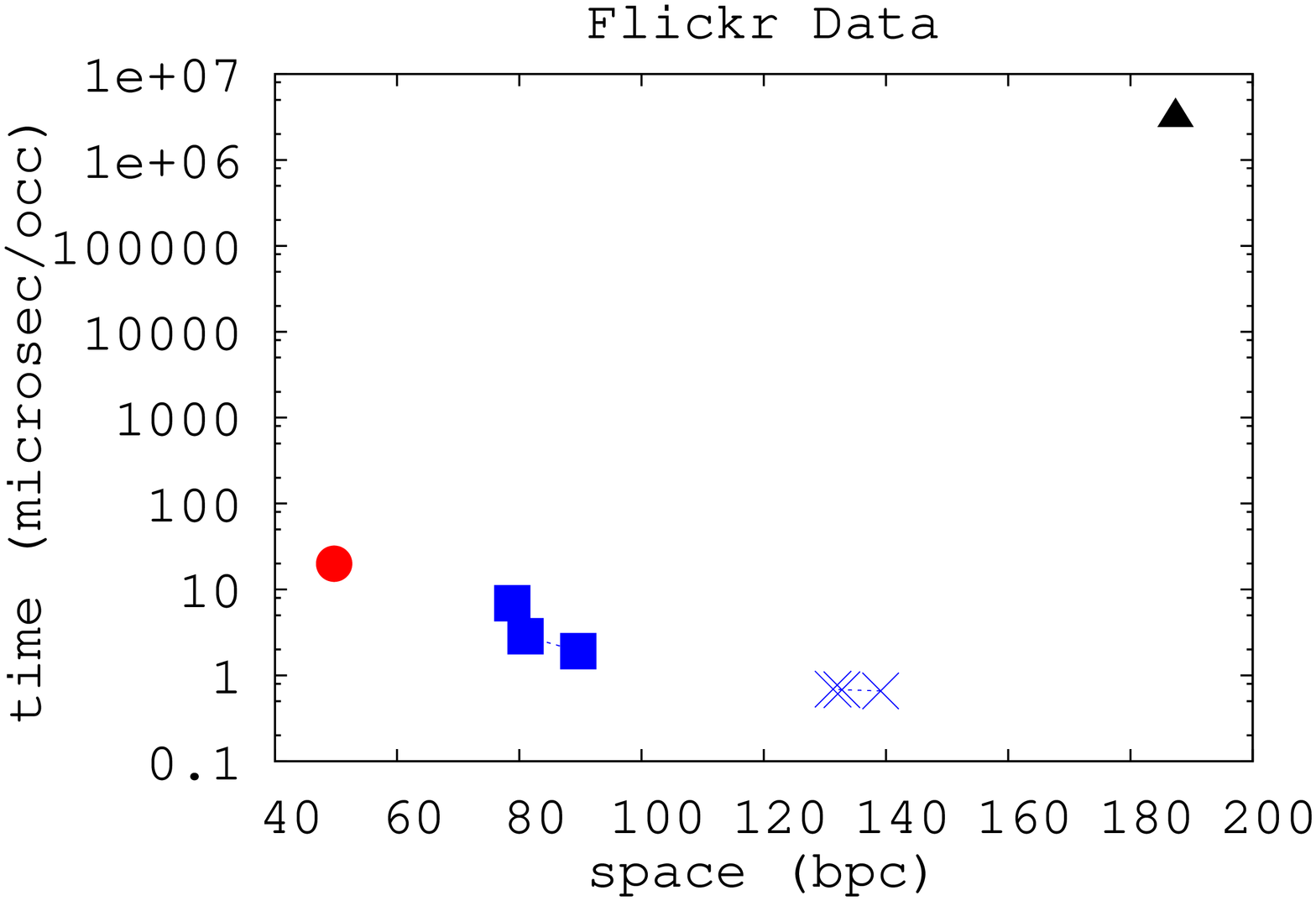}
\includegraphics[angle=-90,width=0.32\textwidth]{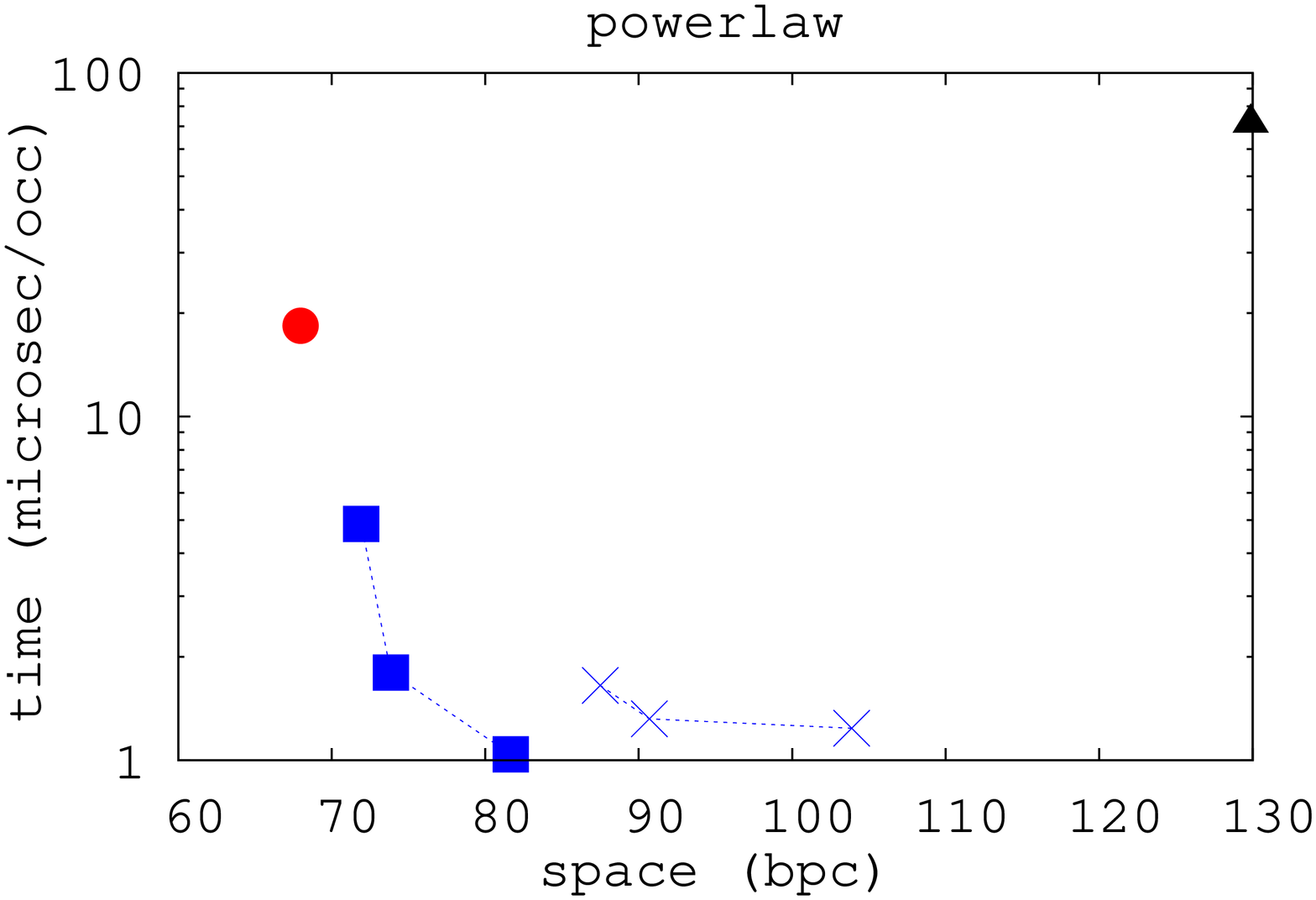}
\includegraphics[angle=-90,width=0.32\textwidth]{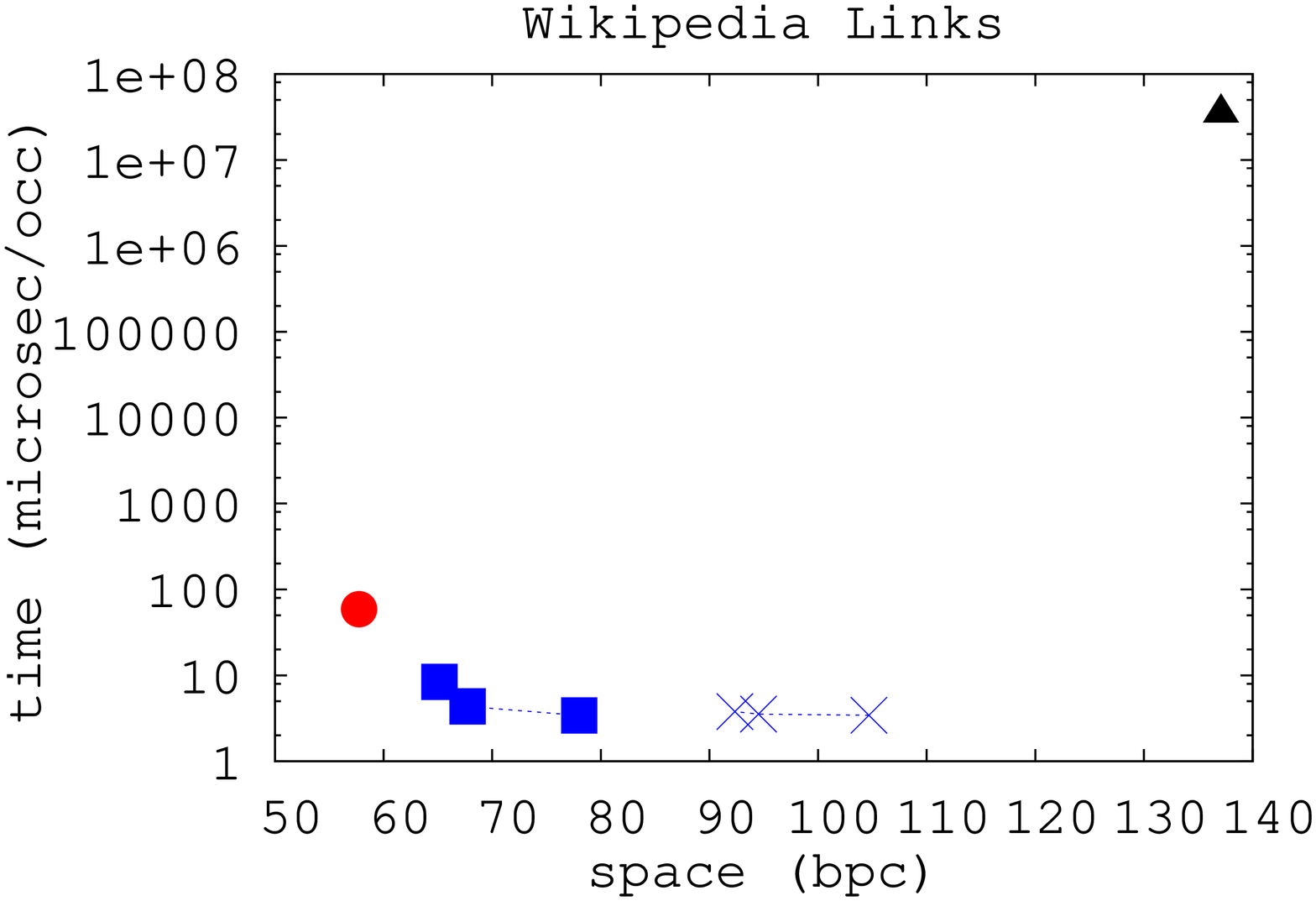}
\includegraphics[angle=-90,width=0.32\textwidth]{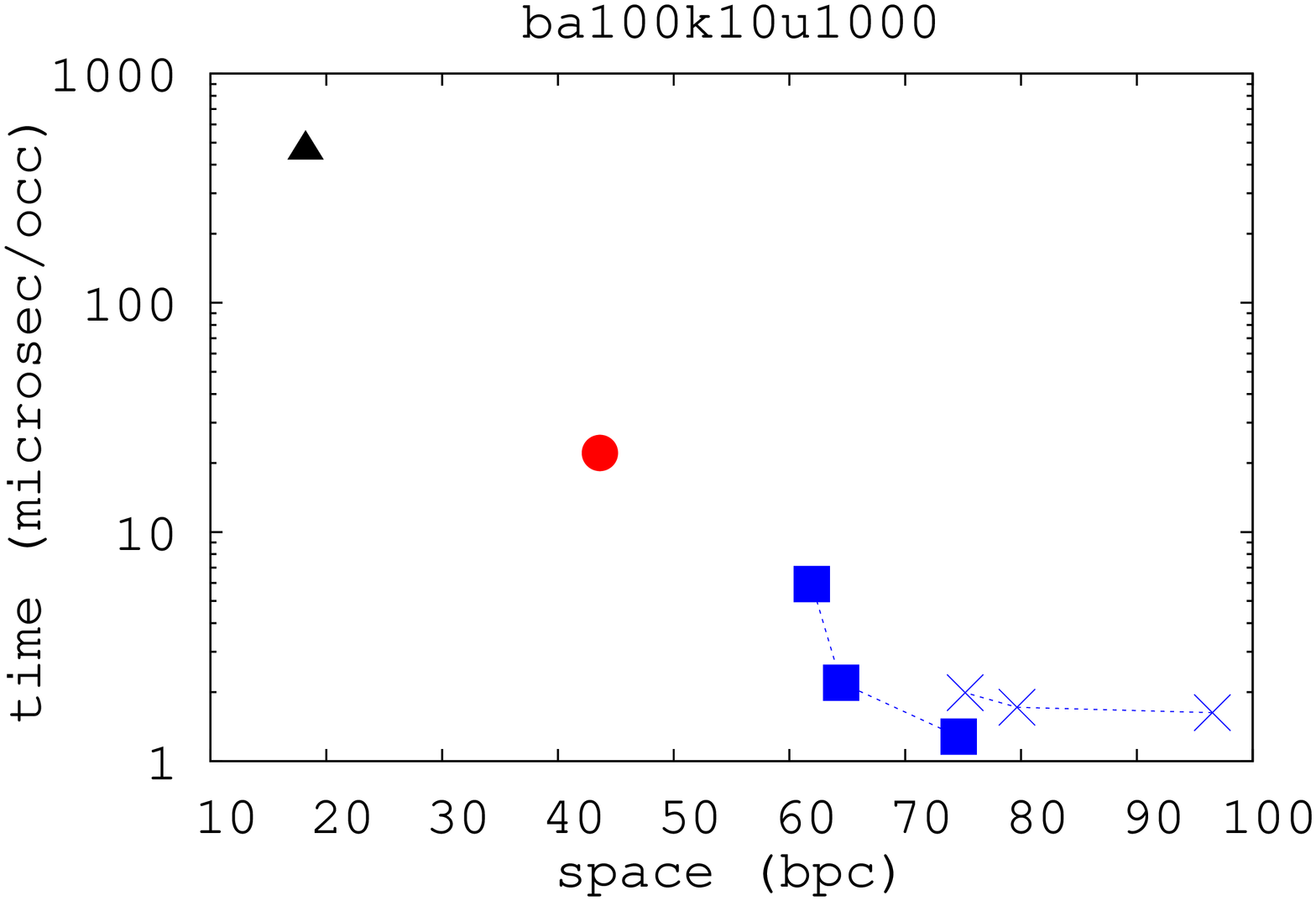}
\includegraphics[angle=-90,width=0.32\textwidth]{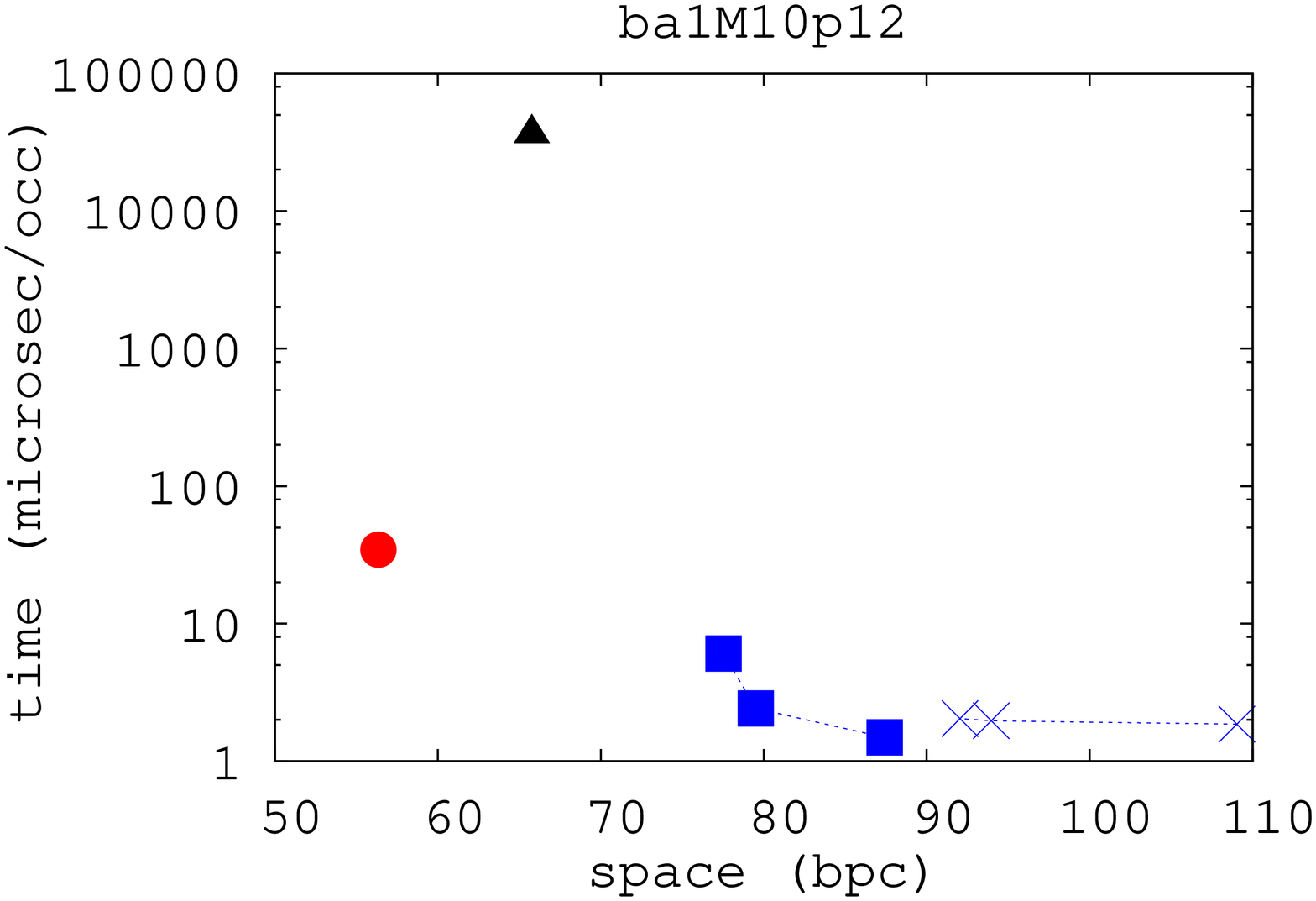}
\includegraphics[angle=-90,width=0.32\textwidth]{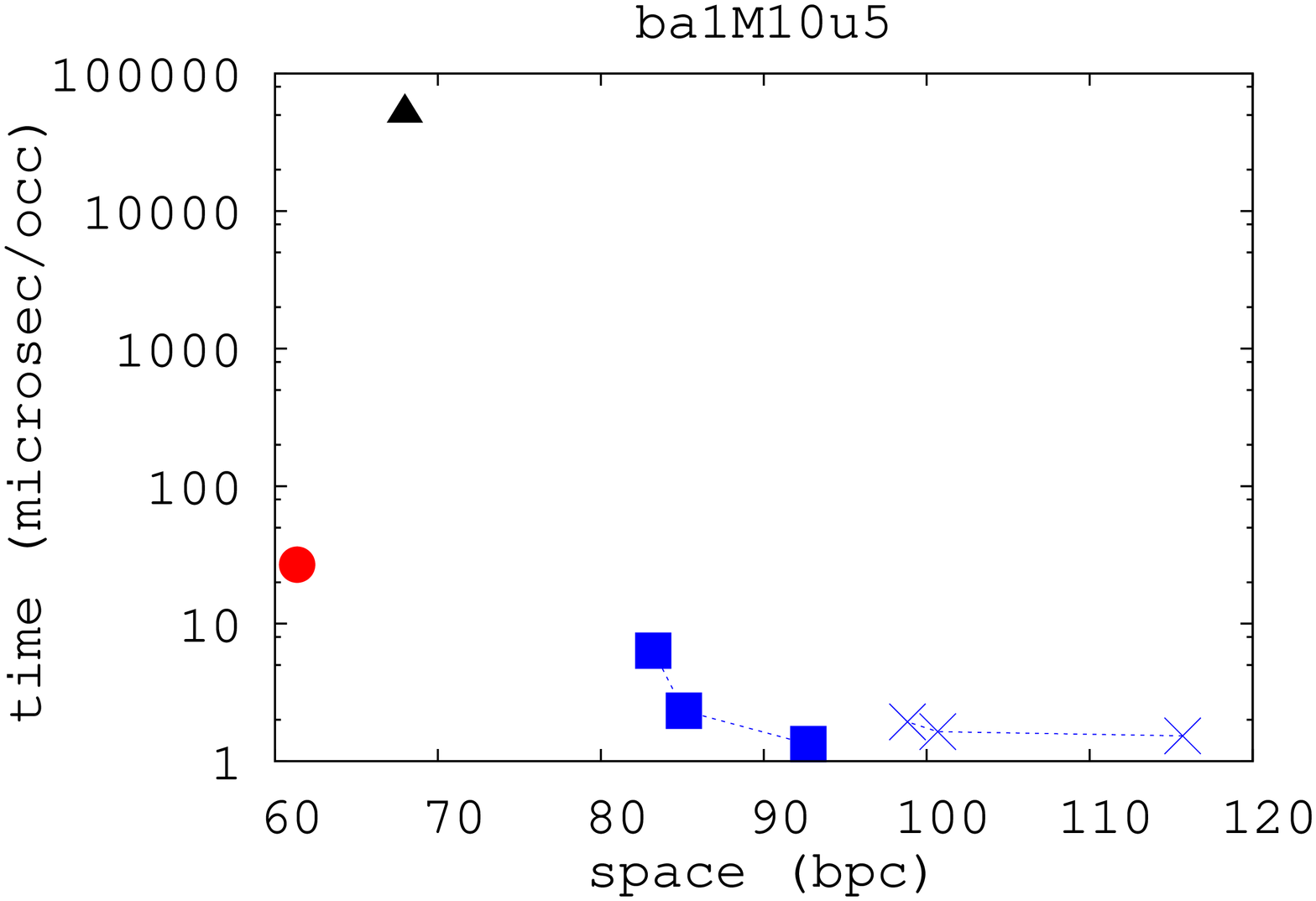}
\includegraphics[angle=-90,width=0.32\textwidth]{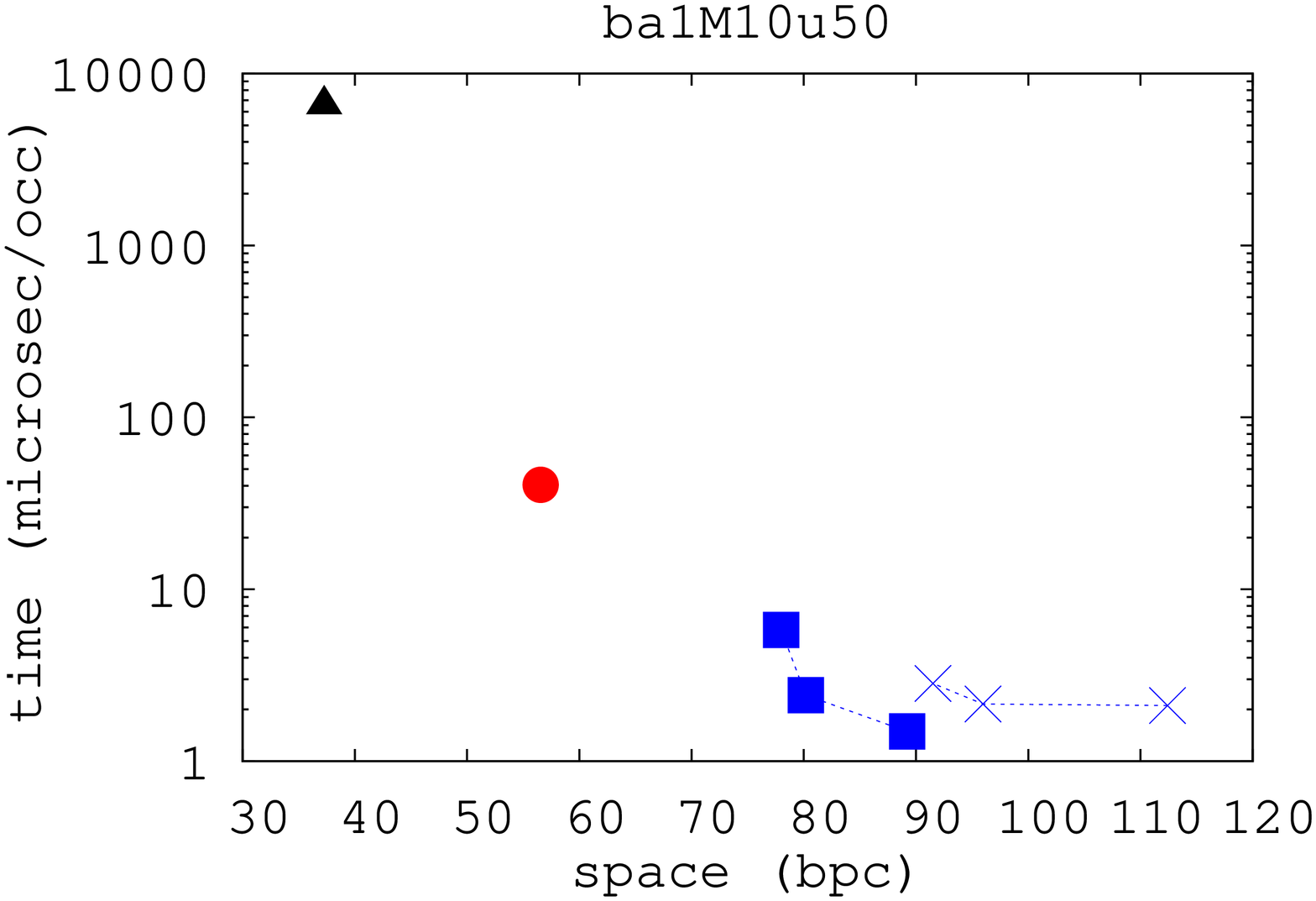}
\includegraphics[angle=-90,width=0.32\textwidth]{dirnei-9legend.eps} 
  \end{center}
  \vspace{-0.3cm}
\caption{ Space/time trade-off for $\deactivedEdge$ operations.}

\label{fig:deactive}
\end{figure}

Figures \ref{fig:active} and \ref{fig:deactive} show the results. We can see that these types of operations
are probably the best scenario for \tgcsa because they are solved by a single binary search to find the
given time instant. For example, in the case of $\deactivedEdge$  queries at time $t$,
the binary search returns an interval $[lt, rt]$ corresponding to all the contacts that are
deactivated at time $t$. Therefore, for each $i \in [lt,rt]$, we apply $\Psi$ circularly to recover the corresponding
source vertex ($u \leftarrow \Psi[i]$) and target vertex ($v \leftarrow \Psi\Psi[i]$). Similarly,
for $\activedEdge$ queries at time instant $t$,  we apply $\Psi$ circularly from a starting interval
within the third part of the suffix array in \tgcsa.


Note that the time per contact  reported of  \tgcsa for these operations is much better than for the $\directNeighbor$ and $\reverseNeighbor$ operations because now the traversal of the starting range and the application of $\Psi$  always
recover one contact.  For the $\directNeighbor$ and $\reverseNeighbor$ 
 operations, however, many checks (that implied applying
$\Psi$ to reach a starting or ending time instant) could discard a candidate contact and, consequently,
\tgcsa was doing unsuccessful work that increases the reported time per occurrence.

As expected, \tgcsa reports the best time performance for $\activedEdge$ and $\deactivedEdge$ operations. 
With the densest configuration, \tgcsa slightly overcomes \tgcsavb (being 0-40\% faster). Yet,
when we set $t_{\Psi} = 256$, \tgcsavb becomes around $2-4$ times faster than \tgcsa.

\cet\ still draws good results, yet it is is clearly overcome by \tgcsa. We can also
see that \edgelog\ is by far the slowest technique. Finally, it is interesting to note that in the
\texttt{Flickr-Data} graph, \tgcsavvt improves the times of \tgcsa by around one third in
$\activedEdge$ queries. This is clearly expectable because \tgcsa has to apply $\Psi$ three
times to recover the source and target vertexes of the edge, whereas \tgcsavvt requires only
two $\Psi$ applications.

\subsection{Time comparison: Snapshot operation}

We  studied the performance obtained when retrieving the set of all the active edges at a certain time instant ($\snapshot$ operation).
We compared the average retrieval time at five  instants of the lifetime of the temporal graphs:
the first and last ones, and those at the 25\%, 50\%, and 75\% of the lifetime in each graph. 
Table~\ref{tab:snapsh} provides the average number of active edges per time instant, that is, the expected output size.


\begin{table}[htbp]
\scriptsize
\centering
  \begin{tabular}{r||r|r|r|r|r}
  \hline  
  Timeline             &      0\%&      25\% &       50\% &        75\% &       100\%\\
  \hline    \hline  
  I.Comm.Net           &   19,997&    19,991 &     19,997 &      19,999 &      19,996\\
  Flickr-Data          &        2&    17,428 &  2,313,193 &  17,586,575 &  71,345,977\\
  Powerlaw             &2,914,527& 2,925,980 &  2,931,495 &   2,934,810 &   2,931,023\\
  Wikipedia-Links      &        1& 5,360,597 & 80,291,698 & 206,020,758 & 307,690,159\\
  ba100k10u1000        &   18,847&   470,948 &    470,824 &      18,786 &     470,061\\
  ba1M10p12            &       90& 4,121,832 &  4,866,245 &   4,121,871 &          95\\
  ba1M10u5             &       90& 4,864,776 &  4,866,275 &   4,863,160 &          94\\
  ba1M10u50            &      988& 4,866,241 &  4,866,821 &   4,866,351 &         937\\
  \hline
  \end{tabular}
\caption{Number of contacts reported at each instant of the timeline for $\snapshot$ operations per each temporal graph.}
\label{tab:snapsh}
\end{table}

\begin{figure}[tbp]
\begin{center}

\includegraphics[angle=-90,width=0.32\textwidth]{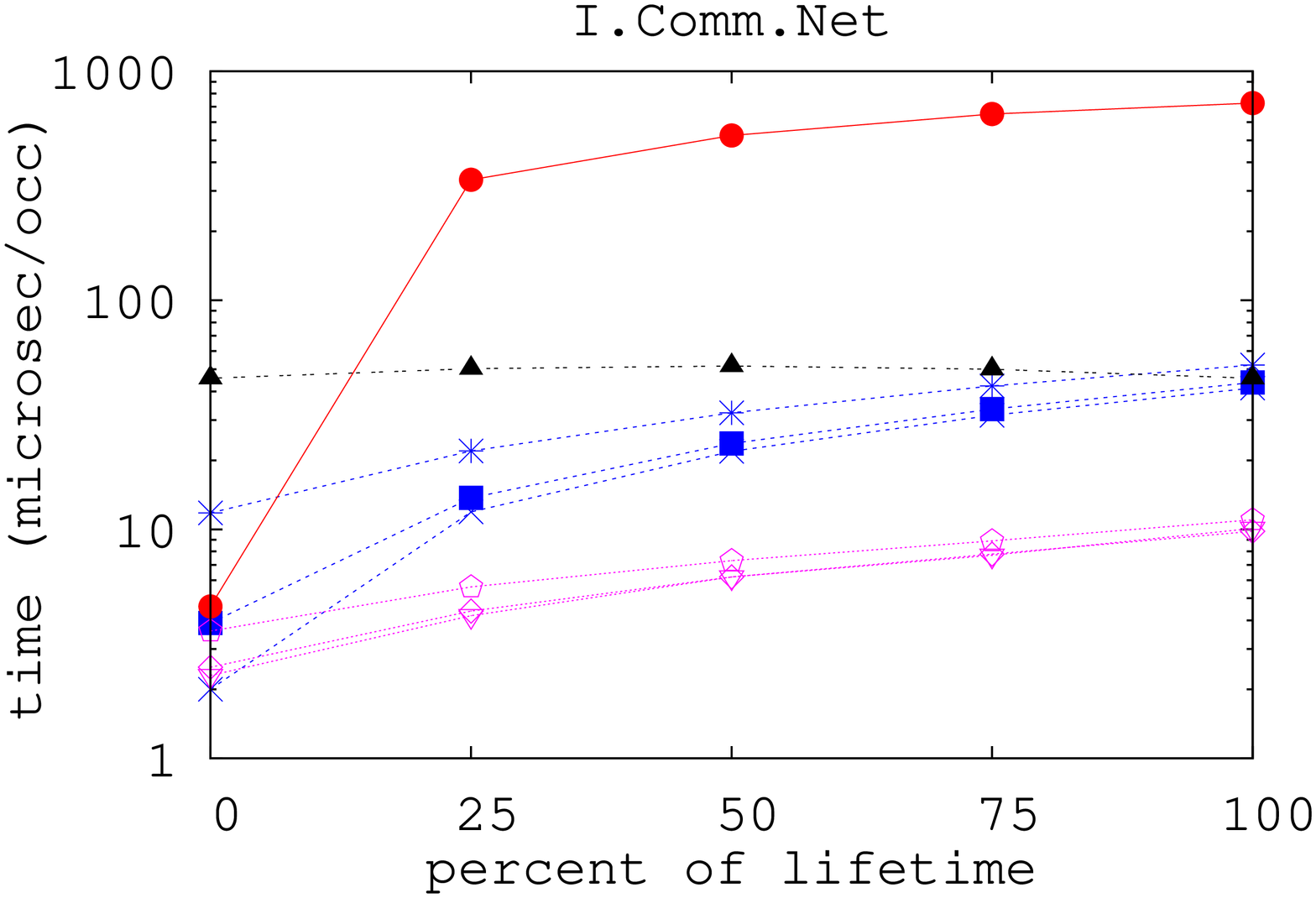}
\includegraphics[angle=-90,width=0.32\textwidth]{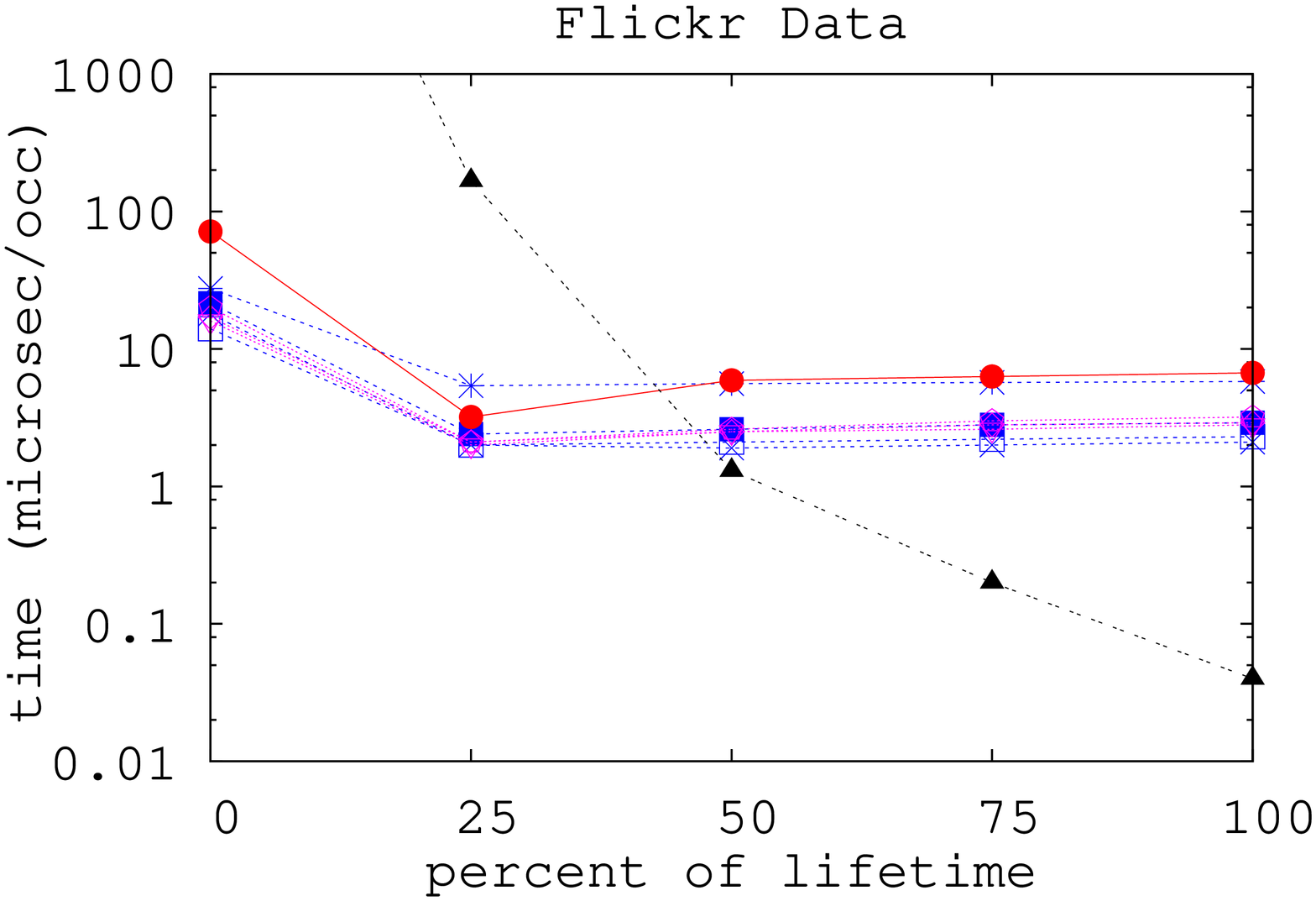}
\includegraphics[angle=-90,width=0.32\textwidth]{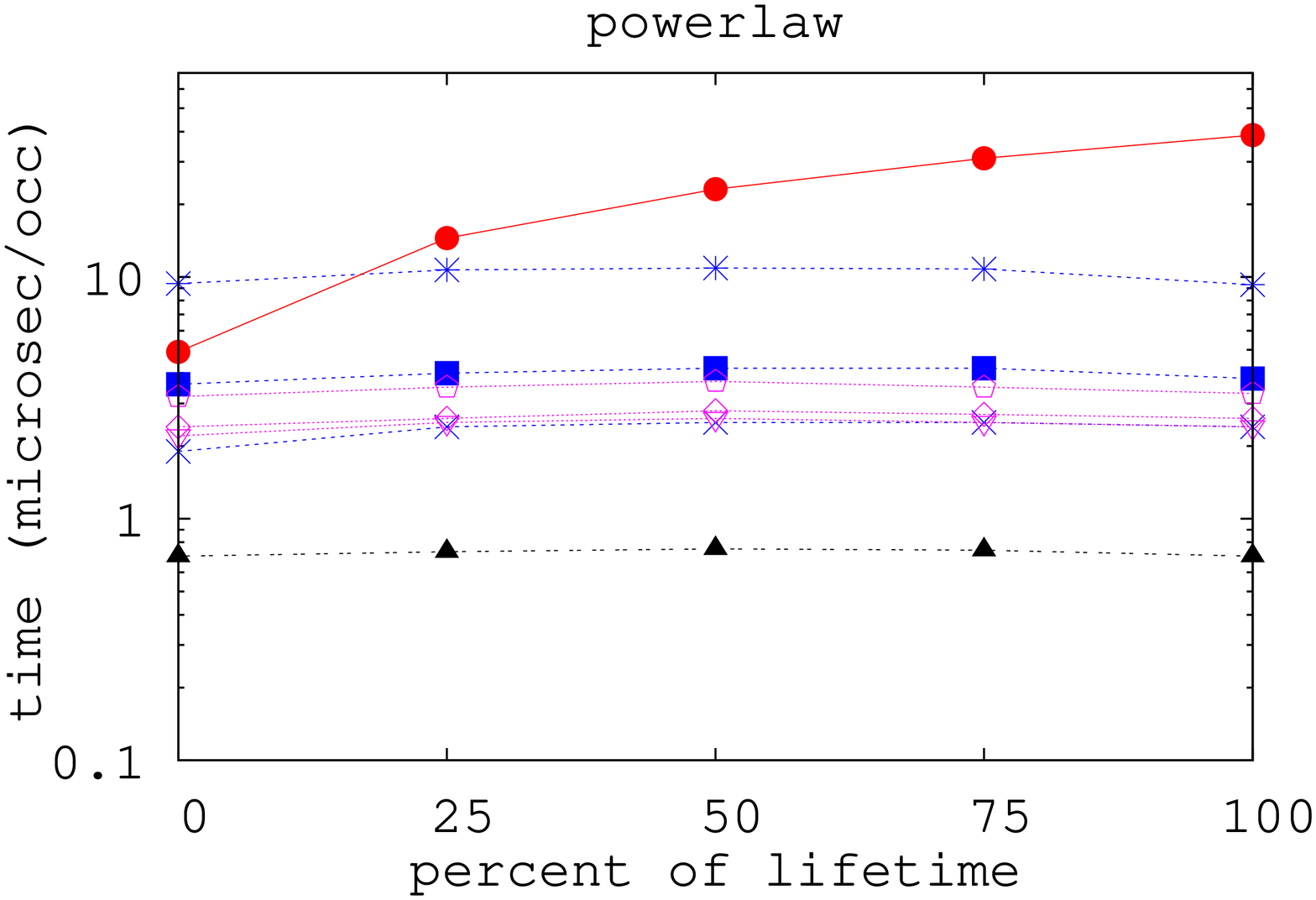}
\includegraphics[angle=-90,width=0.32\textwidth]{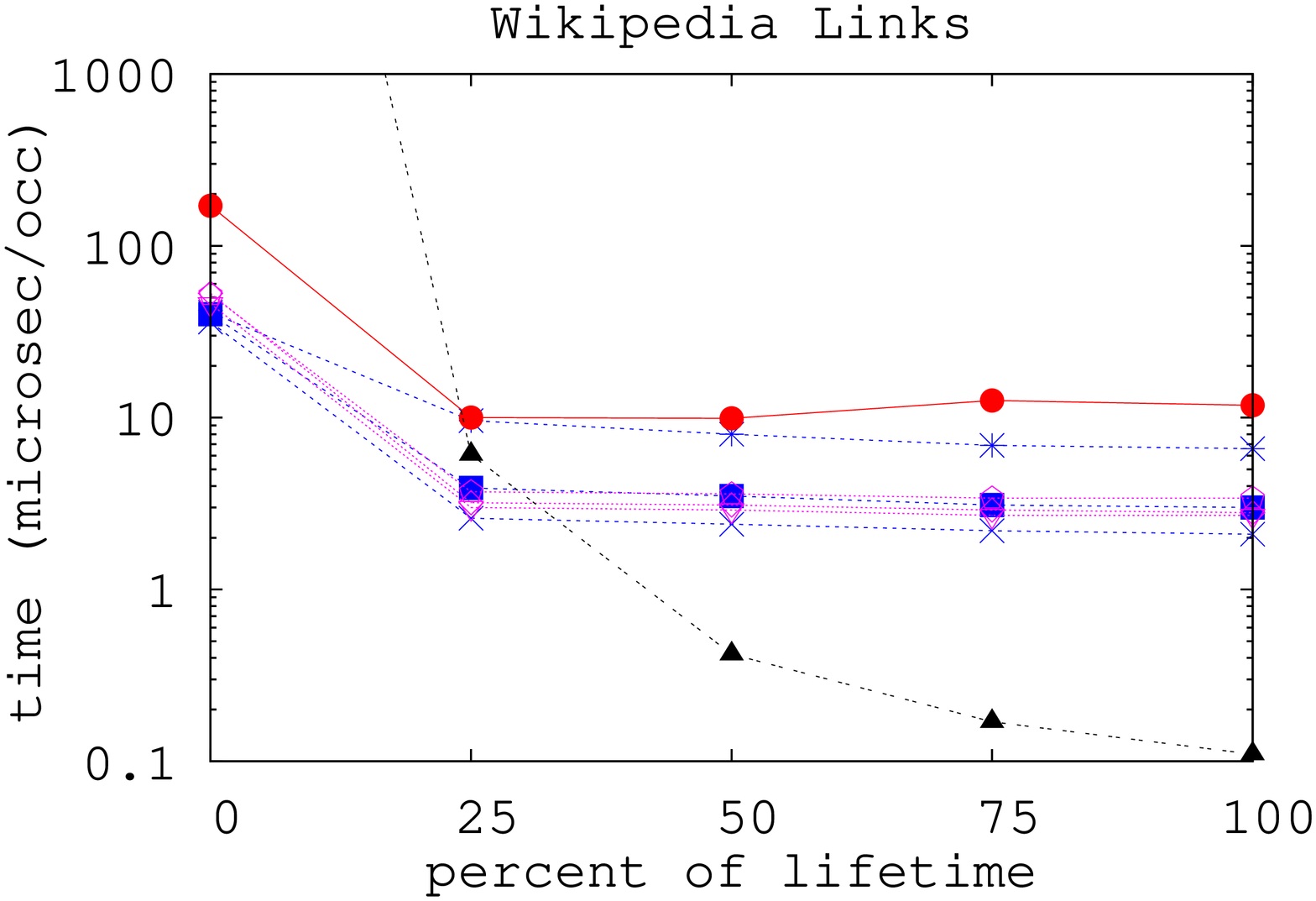}
\includegraphics[angle=-90,width=0.32\textwidth]{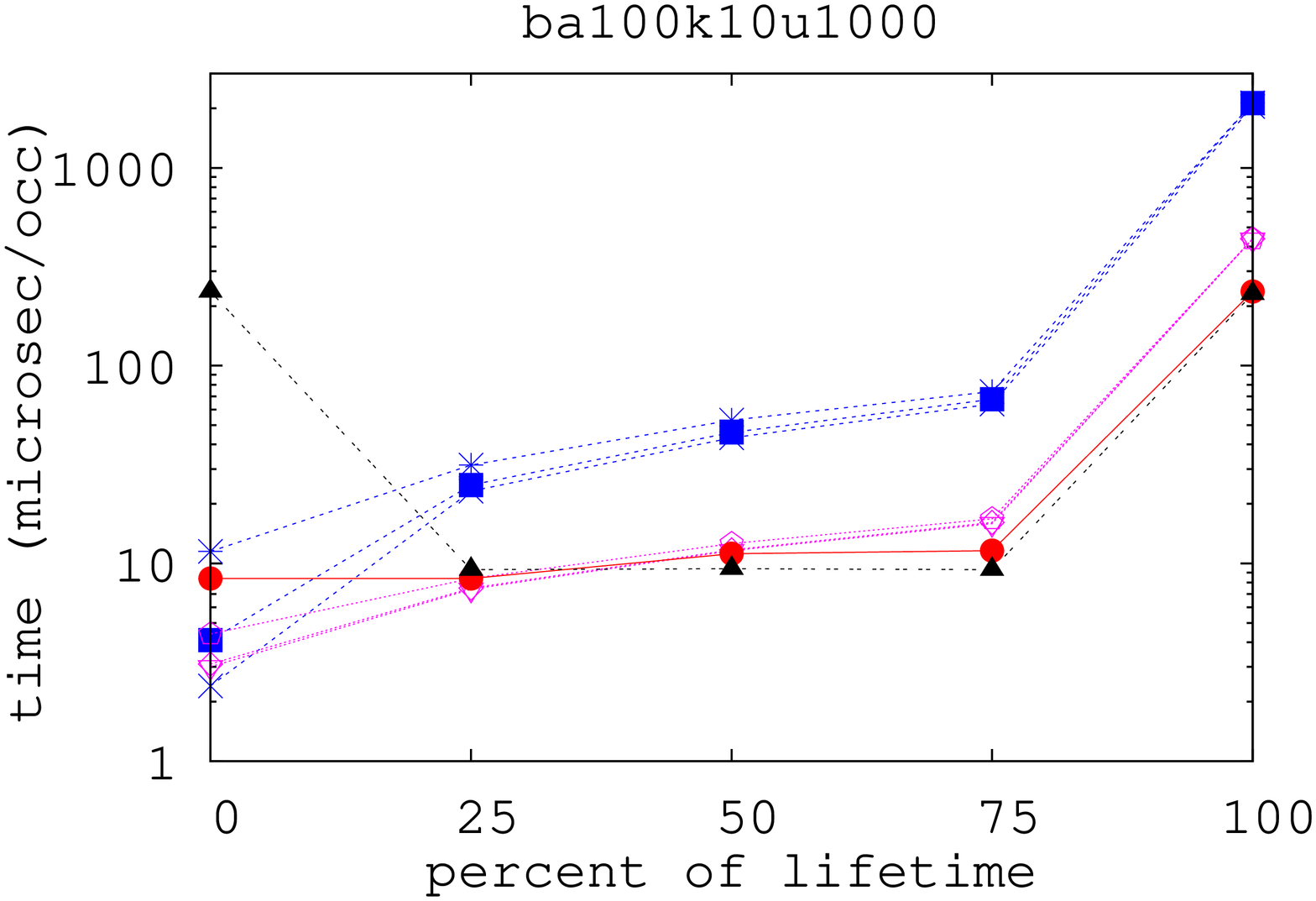}
\includegraphics[angle=-90,width=0.32\textwidth]{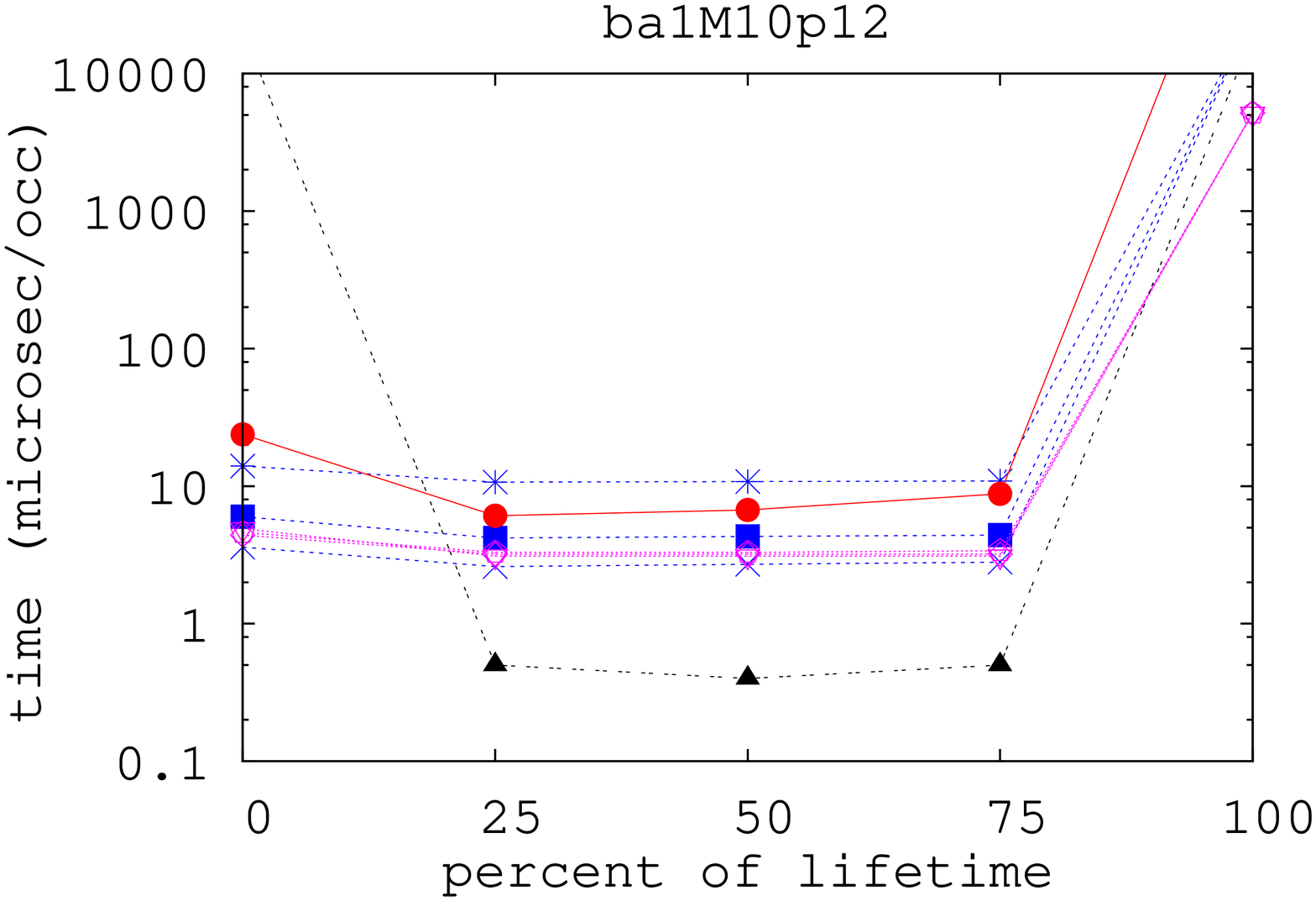}
\includegraphics[angle=-90,width=0.32\textwidth]{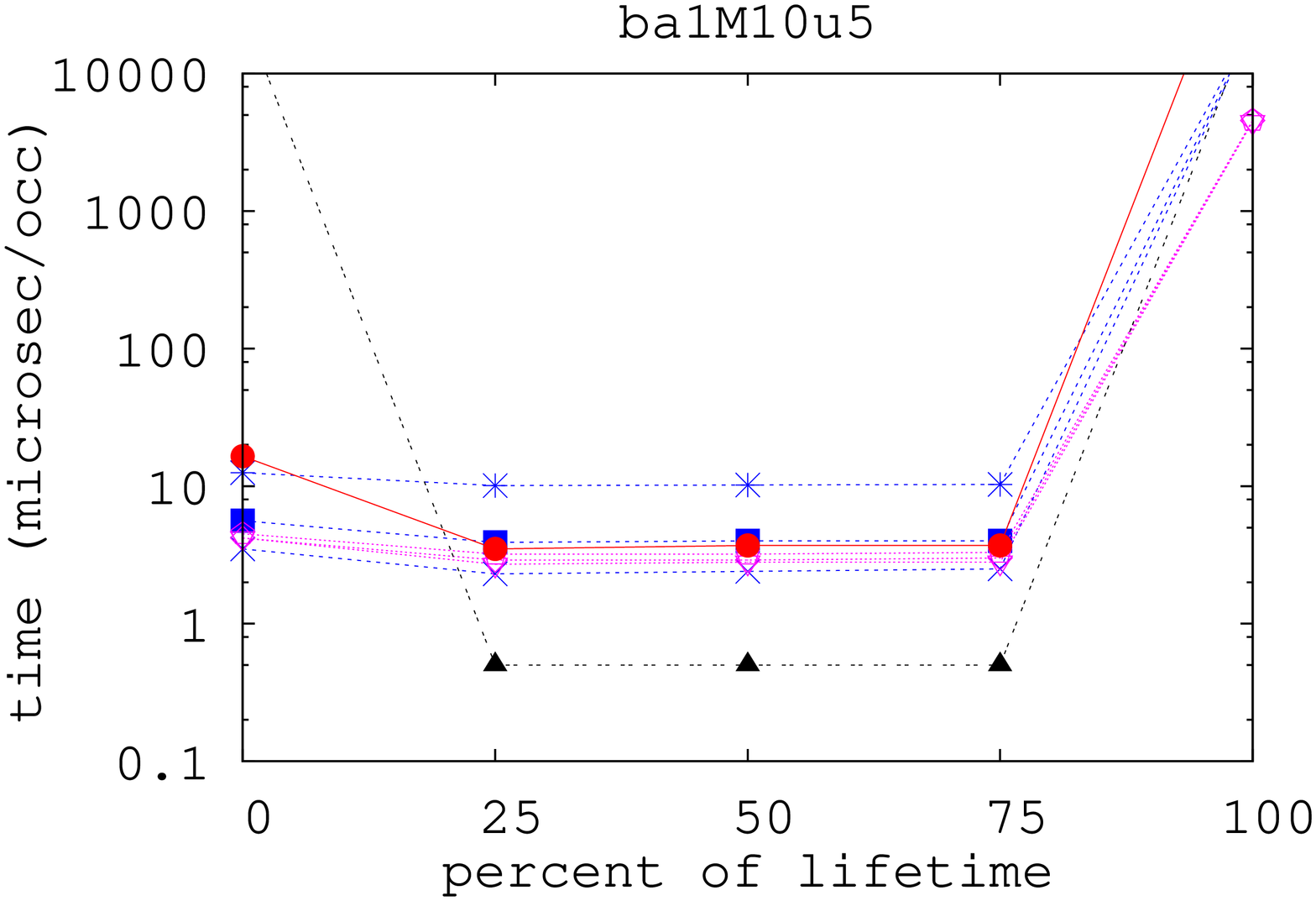}
\includegraphics[angle=-90,width=0.32\textwidth]{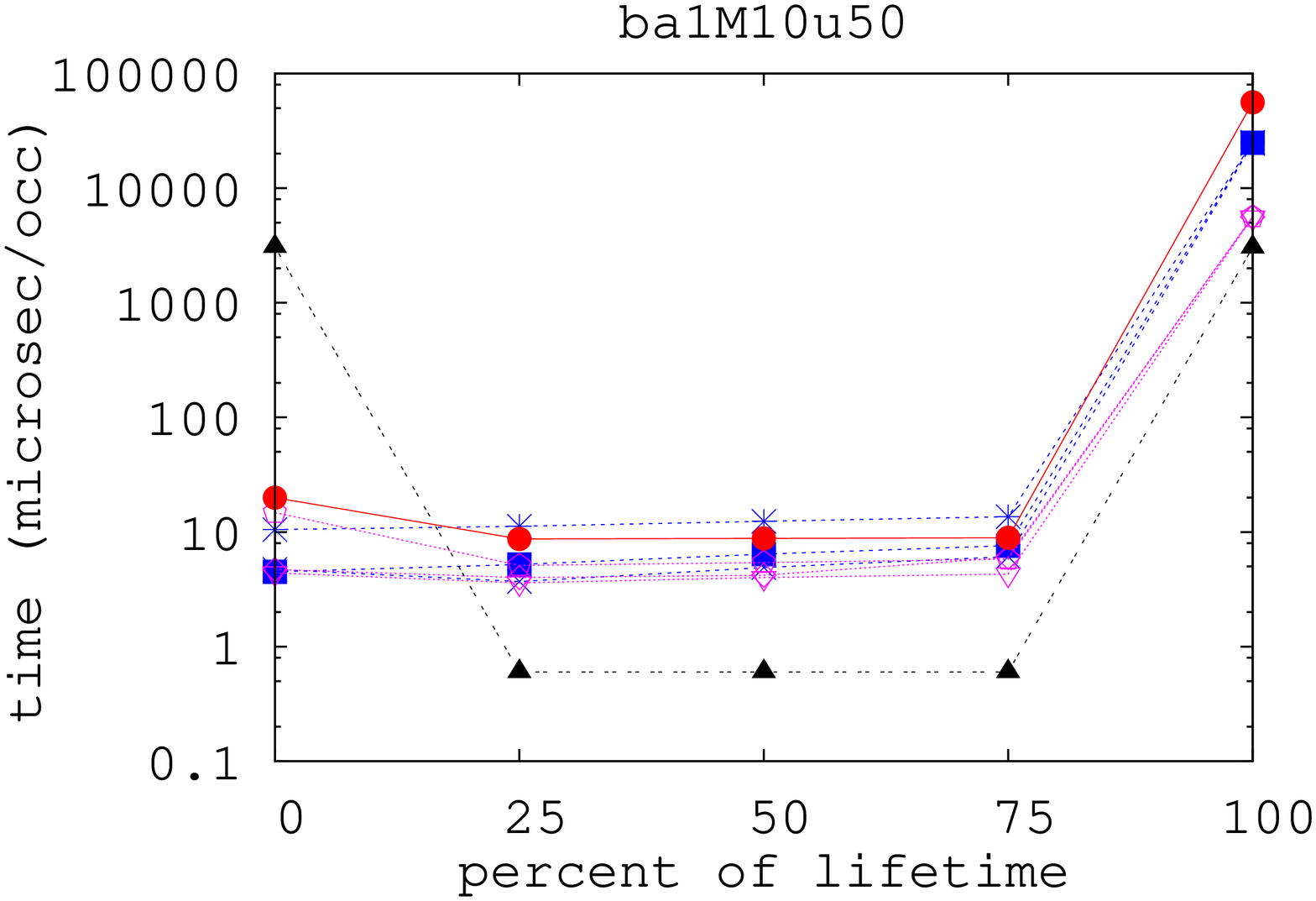}
\includegraphics[angle=-90,width=0.32\textwidth]{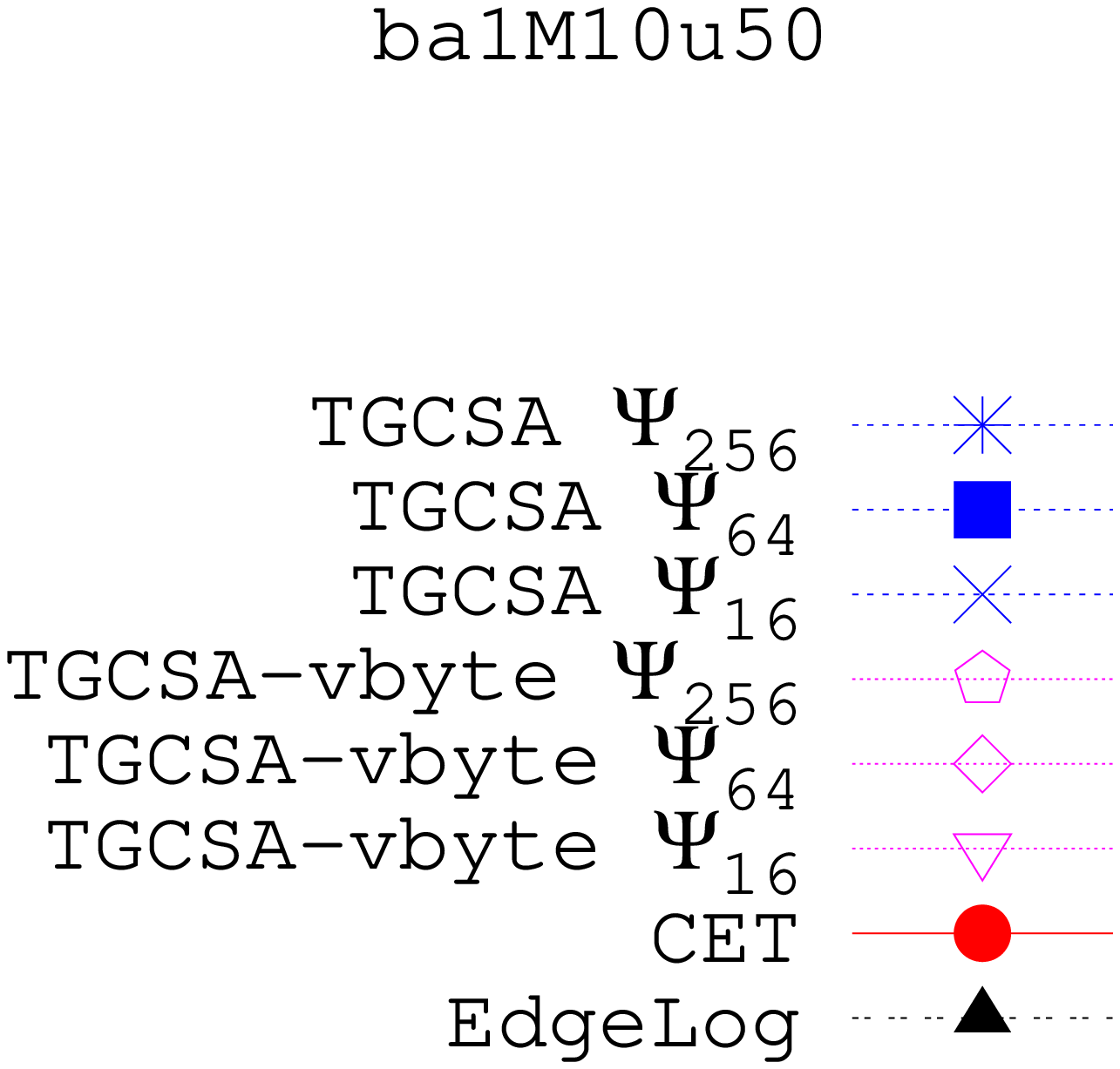} 
  \end{center}
  \vspace{-0.3cm}
\caption{Performance of $\snapshot$ operations at different time instants (percent of lifetime).}

\label{fig:snap}
\end{figure}

Note that  \edgelog\ computes the $\snapshot$  operations  with the application of 
$\directNeighbor$ queries over all the vertexes in the graph. \cet\ computes this operation as a
$\rangereport$ operation in the underlying Wavelet Matrix~\cite{CNOis14} and its cost
is logarithmic with respect to the total number of edges in the graph.
 \tgcsa, instead, must check which contacts match the time constraints
of the query for all the candidate contacts. As shown, this is done with a binary search to find the
ranges within the suffix array with possible both valid starting and ending time instants. That is followed
by a traversal of the valid starting times (buffered access to $\Psi$) to check if the end-time constraint is matched.
In that case, we recover the source and target vertexes with one and two applications of $\Psi$, respectively.

Figure~\ref{fig:snap} shows the results. The time measures are shown in $\mu s$ per edge reported.
 Overall, the  results show  that \tgcsa overcomes \cet\ in most cases and, in particular, in the non-synthetic
datasets. \tgcsavb draws also very good performance for  {\em snapshot} operations and, as expected, it excels in 
\texttt{ba100k10u1000} dataset due to its small vocabulary (few vertexes and short lifetime). This allows
\tgcsavb to exploit the faster sequential decoding of $\vbyte$ when compared with the $\huffrle$ that is
used in \tgcsa. 
Note that, in this particular dataset, where \cet\ clearly overcomes \tgcsa, now 
\tgcsavb is able to reach the same performance as \cet.


For these types of queries,  \edgelog\ has  a fast decoding of posting lists based on the use of 
$\pfordelta$, but it must traverse all these lists for each source vertex. This leads to a very fast $\snapshot$  performance
when the number of retrieved contacts is high, but it becomes very slow when we recover only a few contacts.

\no{
\subsubsection{Flexibility to support special queries.}
\tgcsa\ can give support to others query types that could be
interesting in some domains. In particular, those queries including
exact time instants or edges, can benefit of searching more than one
term in the initial binary search in the \tgcsa\. For example, in a
temporal graph representing phone calls from a given user to
another, starting and ending at a given time it could be interesting
queries such as: (i) {\em who phoned user $A$ exactly at time
$t_s$?}, or (ii) {\em who received a phone call from $B$ that
started at time $t_s$ and ended at $t_e$?}. They could be
implemented in the \tgcsa\ as an initial binary search for ($A \cdot
t_s$) and ($t_s \cdot t_e \cdot B$) respectively. Then for the
entries in the returned ranges $A[l,r]$ two or one accesses to
$\Psi$ respectively would be needed to retrieve the caller for the
first query, and the receiver in the latter one. Note also that, in
the second query, the initial binary search would report a unique
entry in $A$, hence the query is answered almost instantaneously.
}
%

\section{Conclusions and future work} \label{sec:conclusions}

We  presented \tgcsa, a new representation for temporal
graphs based on the well-known \csa. We showed how we can 
adapt the temporal graph so that it can be indexed with an
\icsa\ self-index. Then, we proposed a modification of the 
regular $\Psi$ structure in \icsa\ in such a way that it allows
us to move circularly from one term to the other within each contact.
This  modification solves queries using the \csa\ mechanism
to search for one or more terms of the contacts. This is both fast
and flexible.

In addition, we  explored a new way to increase the performance of \icsa\ based on
replacing its traditional $\huffrle$  compressed representation of $\Psi$ by
a new representation that we called $\vbyte$. To improve access to $\Psi$ values, 
our new technique uses byte-aligned codewords instead of bit-oriented Huffman 
(other traditional representations used delta and 
gamma codes, see~\cite{FBNCPR12} for more details). We  also avoided 
sampling $\Psi$ at regular intervals because it is done in traditional compressed representations of $\Psi$.
In our case, since many operations in  \tgcsa imply recovering a sequence of consecutive values
$\Psi[l_c, r_c]$ related to a given symbol $c$, we  sampled the starting positions of $\Psi$  ($\Psi[l_c]$)
for all the different symbols $c$. We  ran experiments that verified that
our new representation is typically much faster than $\huffrle$ when we want to retrieve
a buffer with consecutive values from $\Psi$. Yet, it is not so advantageous when accessing values
at random positions.  We created a variant of \tgcsa, named \tgcsavb, that
uses the $\vbyte$ approach to represent $\Psi$. \tgcsavb is up to $5$ times faster than \tgcsa
in some operations; however, it uses around $20$-$30$\% more space. Finally,
we also adapted \tgcsa to the particular case of temporal graphs where contacts have only
three terms (an edge is never deactivated). This is the particular case of the  \texttt{Flickr-Data} dataset.
The resulting variant (referred to as \tgcsavvt) improved the results of \tgcsa in both space and time.

The experimental results showed that \tgcsa behaves reasonably well
in space. In general, space needs are between $50$-$90$ bits per
contact. 
With respect to time performance, 
\tgcsa is very successful for queries that can filter out many contacts from the dataset with an
initial binary search in the \tgcsa. This avoids the need for sequentially checking a large number of contacts.

\no{
As expected the best trade-off between space and query performance
was obtained in the real  data set such as {\em Flicker-Days} or
{\em Wikipedia} because they can be considered \textit{Incremental
Temporal Graphs} and therefore the final temporal point of the
contacts do not need to be represented because it is always the
present time.
}

We  compared \tgcsa with \cet and \edgelog.
In $\directNeighbor$ and $\reverseNeighbor$ queries, \edgelog
is a hard rival because it is an inverted index designed to answer
$\directNeighbor$ queries in a very efficient way and it also uses
a reverse aggregated graph to support $\reverseNeighbor$ queries efficiently.
However,   even in this
case, \tgcsa solves most queries in less than 1 millisecond per
contact reported.  For queries about events (i.e., $\activedEdge$ or $\deactivedEdge$), in constrast,
\edgelog performs poorly and \tgcsa is clearly the fastest
alternative. With respect to \cet, we have shown that, even though \cet\ typically
uses less space than \tgcsa, it is also usually slower. In particular,
in $\activedEdge$ and $\deactivedEdge$ queries \cet\ is around one order of magnitude
slower than \tgcsa.

An important feature of \tgcsa is its expressive power. We can
use it to represent any set of contacts without any limitation. For example,
we could deal with  contacts of an edge with overlapping time intervals. Also, as it was indicated above,
the indexing capabilities of the \csa\ allow us to perform most operations
following the same structure: (i) performing an initial binary search in \csa\ to 
obtain one range (or more) $[l,r]$ corresponding either to the vertexes or the
times in the contacts, and (ii) for all the entries in such
range (each one corresponding to a different contact), we can apply $\Psi$ circularly
to either recover the other terms of the contacts, or to check a constraint about them.
\medskip

As future work, we consider that there are two interesting lines we would like to explore in the scope
of temporal graphs.
On the one hand, 
our new $\vbyte$ allows us to improve the
performance of previous $\Psi$ representations~\cite{FBNCPR12}, but it requires a large
amount of extra space. Likewise,  the variant $\vbyteselect$ uses less space but it
also shows  to be slower. Since $\Psi$ is the most
important structure in \tgcsa (it uses around 80-90\% of its space, and it is accessed profusely during 
searches), we still want to try other ways to represent  $\Psi$. On the other hand,
we are also interested in studying the applicability of
other self indexes to the scope of this paper. 

Finally, the variant of \csa\ shown in this paper is not only of interest in the field of 
temporal graphs, but it has also opened new opportunities for the application of suffix arrays
in other fields. For example, it has obtained very good results when representing
{\em RDF datasets}~\cite{BCFNspire15,Cerdeira-Pena16}. %
%
In the future we are also planning to study its applicability to represent other types of networks.
For example, we have obtained promising results when using a \csa-based
approach to represent trajectories of moving objects constrained to a network \cite{Brisaboa2016:TRAJ}. We would
expect that the flexibility of our approach could make it successful in other contexts.



\bibliographystyle{elsarticle-harv}



\bibliography{refs}

\end{document}